\RequirePackage{fix-cm}

\documentclass[numbook]{svjour3} 
\journalname{Empirical Software Engineering}
\counterwithin*{section}{part}
\usepackage[a4paper,bindingoffset=0mm,left=30mm,right=30mm,top=20mm,bottom=20mm]{geometry}

\usepackage{array}
\newcolumntype{L}{>{\arraybackslash}m{12cm}}

\newcolumntype{P}{>{\arraybackslash}m{6cm}}
\newcolumntype{C}{>{\centering\arraybackslash}m{12cm}}

\def\BibTeX{{\rm B\kern-.05em{\sc i\kern-.025em b}\kern-.08em
    T\kern-.1667em\lower.7ex\hbox{E}\kern-.125emX}}

\makeatletter
\renewcommand\paragraph{\@startsection{paragraph}{4}{\z@}%
  {0.5ex \@plus 1ex \@minus .2ex}%
  {-0.5em}%
  {\normalfont\normalsize\itshape}}
\makeatother

\usepackage{amsmath,amssymb,amsfonts} 
\usepackage{times}
\usepackage{url}
\usepackage{relsize}
\usepackage{lscape}
\usepackage{multirow}
\usepackage{ragged2e}
\usepackage{adjustbox, tabularx,booktabs}
\usepackage{dcolumn}
\usepackage{float}
\floatstyle{plaintop}
\usepackage{adjustbox}
\restylefloat{table}
\usepackage{graphicx}
\usepackage{rotating}

\usepackage{algorithm}
\usepackage{algpseudocode}
\usepackage{tcolorbox}

\usepackage[utf8]{inputenc}
\usepackage{listings}
\usepackage{xcolor}
\usepackage{paralist} 
\usepackage{verbatim} 
\usepackage{enumitem}
\usepackage{ifthen}
\usepackage{xspace}
\usepackage[table]{xcolor}
\usepackage{fancybox} 
\usepackage{soul}
\usepackage{balance}

\usepackage{longtable}
\usepackage[comma,authoryear]{natbib}
\usepackage{cleveref}
\usepackage{blindtext}
\usepackage{textcomp}
\usepackage{chngcntr}
\usepackage{framed}
\usepackage{mdframed}
\counterwithout{table}{section}
\counterwithout{figure}{section}
\counterwithout{equation}{section}

\tcbuselibrary{skins} 

\definecolor{custom-gray}{cmyk}{0, 0, 0, 0.7, 1.00}

\newtcolorbox{Summary}[2][]{
top=0.15in,
fonttitle=\bfseries,
colbacktitle=custom-gray,
colback=gray!5,
colframe=gray!40!black,
enhanced,
attach boxed title to top left={xshift=1.5em,yshift=-\tcboxedtitleheight/2},
boxed title style={size=small,colback=custom-gray},
drop shadow={black!50!white},
title=#2,#1}

\usepackage[multiple]{footmisc}

\algnewcommand\algorithmicforeach{\textbf{for each}}
\algdef{S}[FOR]{ForEach}[1]{\algorithmicforeach\ #1\ \algorithmicdo}

\def\Intensity{Intensity\xspace}
\def\intensity{intensity\xspace}

\def\coordinate{synchronize\xspace}
\def\coordinated{synchronized\xspace}

\def\Simultaneous{Simultaneous\xspace}

\def\RAP{Rare Synchronization Pattern\xspace}
\def\IAP{Intermittent Synchronization Pattern\xspace}
\def\FAP{Frequent Synchronization Pattern\xspace}
\def\DAP{Disperse Synchronization Pattern\xspace}
\def\SAP{Sparse Synchronization Pattern\xspace}
\def\DPAP{Dense Partial Synchronization Pattern\xspace}
\def\SDP{Sporadic Disjoint Synchronization Pattern\xspace}
\def\RDP{Rare Disjoint Synchronization Pattern\xspace}

\def\project{PTLM family\xspace}

\def\projects{PTLM families\xspace}

\def\model{PTLM\xspace}
\def\models{PTLMs\xspace}

\def\changes{commit change types\xspace}

\def\contributor{contributor\xspace}
\def\contributors{contributors\xspace}

\def\RQa{How do the changes in the commit activities of upstream and downstream repositories of PTLMs compare to each other?}
\def\RQb{What are the synchronization patterns of commit activities across PTLMs on upstream and downstream?}
\def\RQc{How are the synchronization patterns between upstream and downstream distributed across PTLM families?}

\titlerunning{Synchronization of PTLMs on Hugging Face and GitHub}

\begin{document}

\title{On the synchronization between Hugging Face pre-trained language models and their upstream GitHub repository}

\author{Adekunle Ajibode         \and
        Abdul Ali Bangash \and
        Oussama Ben Sghaier
        \and Bram Adams
        \and Ahmed E. Hassan}

\institute{Adekunle Ajibode \at
              School of Computing, Queen’s University, Kingston, ON, Canada\\
              \email{ajibode.a@queensu.ca}
           \and
           Abdul Ali Bangash \at
              SBASSE, Lahore University of Management Sciences, Lahore, Pakistan\\
              \email{abdulali@lums.edu.pk}
            \and
           Oussama Ben Sghaier \at
              School of Computing, Queen’s University, Kingston, ON, Canada\\
              \email{oussama.sghaier@queensu.ca}
            \and
           Bram Adams \at
               School of Computing, Queen’s University, Kingston, ON, Canada\\
              \email{bram.adams@queensu.ca}
              \and
           Ahmed E. Hassan \at
               School of Computing, Queen’s University, Kingston, ON, Canada\\
              \email{hassan@queensu.ca}}

\date{Received: date / Accepted: date}

\maketitle

\begin{abstract}
Pre-trained language models (PTLMs) have revolutionized the field of natural language processing (NLP), enabling significant advancements in tasks such as text generation and translation. Similar to software package management, which involves centralized version control and distributed consumption, PTLMs are trained using code and environment scripts hosted in an upstream repository (e.g., a GitHub (GH) repository), while the family of model variants trained from a given repository's scripts are distributed using dedicated downstream distribution platforms like Hugging Face (HF). Despite these similarities, coordinating development activities between GH and HF presents several challenges to avoid misaligned release timelines, inconsistent versioning practices, and other obstacles to seamless reuse of PTLM variants. To understand how commit activities are coordinated between these two platforms, we conducted an in-depth mixed-method study of 325 PTLM families consisting of 904 HF PTLM variants. Our analysis reveals that GH contributors typically make changes related to specifying the version of the model, improving code quality, performance optimization, and dependency management within the training scripts, while HF contributors make changes related to improving model descriptions, data set handling, and setup required for model inference. Furthermore, to understand the synchronization aspects of commit activities between GH and HF, we examined three dimensions of these activities—lag (delay), type of synchronization, and intensity—which together yielded eight distinct synchronization patterns. The prevalence of partially synchronized patterns, such as \emph{Disperse synchronization} and \emph{Sparse synchronization}, reveals structural disconnects in current cross-platform release practices. These patterns often result in isolated changes—where improvements or fixes made on one platform are never replicated on the other—and in some cases, indicate an abandonment of one repository in favor of the other. Such fragmentation risks exposing end users to incomplete, outdated, or behaviorally inconsistent models. Hence, recognizing these synchronization patterns is critical for improving oversight and traceability in PTLM release workflows.

\end{abstract}

\section{Introduction}\label{introduction}
In recent years, pre-trained language models (PTLMs) have become integral to the development of Natural Language Processing (NLP) systems, due to their effectiveness in improving performance and reducing the need for task-specific training. These models have driven significant progress in tasks such as text generation, translation, and sentiment analysis~\citep{min2023recent}. Beyond research, PTLMs now influence software engineering practices by powering intelligent applications like chatbots and code assistants, and automating tasks such as documentation generation, bug reporting, and code summarization \citep{hou2024large}. Their ability to capture the structure and semantics of human language makes them particularly effective for building intelligent systems from large text data~\citep{wang2022pre}. As a result, PTLMs have become foundational to modern AI development. Their distribution is facilitated by several public model repositories, including \textit{HF}\footnote{https://huggingface.co/models}, \textit{ONNX}\footnote{https://github.com/onnx/models}, \textit{PyTorch Hub}\footnote{https://pytorch.org/hub/}, \textit{Model-Zoo}\footnote{https://modelzoo.co/}, and \textit{Modelhub}\footnote{http://app.modelhub.ai/}, which enhance accessibility and promote the reuse of models across diverse applications~\citep{zhao2023survey}.

Among these platforms, HF stands out not only for hosting model weights but also for offering tools for fine-tuning and deployment~\citep{castano2024analyzing}, including metadata, tokenizer files, inference scripts, and documentation, which significantly simplify adoption and integration. In contrast, platforms like GH serve as the core infrastructure for collaborative development and version control, with repositories typically containing training code, preprocessing scripts, experiment configurations, and usage documentation. Yet, the interaction between both platforms is not straightforward, as our preliminary exploration in \Cref{methodology} reveals that multiple PTLMs—regardless of whether they share the same base model, such as BERT~\citep{devlin2018bert}, GPT~\citep{openai2023gpt}, and RoBERTa~\citep{liu2019roberta}—are often developed and maintained within a single GH repository. To reflect this structure, we introduce the term `PTLM family', which refers to a group of PTLMs managed within the same upstream GH repository.

Within a PTLM family, the roles of the upstream GH and downstream HF repositories are similar to those of open-source projects and Linux distributions, respectively, with the former performing core development activities such as creating features, resolving bugs, and implementing improvements while the latter phase focuses on packaging, distributing, and preparing models for deployment. In the context of PTLMs development, this lifecycle is part of a broader and more intricate machine learning (ML) supply chain, which encompasses not only software components but also the interdependencies among datasets, model reuse, fine-tuning, and other iterative adaptation processes that make modern ML systems increasingly complex and non-linear \citep{stalnaker2025ml}.

Despite the similarity with traditional open-source ecosystems, the coordination of development activities of PTLMs between GH as the upstream platform and HF as the downstream platform can present unique challenges. Coordination involves managing interdependencies between tasks, aligning goals, negotiating responsibilities, and maintaining shared understanding across individuals and teams \citep{kraut1995coordination}. Since coordination is a complex phenomenon, this study, as a first step, focuses specifically on the synchronization aspect of PTLM release activity coordination. Synchronization is an observable and measurable form of coordination, providing a lower bound on the overall coordination effort. Specifically, in this study, we refer synchronization as the simultaneous occurrence of related activities on both GH and HF. 

The existing problem on synchronizing related activities between upstream GH and downstream HF is not just theoretical; it manifests in real-world issues encountered in PTLM projects. One such example is the PTLM project cahya/bert-base-indonesian-522M\footnote{https://huggingface.co/cahya/bert-base-indonesian-522M}, which demonstrates this disconnect. The HF repository was created on June 23, 2020, while the corresponding GitHub repository\footnote{https://github.com/cahya-wirawan/indonesian-language-models} had been in existence since August 19, 2018. It is likely that the model files were initially transferred from GH. However, after June 23, 2020, activities ceased on HF until September 22, 2020, despite multiple updates occurring on GH throughout July and August. During this period, numerous commits on GH involved updates to the model card, changes to the base model in the codebase, corrections to model names, and the addition of a training script. When activities resumed on HF, only a few updates—mainly related to essential configuration files and converted model weight files—were replicated from GH. Consequently, some model information on HF remained outdated or incomplete, indicating a gap in consistency between the two platforms. This example illustrates how developers may update training code, configurations, or other components on GH that might alter the model's behavior, but fail to propagate these changes to HF. Such a disconnect can result in users accessing outdated or inconsistent model versions, thereby undermining reproducibility, trust, and deployment reliability. 

Despite the practical importance of synchronizing cross-platform development activities, this topic remains underexplored in the context of PTLMs, even as its dynamics differ significantly from those in traditional software engineering. For instance, ongoing community discussions on HF\footnote{\url{https://discuss.huggingface.co/t/github-repo-and-hugging-face-repo-sync/114697}}\footnote{\url{https://github.com/huggingface/huggingface_hub/issues/534}} reveal persistent challenges in synchronizing upstream GitHub and downstream HF repositories. Specifically, little is known about the types of changes reflected in commit activities across GH and HF in PTLM development, the timing patterns of such synchronization, and how these patterns vary and evolve across PTLM families of different ages. 

To address this gap, our study empirically investigates the nature, types, and frequency of changes made across HF and GH as well as how these activities are synchronized in time. This work presents the first systematic analysis of cross-platform commit activity synchronization between the two platforms. Our findings offer insights to help stakeholders understand current synchronization behaviors, identify existing gaps, and explore opportunities to improve model release workflows. By characterizing these practices, the study raises awareness among researchers and practitioners and lays the groundwork for future efforts toward more reliable and better-synchronized model development processes across platforms. Specifically, we address the following research questions:

\begin{itemize}
\item[\textbf{$RQ_1.$}] \textbf{\RQa}

\noindent \textit{\underline{Motivation}}: Understanding the relationship between commits on GH and HF—specifically whether a clear upstream-downstream dynamic exists between the two platforms, is important. To address this, we analyze the prevalent change topics in the commit activities of PTLMs on GH as the upstream and HF as the downstream, identify the dominant change types in commit activities across both platforms, and compare differences in the distribution of similarity in these change types across PTLM families. These insights can raise awareness among practitioners regarding the nature of changes they make and how GH changes can affect HF behavior. This understanding can help them design better methods to manage these changes, preparing them to enhance and maintain models more effectively, leveraging collaborative insights and iterative development practices.

\noindent \textit{\underline{Findings}}: Our analysis reveals that commit change topics of the upstream GH repositories, emphasize PTLM version specification, code quality, performance optimization, and dependency management. In contrast, commits of the downstream HF, focus on repository setup, model descriptions, dataset handling, and inference setup—reflecting the distinct roles these platforms play in the PTLM release pipeline. At a higher level of abstraction, GH commits center more on model structure, external documentation, and training infrastructure, while HF emphasizes external documentation, preprocessing, and project metadata. These differences are statistically significant and tend to evolve with model maturity. Furthermore, we find that the presence of cross-platform contributors is associated with a higher degree of similarity in change types between GH and HF repositories. This relationship appears to be influenced by model maturity and the extent of collaborative involvement.

\item[\textbf{$RQ_2.$}] \textbf{\RQb}

\noindent \textit{\underline{Motivation}}: Understanding how commit activities of PTLM families synchronize over time across GH and HF platforms is crucial for uncovering current synchronization practices and identifying potential gaps or inefficiencies. This understanding enables practitioners to reflect on existing challenges and consider improvements. To address this, we investigate the underlying characteristics that shape synchronization as well as different  synchronization patterns of activities between GH and HF. These insights can support the development of more efficient and aligned release workflows across platforms.

\noindent \textit{\underline{Findings}}: We identified \textit{lag} (the order in which PTLM family commit activities first appear between GH and HF), \textit{synchronization type} (how commit activities on GH and HF align over time) and \textit{Intensity} (the frequency and concentration of commit activities across ecosystems) as key elements characterizing commit activity synchronization between GH and HF within PTLM families. Based on these characteristics, we uncovered eight distinct synchronization patterns: \emph{Rare}, \emph{Intermittent}, \emph{Frequent}, \emph{Disperse}, \emph{Sparse}, \emph{Dense Partial}, \emph{Sporadic Disjoint}, and \emph{Rare Disjoint}. These patterns reflect varying degrees and types of synchronization between upstream (GH) and downstream (HF) activities, highlighting how existing practices often lack synchronization and may lead end users to access stale or outdated models.

\item[\textbf{$RQ_3.$}] \textbf{\RQc} 

\noindent \textit{\underline{Motivation}}: Understanding the distribution of change types within synchronization patterns and the variations of these synchronization patterns reveals how consistently updates are managed across platforms and highlights patterns of delay, alignment, and divergence between upstream and downstream activities. To address this, we first examine how different change types are distributed across synchronization patterns to determine whether the prevalent activities in each pattern differ from those observed in RQ1. This is important to understand if change types actually define the observed synchronization patterns. Furthermore, we investigate the prevalence of lag and synchronization patterns across PTLM families at different stages of maturity. We also analyze how long it typically takes for change types made on one platform to be reflected on the other, depending on the synchronization pattern and the maturity stage of the model family. This understanding can help open-source communities and end-users plan and implement more effective release and maintenance strategies as their PTLMs evolve.

\noindent\textit{\underline{Findings}}: Our results show that the most prevalent synchronization pattern is \textit{Disperse} (39.4\%), where activities on the two platforms occur with limited overlap and extended delay—often resulting in one platform continuing development while the other remains inactive. The distribution of change types across these synchronization patterns differ significantly, indicating that the synchronization pattern is correlated with the nature of changes made. We also observed that commit activities between platforms are often misaligned: on average, changes made on GH are reflected on HF after a lag of 15.82 days. Although contributor count correlates with increased activity across both platforms, it does not consistently lead to tightly synchronized updates, highlighting the complexity of managing cross-platform collaboration in evolving PTLM families.
\end{itemize}

Our findings highlight substantial variability in how commit activities are \coordinated between GH and HF, revealing inefficiencies in synchronizing model development and release. These synchronization gaps suggest that project maintainers often lack structured workflows to manage changes across platforms, relying instead on ad hoc\footnote{The dominance of \emph{Disperse Synchronization} and \emph{Rare Synchronization} patterns, combined with inconsistent lag times and limited cross-platform contributor overlap, suggests that synchronization between GH and HF is often handled in an ad hoc rather than systematic manner.} updates. The observed lag and divergence in change topics highlight the need for improved release practices. For instance, practitioners could benefit from: (1) GitHub Actions or CI/CD pipelines that automatically trigger model rebuilds on HF when training code changes on GH; (2) Contribution templates that prompt developers to check for cross-platform impacts when submitting changes; and (3) Explicit versioning policies that link model weights on HF to specific code commits on GH. Such concrete mechanisms would explicitly support synchronization across both development and deployment stages.

Importantly, our analysis of synchronization captures only the potential temporal co-occurrence of upstream and downstream changes, rather than tracking exact change propagation across platforms. As such, the measured lags and topic divergences represent a lower bound on the true coordination gap, suggesting even greater misalignment at deeper levels of the release process. While HF provides robust tools for inference integration and model packaging, there is limited support for the kind of upstream/downstream synchronization required for end-to-end PTLM releases. Practitioners could benefit from release engineering practices that unify GH training pipelines with HF deployment workflows—especially as projects mature and contributor engagement evolves. Specifically, our study provides the following contributions:
\begin{itemize}
    \item We pioneer the empirical study of the relationship between GH (as the upstream platform) and HF (as the downstream platform) in the context of pre-trained language model (PTLM) development.

    \item We identify key characteristics—such as lag, synchronization type, and Intensity—that shape eight distinct synchronization patterns in cross-platform model development. These patterns reveal that PTLM practitioners often rely on unstructured, ad hoc synchronization strategies. This points to the need for automated mechanisms to support more effective synchronization and release engineering workflows.

    \item We provide a publicly available dataset and replication package to support future research on cross-platform development synchronization and release practices within the PTLM ecosystem \citep{replication}.
\end{itemize}

This paper is structured as follows. \Cref{background} discusses key concepts such as pre-trained language models, GH as upstream and HF as downstream for PTLM management, coordination and synchronization of development activities in PTLMs, and related work. \Cref{methodology} outlines the study setup. \Cref{result} presents the findings of the research questions. \Cref{discussion} covers the study’s discussion and implications, while \Cref{ttv} addresses potential threats to validity. Finally, \Cref{conclusion} summarizes the study and outlines key directions for future research.

\section{Background and Related Work}\label{background}
\subsection{Pre-Trained Language Models}\label{pretrained}
Pre-trained large models are general-purpose models trained on large-scale datasets to learn transferable patterns for downstream tasks. Unlike traditional models built from scratch for specific goals, modern architectures employ pre-training to capture broad representations across domains such as vision, speech, and NLP. Their adaptability and performance stem from combining high-capacity architectures, large datasets, and refined training methods~\citep{mao2020survey}. PTLMs—such as BERT~\citep{devlin2018bert}, GPT~\citep{openai2023gpt}, and RoBERTa~\citep{liu2019roberta}—are trained on large textual corpora to support token prediction, sequence understanding, and semantic representation. They underpin NLP applications like text classification, translation, and question answering. This study focuses on the interaction between GitHub and Hugging Face during the development of these PTLMs.

\subsection{GitHub as Upstream and Hugging Face as Downstream for PTLM Management}

In software development, upstream refers to the entities or stakeholders responsible for ongoing development and maintenance of a software project in terms of source code, patches, and fixes~\citep{adrian}, while downstream projects depend on or reuse these assets. For example, (downstream) Linux distributions like Fedora or Debian build and distribute upstream open-source project as packages~\citep{lin2022upstream}. Similarly, HF provides access to models whose scripts and related assets are developed upstream.

GH and HF play complementary roles in PTLM development and distribution. GH typically hosts model source code, training scripts, datasets, configuration files, and supports collaborative practices such as version control, issue tracking, and pull requests~\citep{loeliger2012version, dabbish2012social}. HF focuses on distribution, fine-tuning, and deployment through APIs, model hubs, and libraries that enable reuse and adaptation. Based on observed artifacts and activities, we consider GH the upstream platform—where development and synchronization occur—and HF the downstream platform—where models are published and integrated into applications.

\begin{figure*}[t]
\centering
\includegraphics[width=1\textwidth]{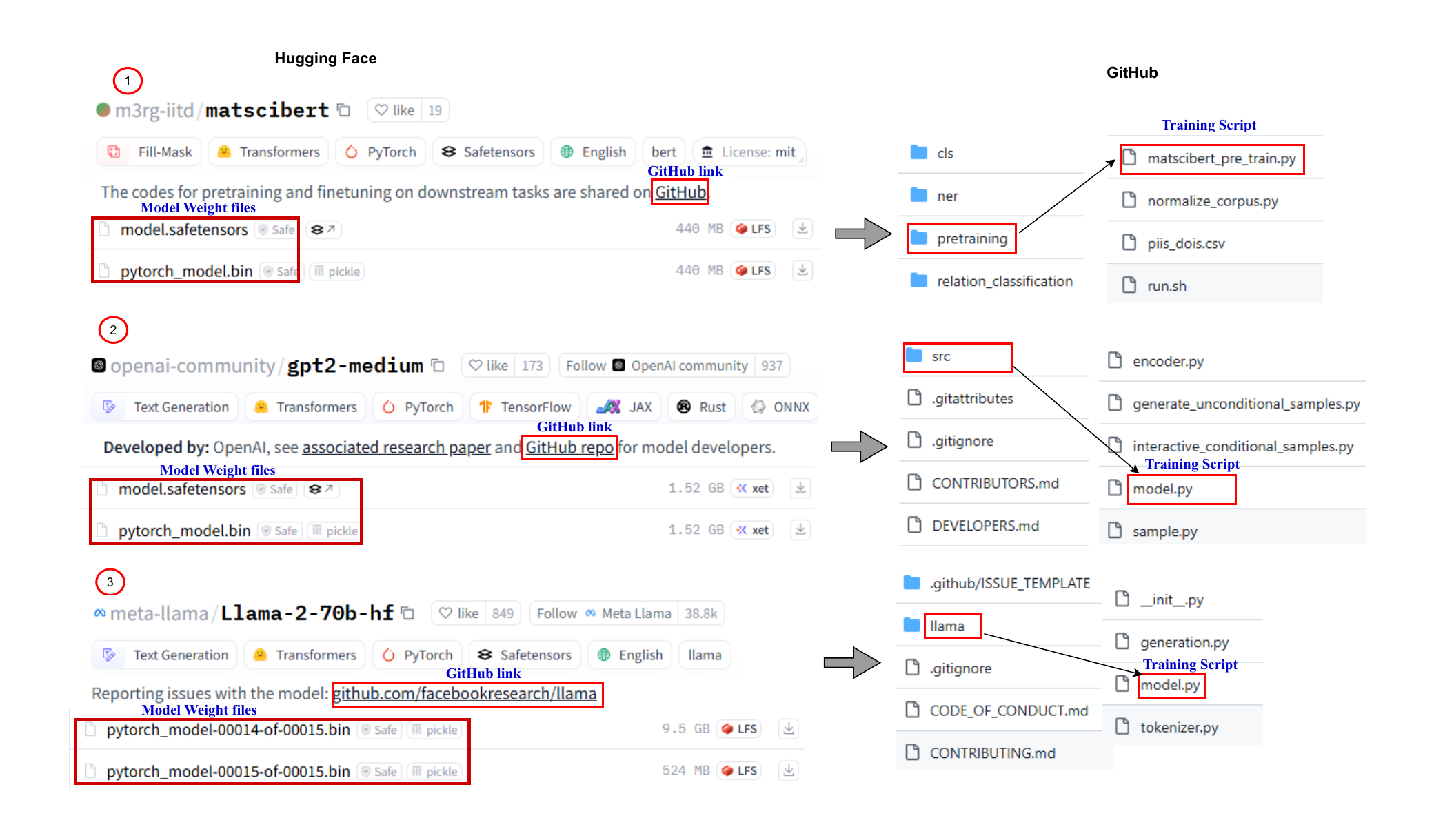}
\caption{Examples of PTLM GH repositories and their corresponding HF counterparts}
\label{updown}
\end{figure*}

Even though, as shown in \Cref{updown} many PTLMs link script, no off-the-shelf tool automates synchronization between GH and HF, leaving practitioners to manually coordinate updates—an issue noted in community discussions\footnote{\url{https://discuss.huggingface.co/t/github-repo-and-hugging-face-repo-sync/114697}}. For instance, m3rg-iitd/matscibert\footnote{\url{https://huggingface.co/m3rg-iitd/matscibert}} links to github.com/m3RG-IITD/MatSciBERT, openai-community/gpt2-medium\footnote{\url{https://huggingface.co/openai-community/gpt2-medium}} to github.com/openai/gpt-2, and meta-llama/Llama-2-70b-hf\footnote{\url{https://huggingface.co/meta-llama/Llama-2-70b-hf}} to github.com/facebookresearch/llama. Understanding these practices is essential to improving synchronization workflows. As shown in \Cref{updown}, many PTLMs link their HF repositories to corresponding GH repositories containing training code, datasets, and scripts.

Our study examines this upstream–downstream interplay to understand how PTLMs evolve across both environments, identify synchronization practices, and suggest opportunities to improve coordination between model development and distribution workflows.

\subsection{Coordination and Synchronization of Development Activities in PTLMs}

In software engineering, coordination refers to how individuals working on a shared project align efforts toward common goals—agreeing on software definitions, sharing information, integrating components efficiently, and avoiding redundancy~\citep{kraut1995coordination}. Studying coordination in PTLM ecosystems is challenging due to the diversity and scale of activities involved. PTLM variants extend beyond model code to include datasets, evaluation scripts, training configurations, and documentation. These interconnected components form a loosely coupled supply chain~\citep{wang2025large} spanning both model and dataset ecosystems, complicating the tracing of changes across platforms and artifacts.

As an initial step, we focus on a lower bound of coordination: \textit{synchronization}, which considers only the temporal co-occurrence of activities across platforms, regardless of content. Synchronization captures the time-based alignment of commit activities between GH and HF, serving as a proxy for whether updates occur within the same timeframe and in a coordinated manner. Effective synchronization supports model version consistency, enables seamless collaboration, and ensures that HF users access timely improvements and fixes.

\begin{figure*}[t]
\centering
\includegraphics[width=\textwidth]{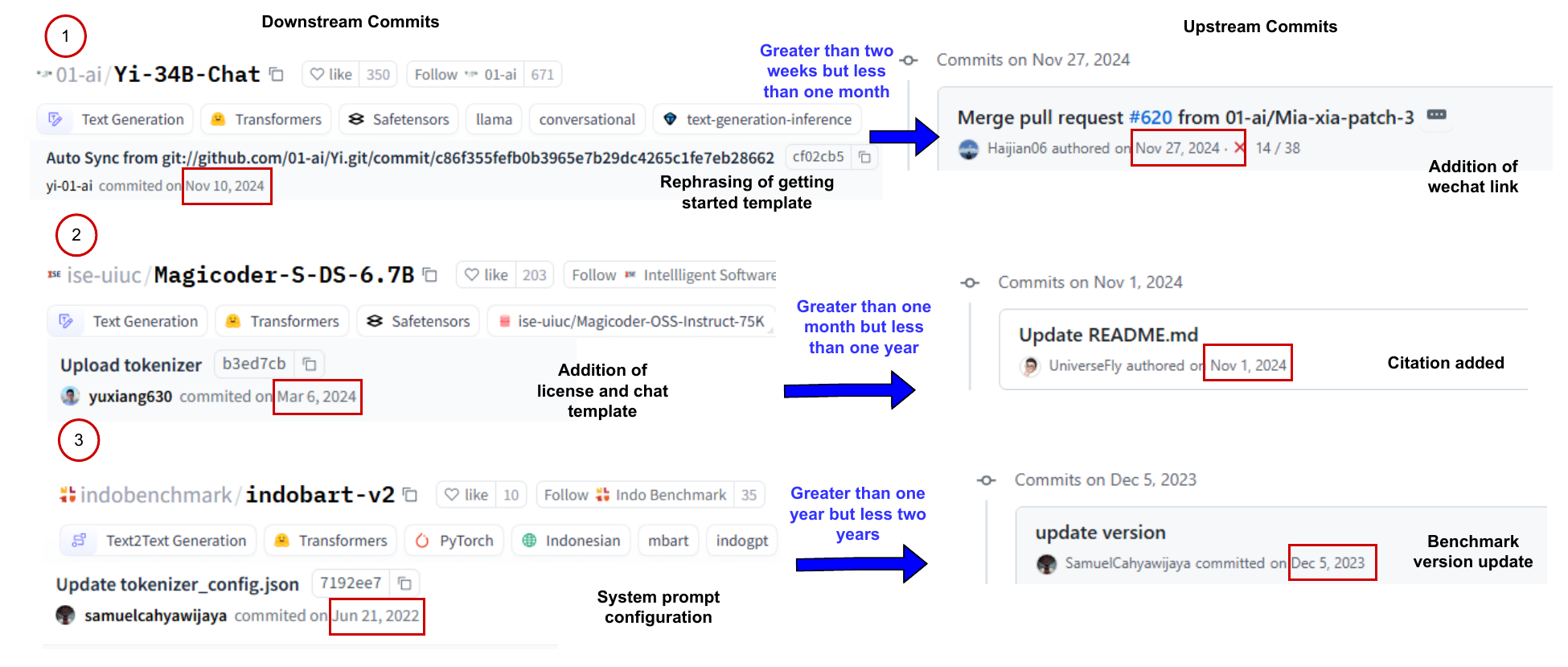}
\caption{Examples of delays and inconsistencies in synchronizing development activities between upstream and downstream repositories across different PTLM families. Each example shows the most recent commit on each platform at the time of data collection.}
\label{sync_prob}
\end{figure*}

However, synchronization between GH and HF repositories is often inconsistent. As shown in \Cref{sync_prob}, discrepancies frequently arise across PTLM families (groups of PTLM variants managed under a common upstream GH repository), where updates fail to propagate in a timely or consistent manner. This misalignment creates release-time inconsistencies and can leave one platform outdated. For instance, in the YI family, 01-ai/Yi-34B-Chat\footnote{\url{https://huggingface.co/01-ai/Yi-34B-Chat}}
 received a template update on HF on November 10, 2024, while the corresponding GH repository remained inactive for over two weeks. In the Magicoder family, ise-uiuc/Magicoder-DS-6.7B\footnote{\url{https://huggingface.co/ise-uiuc/Magicoder-DS-6.7B}}
 had its license and chat template modified on HF on March 6, 2024, whereas a citation update appeared on GH only on November 1, 2024—a gap of nearly eight months. Similarly, in the Indobart family, indobenchmark/indobart-v2\footnote{\url{https://huggingface.co/indobenchmark/indobart-v2}}
 received a system prompt update on HF on June 21, 2022, while the corresponding benchmark version change on GH occurred more than a year later, on December 5, 2023.

These examples reveal a lack of systematic coordination between GH and HF repositories, with updates often appearing on one platform long before—or after—they are reflected on the other. The extent of these delays varies across PTLM families, indicating an uneven flow of development activities between platforms. This study investigates these patterns by identifying factors influencing coordination and delays, examining synchronization types that emerge, and assessing their prevalence across PTLM families.

\subsection{Related Work} 
\subsubsection{Commit classification and taxonomy}
Understanding software changes through commits has long been a key research focus, with early studies proposing taxonomies that classify changes by purpose. Our work also relates to research on detecting inconsistencies across software artifacts, particularly between source code and its documentation \citep{ciraci2012approach, stulova2020towards, ray2013detecting} model \& dataset documentation \citep{oreamuno2024state, donald2025semantic, ajibode2025towards}. However, unlike these studies, which focus on intra-repository inconsistencies, we examine temporal synchronization inconsistencies across distinct platforms in the PTLM ecosystem—spanning code, model weights, and configuration files.

\citet{hindle2008large} observed that large commits often reflect architectural or perfective changes, while smaller ones tend to be corrective. Extending this work, \citet{bhatia2023towards} introduced two ML-specific categories—Data and Dependency Management—with 16 subcategories such as input data, parameter tuning, and model structure. Similarly, \citet{janke20247} analyzed 1,000 Java projects and 45,000 change patterns, grouping them into seven categories and showing that while many patterns are project-specific, a few recur across projects. These studies reveal how commit patterns characterize software evolution and maintenance practices.

Subsequent research shifted toward automated commit classification. Early approaches used metadata such as commit messages and author information. For example, \citet{mockus2000identifying} trained classifiers to detect maintenance types from textual descriptions, finding perfective changes prevalent in legacy systems. \citet{yan2016automatically} proposed a discriminative probabilistic model (DPLSA) that improved accuracy and recall across multiple projects. These machine learning methods laid the foundation for later deep learning approaches that achieved higher precision in commit categorization.

Commit classification has recently extended to pre-trained model repositories. \citet{castano2024machine} applied Gemini-1.5 Flash \citep{google2024gemini} to categorize HF commits using \citet{bhatia2023towards}’ taxonomy, identifying Training Infrastructure, Output Data, and Project Metadata as dominant types. While their study focused on internal HF dynamics, we adopt the same taxonomy to analyze cross-platform coordination between GH and HF. Although developed for general ML projects, the taxonomy provides a practical framework for distinguishing commit types in PTLM development. Using it, we classify commits to compare update patterns, apply topic modeling within each change type to uncover thematic focus and directional relationships, and identify eight synchronization patterns between GH and HF. These patterns highlight the distributed and interdependent nature of PTLM maintenance across platforms.

\subsubsection{Coordination strategies in software development}
Coordination has long been a central concern in software engineering, dating back to foundational studies on task interdependencies and communication bottlenecks \citep{malone1994interdisciplinary, herbsleb2001empirical}. For broader discussions, we refer readers to surveys such as \citet{talukder2017coordination}. Here, we focus on research most relevant to coordination during PTLM release across collaborative platforms.

Early work examined coordination in agile and co-located teams. \citet{strode2012coordination, strode2015coordination} developed a theoretical model identifying synchronization, structure, and boundary spanning as key coordination components. \citet{kanaparan2025investigating} validated this model through a survey of 340 practitioners, showing these strategies significantly influence coordination outcomes, with customer involvement moderating their effects. These studies provide a strong theoretical basis for understanding coordination in dynamic, collaborative projects.

In distributed and global contexts, researchers have explored how digital tools and asynchronous practices enable coordination. \citet{giuffrida2015conceptual} proposed a framework combining communicative genres and coordination mechanisms to study distributed teams, emphasizing the role of social software in maintaining team protocols. \citet{stray2020understanding} found that tools such as Slack improve collaboration but cannot fully mitigate participation challenges. Likewise, \citet{li2012formulating} showed how agile coordination practices replace rigid plan-driven methods in cross-boundary projects.

At scale, inter-team coordination has gained attention. \citet{berntzen2021coordination} studied a 16-team agile program and identified four strategies: aligning autonomous teams, maintaining project awareness, managing prioritization, and handling architectural dependencies—extending coordination research beyond team-level boundaries.

Broader empirical and technical perspectives also enrich the discussion. \citet{foundjem2021release} analyzed release coordination in OpenStack, identifying ten key coordination activities. \citet{bock2022synchronous} linked communication threads to software features, showing stronger associations at higher coordination levels. \citet{krause2022collaborative} developed a real-time collaborative visualization system that supports coordination through synchronized user actions. Beyond traditional software contexts, \citet{magelinski2022synchronized} proposed a framework for detecting covert synchronized behaviors on Twitter, illustrating how coordination concepts apply to digital ecosystems more generally.

\citet{talukder2017coordination} synthesized 50 studies on distributed agile coordination, categorizing work into theory, tools, and challenges, and highlighting the fragmented state of current research—underscoring the need for more integrated frameworks.

Our study extends this body of work by examining synchronization of commit and release activities for PTLMs across GH and HF. We focus on how practitioners align actions across platforms to maintain consistency, traceability, and reliability in PTLM evolution. Unlike prior work emphasizing team-based or organizational coordination, we analyze temporal synchronization between repositories using a repository-mining approach, focusing specifically on timing relationships among commit activities rather than broader workflow processes.

\begin{figure*}[t]
\centering
\includegraphics[width=\textwidth]{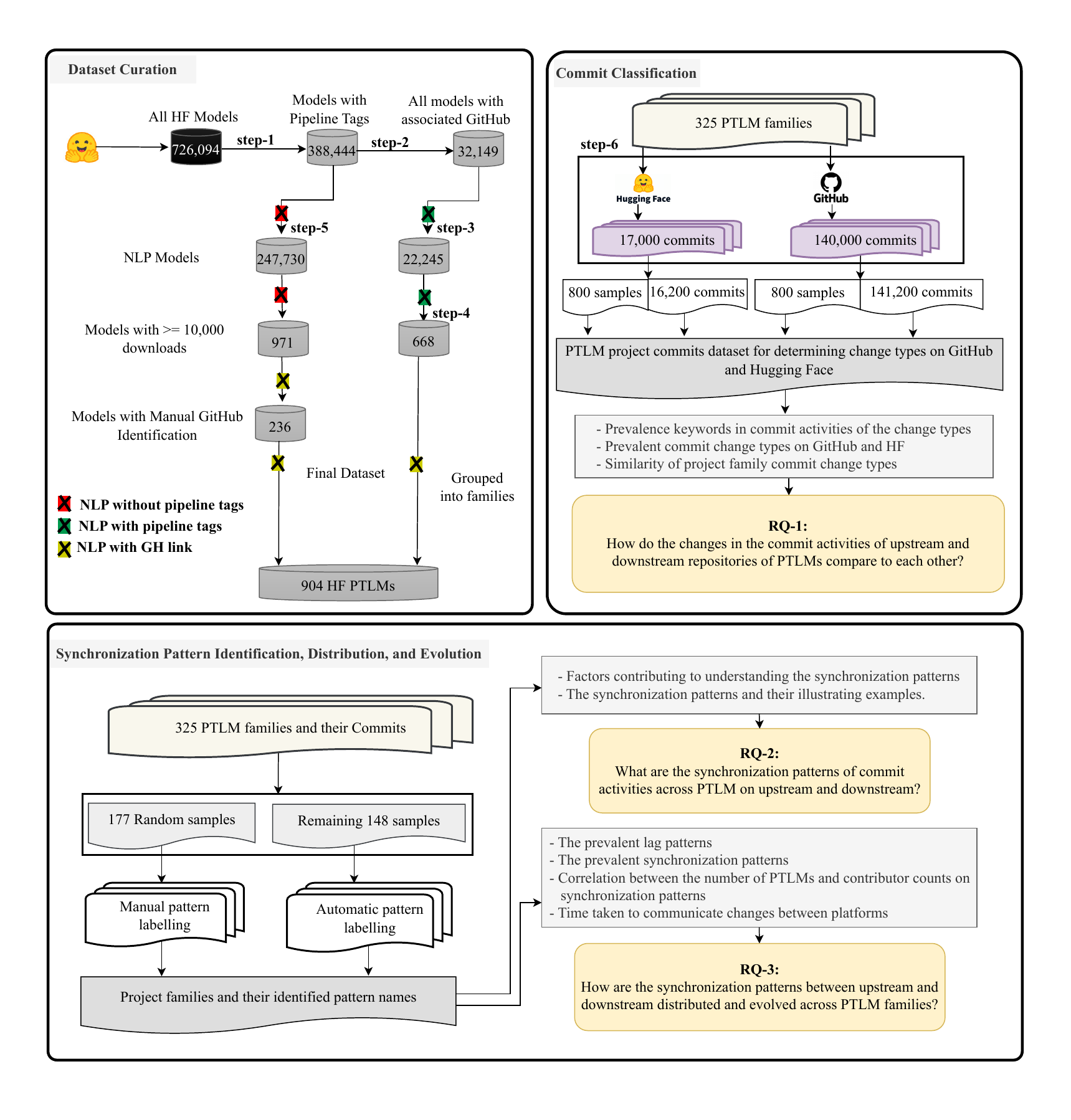}
\caption{Data collection procedure}
\label{framework}
\end{figure*}

\section{Study Setup}\label{methodology}

This section presents the design of our empirical study, which investigates the synchronization of commit activities for PTLMs between GH (the upstream platform) and HF (the downstream platform). To explore these synchronization patterns, we address the three interconnected research questions introduced earlier. We begin by describing our data collection methodology in \Cref{data_collection}, outlining the step-by-step process used to curate the dataset forming the basis of our analysis in \Cref{result}. \Cref{framework} provides a visual overview of the extraction, refinement, and analysis workflow, showing how each stage aligns with our research questions.

\subsection{Data Collection Methodology}\label{data_collection}

We selected HF as the downstream platform due to its widespread adoption and central role in PTLM distribution. Compared to alternatives such as TensorFlow Hub or ONNX Model Zoo, HF provides broader model coverage, richer metadata, and stronger community support~\citep{ait2025suitability}. This prominence is evident in its hosting of over one million models across diverse tasks. GH was chosen as the upstream platform, as it remains the primary open-source environment for core development and collaborative commit activities.

Our dataset construction followed six structured steps, capturing PTLM commit activities from both platforms while accounting for their interactions. Steps 1–4 and Step 6 involved automated filtering, while Step 5 relied on manual inspection and refinement. Outputs from both streams were merged to produce the final dataset.

\begin{itemize}
\item \textbf{Step 1: Extracting all models from HF:} Using the HfApi Client\footnote{\url{https://huggingface.co/docs/huggingface\_hub/package\_reference/hf\_api}}
 (see \Cref{framework}), we retrieved all models uploaded to HF as of June 10, 2024. The extraction yielded 726,094 models, with 388,444 (53.5\%) categorized into groups based on HF’s ML pipeline tag taxonomy: Audio (27,040), Computer Vision (46,788), Multimodal (853), Natural Language Processing (PTLMs) (269,975), Reinforcement Learning (43,457), and Tabular (331). The remaining 337,650 models (46.5\%) lacked identifiable pipeline tags. To maintain consistency, subsequent filtering focused only on models with identifiable pipelines.

\item \textbf{Step 2: Automatically identifying GH links for the 388,444 models:}
To extract GH repository links from HF model cards, we developed a script (available in our replication package~\citep{replication}). The main challenge was that many model cards include multiple GH links, some unrelated to the model itself (e.g., external references or other repositories). Our script distinguishes and retrieves only the links directly corresponding to each model.

\begin{algorithm}[t]
\caption{Identifying the Correct GH Link for a PTLM}
\begin{algorithmic}[1]
\State \textbf{Input:} List of GH links, Model name in the form `owner/model\_name` (e.g. ProsusAI/finbert)
\State Split the model name into two segments: 
\Statex \hspace{1cm} \texttt{left\_segment} (owner), \texttt{right\_segment} (model name after `/`)
\State Initialize an empty list \texttt{valid\_links}
\For{each GH link in the list}
    \If{GH link contains \texttt{left\_segment}}
        \State Add the link to \texttt{valid\_links}
    \EndIf
\EndFor
\If{\texttt{valid\_links} is empty}
    \For{each GH link in the list}
        \If{GH link contains \texttt{right\_segment}}
            \State Add the link to \texttt{valid\_links}
        \EndIf
    \EndFor
\EndIf
\For{each link in \texttt{valid\_links}}
    \If{link does not contain \texttt{left\_segment} or \texttt{right\_segment}}
        \State Remove the link from \texttt{valid\_links}
    \EndIf
\EndFor
\State Remove duplicates from \texttt{valid\_links}
\State \textbf{Output:} List of unique and valid GH links
\end{algorithmic}
\label{link_finder}
\end{algorithm}

To ensure this accurate link identification, we applied several heuristics outlined in \Cref{link_finder}. First, we split each model name into two segments using the forward slash (/), as HF structures model names by separating the owner (left) from the model (right) \citep{ajibode2025towards}. We then searched all GH links in the model card for either segment, discarding those containing neither and removing duplicates. No model owner had more than one link affiliated with their models. Applying this method to the 388,444 models with pipeline tags, we found that 32,149 models (8.27\%) have GH links, corresponding to 3,702 unique GH repositories. The dataset includes models from major families such as Gemma (Google), Phi-3 (Microsoft), and LLaMA (Meta), supporting its representativeness and relevance. We acknowledge that this heuristic approach may exclude valid repositories when owners use naming conventions differing from the HF identifier (e.g., project-specific names or centralized repositories like “research-models”). The potential impact of this limitation is further discussed in \Cref{ttv} (Threats to Validity).

\item \textbf{Step 3: Selecting PTLMs:}
Among the 32,149 models with GH links, 22,245 (69\%) are NLP models (i.e., text-modality models), covering 74\% of the 3,702 unique GH repositories. This aligns with the findings of \citep{castano2024analyzing}, who reported that NLPs tend to be better documented and more widely adopted. Based on these insights, we focused subsequent filtering on the NLPs. For transparency, our replication package includes GH links for both NLP and non-NLP models.

\item \textbf{Step 4: Selecting popular \models based on \#downloads:}

To focus on high-quality, well-maintained models, we applied a 10,000-download threshold \citep{jiang2023peatmoss}, reducing the dataset to 668 NLPs linked to 271 GH repositories. Although 22,245 NLPs had associated GH repositories, not all were directly related to model development. To assess this, the first author manually reviewed 50 randomly selected models and their repositories, checking for training code or replication packages. This review showed that 22\% of the sampled models lacked such resources—and all 22\% had fewer than 10,000 HF downloads. In contrast, models with 10,000+ downloads consistently provided well-documented replication materials. For instance, altsoph/xlmr-AER\footnote{https://huggingface.co/altsoph/xlmr-AER}\footnote{https://github.com/altsoph/BAER}
 (27 downloads) contained only PDF files, while izhx/udever-bloom-560m\footnote{https://huggingface.co/izhx/udever-bloom-560m}\footnote{https://github.com/manueltonneau/twitter-unemployment/tree/main}
 (5,139 downloads) included only a dataset. Conversely, highly downloaded models such as google-bert/bert-base-uncased\footnote{https://huggingface.co/google-bert/bert-base-uncased}\footnote{https://github.com/google-research/bert}
 and wukevin/tcr-bert\footnote{https://huggingface.co/wukevin/tcr-bert}\footnote{https://github.com/wukevin/tcr-bert}, with 67M+ and 6M+ downloads respectively, provided full training resources. We acknowledge that this threshold excludes about 96\% of available NLPs, introducing potential sampling bias. Excluded models may represent recent releases, specialized domains, or smaller research teams. Consequently, our findings are most generalizable to well-established, community-recognized models. We revisit this limitation in \Cref{ttv} and encourage future work to test whether similar trends apply across the broader model ecosystem.

Following this filtering, the first author manually reviewed all the GitHub links and confirmed that they are correct—ensuring that repository references correspond to the intended models, particularly for organizations (e.g., EleutherAI) managing multiple repositories. For example, EleutherAI’s pythia-160m\footnote{https://huggingface.co/EleutherAI/pythia-160m}\footnote{https://github.com/EleutherAI/pythia}
 and llemma\_7b\footnote{https://huggingface.co/EleutherAI/llemma\_7b}\footnote{https://github.com/EleutherAI/math-lm}
 are maintained in separate repositories. This verification confirmed the accuracy of repository-to-model mapping.

\item\textbf{Step 5 Manually identifying GH links for the top popular \models:} While the previous step focused on the 22,245 PTLM variants with GH links identified in Step 3, this step addressed the remaining 247,730 variants without GH links in their model cards. To expand our dataset, we again applied a 10,000-download threshold, prioritizing popular models for manual link identification. This yielded 971 high-download \models manually analyzed for the presence of GH links, using the following process:

\begin{itemize}
\item \textbf{Exploring the \model owner's HF profile to locate a GH link, if available.} If the owner’s profile included a GH homepage link, we reviewed the repositories for HF model training resources such as training scripts, configuration files, or fine-tuning setups. For profiles with fewer than 50 repositories, each was checked individually. For those with more than 50, we conducted a targeted search using the HF model name and verified whether the repository contained training code or replication packages. This ensured we excluded repositories unrelated to model management. For example, YituTech/conv-bert\footnote{https://huggingface.co/YituTech/conv-bert-base} lacked a GH link in its model card. Visiting the owner’s GH profile\footnote{https://github.com/yitu-opensource}, we found the ConvBERT repository\footnote{https://github.com/yitu-opensource/ConvBert} containing relevant training scripts. In contrast, papluca/xlm-roberta-base-language-detection\footnote{https://huggingface.co/papluca/xlm-roberta-base-language-detection} had a GH profile\footnote{https://github.com/LucaPapariello}, but no repository matched the model name or contained relevant files, so it was excluded.

\item \textbf{Examining academic publications referenced in the model card:} When the owner’s GH repository could not be identified through their HF profile, we examined any publication referenced in the model card. If it explicitly mentioned a GH repository containing training or related information, we manually verified it.  
For instance, medicalai/ClinicalBERT\footnote{https://huggingface.co/medicalai/ClinicalBERT} did not include a GH link in its model card or profile. However, its publication\footnote{https://doi.org/10.1038/s41591-023-02552-9} stated: ``The codes are available for academic research and non-commercial use on GH\footnote{https://github.com/rlditr23/RL-DITR}." In such cases, we used the publication to identify the corresponding GH repository.
\end{itemize}

At the end of Step 5, we identified 698 PTLMs with links to their GH repositories, drawn from 971 well-downloaded models. However, 462 model repositories lacked essential training-related resources and were excluded, leaving 236 models associated with 54 unique GH repositories, in addition to those from Step 4. \par\vspace{0.5em}

Combining the manual and automated methods from Steps 4 and 5, we identified 904 HF \models associated with 325 unique GH links, forming our final dataset. To consolidate the dataset, we grouped multiple HF models under the same GH repository into a ``family''—a set of PTLMs managed within a single GH repository—reflecting coordinated development and reuse practices. For example, neulab/codebert-python\footnote{https://huggingface.co/neulab/codebert-python}
, neulab/codebert-cpp\footnote{https://huggingface.co/neulab/codebert-cpp}
, neulab/codebert-java\footnote{https://huggingface.co/neulab/codebert-java}
, and neulab/codebert-c\footnote{https://huggingface.co/neulab/codebert-c}
 are all maintained within a single GH repository\footnote{https://github.com/neulab/code-bert-score}

\item\textbf{Step 6 Extracting Commits from 904 HF \models and their respective 325 GH repositories:} To understand activities on GH and HF and how they differ or align, we extracted commit data from both platforms. Commits were chosen as the focus because they are fundamental to software development and have been widely used to study software evolution \citep{lin2013empirical}, maintenance activities \citep{herivcko2023commit}, and developer collaboration \citep{tian2022makes}. Commits represent finalized work, while pull requests and issues often capture discussions or work in progress. Moreover, HF hosts relatively few pull requests and issues compared to GH, further justifying our focus on commits. Commit information was extracted from both HF and GH on October 13, 2024.

To extract commit activities from HF, we developed a script (available in our replication package \citep{replication}) using the HF HfAPI Client to retrieve repository metadata, including commit titles, messages, authors, and timestamps. Before outputting results, we applied preprocessing steps such as Unicode normalization\footnote{https://unicode.org/reports/tr15/}, removal of HTML tags and code blocks, and collapsing excessive whitespace. These ensured data consistency, readability, and the removal of formatting artifacts while preserving essential content. This process yielded 17,000 commits from HF repositories.

Similarly, for GH, we developed another script (available in our replication package \citep{replication}) that uses the GH REST API\footnote{https://docs.github.com/en/rest?apiVersion=2022-11-28} to retrieve commit messages, authors, and timestamps. We applied the same preprocessing steps as for HF, including the removal of HTML tags and code snippets, and whitespace normalization. To manage API rate limits, we implemented automatic throttling and retries. This process yielded 140,000 commits from GH repositories.

The remaining components of our methodology, as outlined in \Cref{framework}, are discussed within the context of each research question. Rather than presenting all methodological steps together, we integrate them into each research question to maintain clarity and provide focused explanations aligned with specific RQ objectives.
\end{itemize}

\section{Results}\label{result}
\subsection{\textbf{RQ$_1$:} \RQa}\label{RQ0}
Downstream \model owners reference GH repositories for various purposes, as specified in model cards, including providing code for pretraining and fine-tuning\footnote{https://huggingface.co/m3rg-iitd/matscibert}, reporting issues\footnote{https://huggingface.co/meta-llama/Llama-2-70b-hf}, and sharing additional resources\footnote{https://huggingface.co/openai-community/gpt2-medium}, such as usage examples and tutorials. However, it is unclear to what extent \model stakeholders—including developers, maintainers, and users—ensure that changes on one platform are communicated to the other. To address this, we investigate the prevalent change topics (subcategories) in PTLM family commits on GH and HF, quantify change types, and measure their similarity. This analysis reveals the extent of the upstream-downstream relationship between GH and HF, providing insights into cross-platform management and guiding efforts to streamline PTLM release processes.

\subsubsection{Approach}\label{RQ0_approach}
\paragraph{Classifying the change types of PTLM families on GitHub (GH) and HuggingFace (HF) using manual and automated methods.} To classify \model \changes on HF and GH, we combined manual and automated methods. First, the authors held a roundtable to collect and analyze commit categories based on prior work \citep{castano2024machine, bhatia2023towards}, ensuring their applicability to PTLMs. To establish a ground truth for evaluating the LLM-based labeling, the authors then manually labeled 800 commits per platform (1,600 total). Next, this ground truth was used to evaluate how well Gemini-1.5 Flash \citep{google2024gemini} was able to automatically label the 1,600 commits. Based on the high accuracy of 82.6\% for GitHub and 78.8\% for Hugging Face, we then used Gemini-1.5 Flash to label all the remaining 16,200 HF and 141,200 GH commits. We visualized the distribution of \changes using bar charts. The complete process involves the following steps:


\noindent\textbf{Step 1.1: Selection of representative samples of commits from GH and HF.} Given the large volume of commits on both platforms (17,000 from HF and 140,000 from GH), we used stratified random sampling \citep{aubry2023using} to obtain 800 commits per platform, keeping manual labeling manageable. Preliminary observations of the temporal relationship between GH and HF activities identified eight distinct temporal patterns, later formalized as synchronization patterns in RQ2.

To represent all eight patterns, we randomly selected 10 of the 325 PTLM families (see Step 5 in \Cref{data_collection} for family consolidation) and sampled 10 commits per pattern per family, yielding 800 commits per platform. This relatively small family set ensured feasible manual inspection while maintaining diversity across ownership and release activity profiles. The sample was merely intended to (i) validate our taxonomy and to manually label a ground truth dataset used solely for training and (ii) validating the automated LLM classifier that processed the remaining 315 families, not to statistically represent all PTLM families. Stratified sampling ensured proportional representation of synchronization patterns, capturing diverse commit behaviors while keeping the sample size practical for manual labeling.

\noindent\textbf{Step 1.2: Manual and automatic labeling of representative samples of commits.} Following the selection of 800 commits, in order to establish a reliable ground truth for training and validating the automatic labeling process, the first and second authors manually labeled 20\% of them according to the 15 change types (taxonomy) from \citet{castano2024machine, bhatia2023towards}. This taxonomy, developed for AI model commit activities, provides a structured framework to understand PTLM evolution. While \citet{castano2024machine} focused on HF, it was adapted from \citet{bhatia2023towards} for GH, showing its applicability to both platforms and relevance for PTLMs.

This taxonomy categorizes commit activities into the following 15 types:
\begin{itemize}
\item Preprocessing: Changes to data transformation before model training (e.g., tokenizer fixes, normalization).
\item Parameter tuning: Adjustments to hardcoded hyperparameters (e.g., learning rate, batch size).
\item Model structure: Modifications to the model’s architecture or code (e.g., layer changes, bug fixes).
\item Training infrastructure: Updates to training logic (e.g., checkpointing, distributed training setup).
\item Pipeline performance: Optimizations for runtime efficiency (e.g., faster data loading, memory fixes).
\item Sharing: Changes enabling collaboration (e.g., Git hooks, shared configs, CI/CD workflows).
\item Validation infrastructure: Modifications to evaluation logic (e.g., new metrics, benchmark updates).
\item Internal documentation: Developer-facing documentation (e.g., code comments, merges.txt).
\item External documentation: User-facing documentation (e.g., README.md, model cards).
\item Input data: Changes to data loading/ingestion (e.g., new datasets, correct a path to datasets).
\item Output data: Adjustments to output storage (e.g., saving predictions in a new format).
\item Project metadata: Non-functional updates (e.g., versioning, licensing, initial commits).
\item Add dependency: Introduction of a new library/package (e.g., adding transformers).
\item Remove dependency: Removal of a library/package (e.g., dropping flax).
\item Update dependency: Updates in version/metadata (e.g. transformers$>=4.29$).
\end{itemize}

To measure agreement between authors, we used Cohen’s Kappa \citep{vieira2010cohen}, accounting for chance agreement. The agreement scores between the authors are 0.814 (GH) and 0.944 (HF), indicating almost perfect agreement \citep{perez2020systematic}. In the few cases where the two authors disagreed, such as with a Hugging Face commit that mentioned ``update pytorch model.bin" (initially labelled “output data” by the first author and ``model structure" by the second), we resolved the difference after deliberation by adopting “model structure” as the appropriate label. The high agreement of 0.61 exceeded the commonly accepted threshold for substantial inter-rater reliability \citep{el1998benchmarking}, justifying a single author to label the remaining 80\% of commits.

Since manually labeling the full set of 157,400 commits is infeasible, we developed a Python script (available in our replication package \citep{replication}) that embeds Gemini-1.5 Flash \citep{google2024gemini} to automatically label the remaining 16,200 HF and 141,200 GH commits using the 15 taxonomies. We provided manually labeled examples as prompts (\Cref{prompt}). We selected Gemini-1.5 Flash for its efficiency, cost-effectiveness, and speed, delivering GPT-4–comparable performance at lower cost \citep{castano2024machine}. We configured the model to return a single candidate output, with a maximum of 1,024 output tokens and a temperature of 1.0, ensuring reproducible and coherent classifications across diverse commit messages.

For GH commits, we provided the raw commit messages directly to the model. In contrast, for HF commits, we concatenated the commit titles and messages into a single entry. This was done to accommodate cases where commit messages were empty and to ensure that the LLM had sufficient context for accurate classification. At the time of data preparation, we treated GH commit messages as standalone entries. Although the GH API provides a single commit message string, it is a standard convention that the first line serves as the title and the subsequent lines form the body. In contrast, the HF API distinguishes titles and messages explicitly.

To assess the reliability of the LLM-based labeling, we repeated automatic classification 10 times for each platform, producing a complete labeling of all commits in each run. We did not aggregate the labels, focusing instead on evaluating the LLM’s consistency across these 10 runs, following the approach of hallucination-detection techniques such as SelfCheckGPT \citep{manakul2023selfcheckgpt}, which measure variability in outputs across multiple LLM responses. We compared each of the 10 labeled sets against the manually labeled subset of 800 commits using Cohen’s Kappa \citep{vieira2010cohen}. To mitigate bias from our use of few-shot prompting—where manually labeled commit messages serve as examples in the LLM prompt—we excluded these commits from the Kappa calculation. Specifically, we removed the 8 GitHub and 65 Hugging Face commits used as prompt examples. To maintain a sample size of 800 for the calculation, these excluded commits were replaced with an equal number of randomly selected, manually annotated commits that were independent of both the original sample and the LLM's prompt examples. The resulting Kappa scores were $\kappa$ = 0.744 for Hugging Face and $\kappa$ = 0.787 for GitHub, both indicating substantial agreement between human and LLM labels \citep{perez2020systematic}.

To further validate the LLM’s classification performance against the human ground truth, we computed precision, recall, and F1-scores across ten labeling iterations. The LLM achieved strong overall accuracy (82.6\% for GitHub and 78.8\% for Hugging Face). Performance was particularly robust for categories that are frequent and clearly distinguishable in software repositories: on GitHub, External Documentation (F1 = 0.93), Model Structure (F1 = 0.79), and Training Infrastructure (F1 = 0.74) showed high reliability; on Hugging Face, External Documentation (F1 = 0.95), Preprocessing (F1 = 0.81), and Model Structure (F1 = 0.66) performed well. Most disagreement between human and LLM labels occurred between semantically overlapping categories (e.g., Training Infrastructure vs. Pipeline Performance) or in classes with fewer examples, such as Internal Documentation on Hugging Face. Given the strong and balanced performance across key categories, we conclude that the LLM-based labeling provides a reliable foundation for our large-scale analysis.

\noindent\textbf{Step 1.3: Automatic labeling of the remaining commit messages.} After confirming agreement between Gemini-1.5 Flash and human labeling via Cohen’s Kappa, we applied the LLM to categorize the remaining 16,200 HF and 141,200 GH commits from 315 PTLM families, using the same 15 taxonomy categories to ensure consistency across the dataset.

\noindent\paragraph{Quantifying the distribution of PTLM family change types on GH and HF.} To quantify change type distributions in \project commits on GH and HF, we wrote a Python script (available in our replication package \citep{replication}) to aggregate occurrences of each labeled activity and compute their relative proportions. During processing, 41 of 140,000 GH commits received labels not in our 15-category taxonomy—specifically, ``bug fix" and `add model." Manual inspection showed that they were misclassified, all aligned with the ``model structure," and were corrected. These 41 undefined labels, representing 0.03\% of our dataset, were a minor and anticipated outcome of our few-shot prompting strategy. This approach prioritized the LLM's ability to grasp semantic intent, which was crucial for overall accuracy, as confirmed by the high Kappa and F1 scores. A known characteristic of this flexible approach is that the LLM can generate valid paraphrases (e.g., ``bug fix") for our canonical labels (e.g., ``model structure"). We judged this preferable to a more restrictive prompt, which might have forced the model into less accurate classifications for ambiguous commits. All such instances were trivially easy to identify and were manually corrected to ensure final labeling consistency. We then grouped the validated \changes by platform and computed their proportions to assess differences in prevalence. This minimal adjustment further supports the robustness and reliability of our labeling process.

\noindent\paragraph{Automatic identification of prevalent topics in each change type of commit activities on GH and HF.}
To understand dominant development areas within each change type, we applied topic modeling to reveal granular insights beyond the change type taxonomy.

We developed a script—available in our replication package \citep{replication}—that uses BERTopic\footnote{https://github.com/MaartenGr/BERTopic} to group semantically similar topics within each change type. BERTopic captures contextual word relationships, grouping terms like optimize,” enhance,” and ``improve” by semantic similarity rather than simple co-occurrence. We chose BERTopic over traditional methods like bag-of-words or LDA because it identifies nuanced topics and has proven effective in software engineering studies \citep{grootendorst2022bertopic, diamantopoulos2023semantically, gu2023self, zhao2024empirical, chagnon2024benchmarking}.

 To prepare the dataset for BERTopic, we categorized commit messages by change types for both GH and HF repositories. For the commit messages of each change type, we removed URLs, platform-specific stopwords (e.g., ``github", ``huggingface", ``hf", ``org", ``com"), short terms (fewer than three characters), and file extensions.

We let BERTopic dynamically determine the number of topics using HDBSCAN, avoiding pre-specifying k and allowing the model to adapt to the data’s natural clustering. To ensure meaningful results, we inspected topic coherence scores and removed low-value words (e.g., ``joao," ``delete," ``init") after generating topics.

We ranked topics in each change type using BERTopic’s scoring, prioritizing terms frequent within a change type but discriminative across all commits. This ensures we highlight terms that capture unique aspects of each change type. For HF data with fewer than 10 topics or insufficient messages, we used ``No identifiable topic" as a placeholder to maintain consistency.

\noindent\paragraph{Determining the maturity age group for \project.} \label{maturity_calculation}
As a measure of a PTLM family's maturity, we use the average model age, calculated as the arithmetic mean of the ages of all models in that family. We chose the mean because our goal is to capture a family's total temporal longevity—how long it has existed on Hugging Face—not the age of a ``typical" model within the family. The mean is a measure of central tendency that incorporates the full magnitude of every model’s age, including outlier (long-lived) models that contribute meaningfully to a family’s longevity. In contrast, the median suppresses these values. For example, in a family with ages {300, 5, 5}, the median (5) would suggest a very new family, even though it has one model that has persisted on the platform for almost a year. The mean (103) more accurately reflects that the family spans a long time window. This use of the mean—when the goal is to capture cumulative longevity—is consistent with established practices in empirical software engineering \citep{mockus2000case}.

To verify that the mean is appropriate for our dataset, we also computed the temporal span of releases in each family (the time between the earliest and latest release). This analysis showed that 70.9\% of families have a release span of zero weeks, meaning all their models were released at the same time, and this proportion increases to 75\% when considering families with a release span of four weeks or less, and 86\% with a span of 26 weeks (i.e., half a year). For these short-span families, where the variance in age is minimal or zero, the mean is mathematically identical or highly similar to the median, providing a precise measure of longevity. For the remaining families with longer development spans, the mean intentionally captures this extended lifespan by reflecting the contribution of older variants to the family's maturity.

Based on this rationale, we calculated each PTLM ’s age by subtracting its creation date from the dataset extraction date (June 10, 2024), then estimated each PTLM family's maturity using the average age of all PTLMs within the family—for example, a family with PTLMs aged 200, 100, and 30 days has an average age of 110 days. Despite simplifying temporal variation, this measure offers a reasonable approximation of overall maturity.

\noindent\paragraph{Examining the dominant \changes of \project across age group.}
We analyzed the distribution of change types on GH and HF, stratified by \project age group (Section~\ref{maturity_calculation}), to assess whether dominant \changes persist across maturity levels. We visualized relative frequencies using symmetric bar plots to reveal patterns across age groups.

We applied chi-square tests of independence \citep{mchugh2013chi} to evaluate significant differences in change type distributions between platforms. For each age group, we constructed a $15 \times 2$ contingency table (rows: change types; columns: platforms; cell values: frequencies). The null hypothesis ($H_0$) assumed identical distributions across platforms, while the alternative ($H_1$) assumed significant differences. We conducted separate tests for recent, intermediate, and mature projects and applied a Bonferroni correction ($\alpha$ divided by three) to control the family-wise error rate.

To quantify association strength, we calculated Cramér's V \citep{akoglu2018user} for each age group, interpreting values as very strong ($>$0.25), strong ($>$0.15), moderate ($>$0.10), weak ($>$0.05), or negligible (0). This effect size contextualizes statistical significance, highlighting the magnitude of platform-specific differences in \model development activity.

\noindent\paragraph{Calculating the similarity score of change types between GH and HF.}
We quantified activity similarity between GH and HF within PTLM families using the Jaccard similarity coefficient \citep{ivchenko1998jaccard}, which measures overlap relative to the union of change types. Jaccard scores range from 0 (no common activities) to 1 (complete similarity). High scores ($\geq$0.5) indicate similar activities on both platforms, moderate scores suggest partial overlap, and low scores reflect largely independent activities. We visualized the distribution of similarity scores with a histogram showing the percentage of PTLM families per score, highlighting the degree of cross-platform activity overlap.

\noindent\paragraph{Associating maturity and contributor overlap with similarity scores.}
We grouped similarity scores into terciles—low (0.00–0.33), moderate (0.35–0.46), and high (0.50–1.00)—to ensure balanced distribution and analyzed them across model age groups (Recent, Intermediate, Matured) to identify trends in cross-platform activity.

To assess contributor overlap, we identified GH profile links shared with HF. Families with exactly one shared contributor were labeled Single-author; more than one shared contributor were labeled Multiple-authors. We visualized trends using grouped bar charts showing how model age and contributor overlap relate to similarity categories.

Finally, we performed a Chi-Square test of independence to formally assess whether contributor overlap associates with similarity scores. The null hypothesis ($H_0$) assumes independence, while the alternative ($H_1$) posits a significant association. This methodology allowed us to systematically evaluate how model maturity and contributor overlap influence cross-platform activity similarity while maintaining statistical rigor.

\subsubsection{Results.}\label{RQ1_results}
\noindent\textbf{GH (upstream) changes focus most commonly on model structure (29.7\%), external documentation (21.0\%), and training infrastructure (9.0\%), while HF (downstream) changes focus most on external documentation (38.8\%), preprocessing (16.6\%), and model structure (14.4\%).}

\begin{figure*}[t]
\centering
\includegraphics[width=0.8\textwidth]{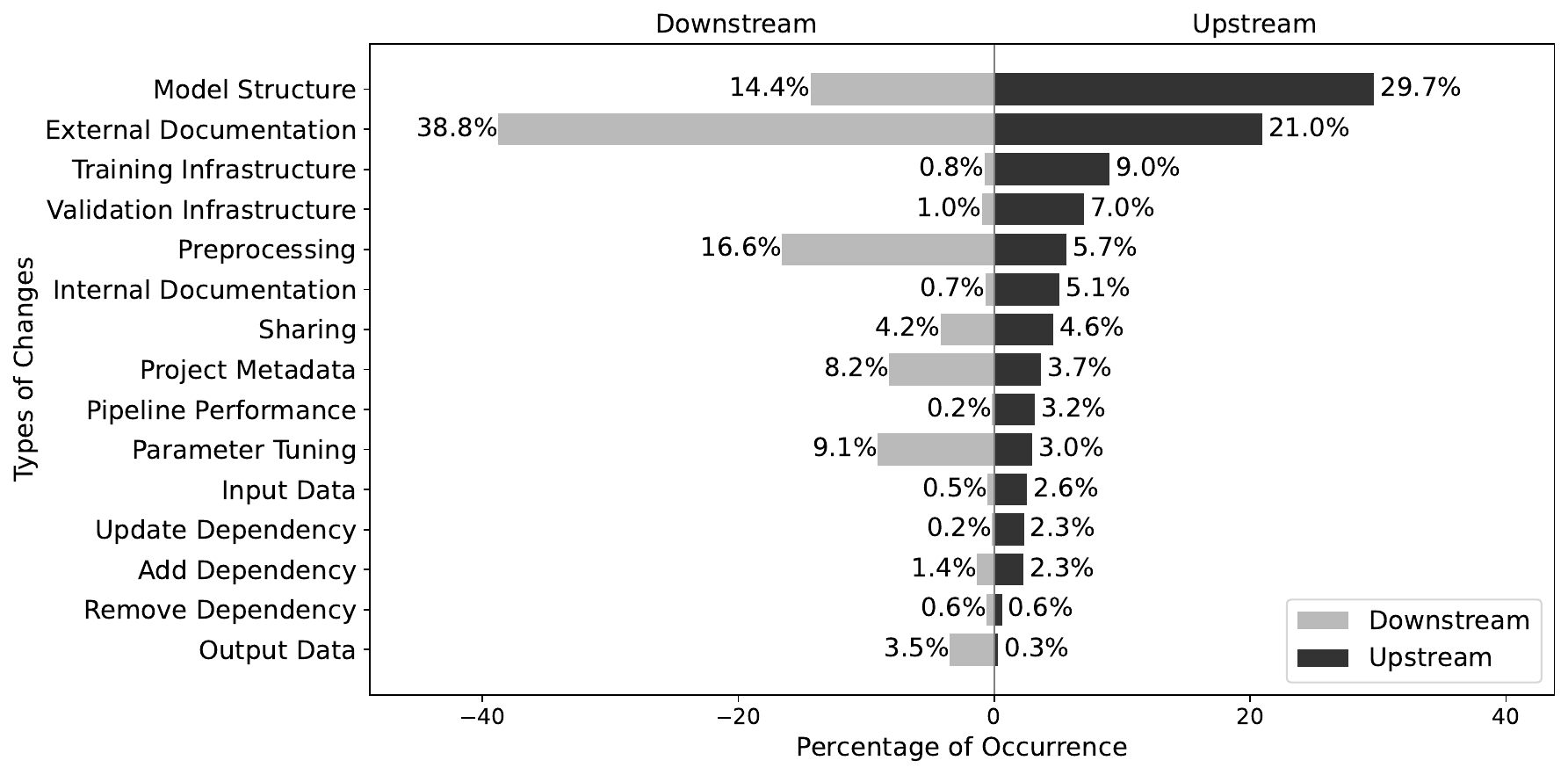}
\caption{Proportion of prevalent commit change types in PTLMs on GH and HF.}
\label{R01}
\end{figure*}

Although similar change types appear on both platforms (Figure \ref{R01}), their frequency reflects the distinct roles of GH and HF. External documentation changes occur almost twice as often on HF, consistent with its role in distributing pre-trained models that require clear user guidance. In contrast, model structure changes are twice as common on GH (29\% vs. 14\%), aligning with upstream development activities where training scripts, architectures, and configurations evolve more frequently. Preprocessing updates are also more common on HF (16.6\% vs. 5.7\%), suggesting greater emphasis on preparing models for deployment or fine-tuning.

Training infrastructure changes occur more frequently on GH (9.0\% vs. 0.8\%), since GH hosts the underlying codebase, including training scripts that require continual updates, while HF primarily distributes finalized models with fewer infrastructure modifications. These trends highlight the complementary roles of GH and HF in the PTLM lifecycle.\\

\noindent\textbf{HF focuses on artifact-specific changes, while GH addresses codebase and infrastructure updates.} Building on our high-level categorization, we now examine the specific topics within each change type. On GH, model structure changes frequently involve large refactoring, task-specific modifications, and architecture revisions. In contrast, HF shows a higher prevalence of external documentation changes, often related to licensing, prompt templates, and publication citations. This topic-level analysis highlights the differing concerns reflected in commit activities across both platforms. The topics are presented below without implying any order of importance or frequency.

\noindent\rule{\linewidth}{0.4pt}
\begin{tabular}{p{0.10\textwidth} p{0.45\textwidth} p{0.45\textwidth}}
    \textbf{Change Types} & \textbf{HF Change Topics} & \textbf{GH Change Topics} \\
    \hline
    & \textbf{Topics:} Checkpoint, Flax, Architecture, SampleFinetunepy, Mistralforcausallm, Huggingfacehub, Weights, Onnx 
    & \textbf{Topics:} Bigrefactor, Greedyuntil, Directory, Alexnet, MultipleChoiceTask, Makefile, Finetune, Multiheadattention, Remotetracking, Indentation \\
    Model Structure & \textbf{Explanation:} The topics characterizing model structure changes primarily involve uploading model weights, defining architecture-specific components, and preparing scripts for model instantiation and integration within the Hugging Face repository.
    & \textbf{Explanation:} These topics are more focused on codebase changes, including large-scale refactoring (Bigrefactor), task-specific adjustments (Multiplechoicetask), and structural changes to specific components of the model training scripts (e.g., multi-head attention, finetuning), along with the management of the codebase itself (Makefile, Directory). \\
    \hline
\end{tabular} 
\begin{tabular}{p{0.10\textwidth} p{0.45\textwidth} p{0.45\textwidth}}
    & \textbf{Topics:} Citation, Paper, Latency, License, Template, Docs, Typo, Terms, Create, Transformers   
    & \textbf{Topics:} Avatar, Custom-moe, Spaces, Acknowledgement, News, Leaderboard, Textdiffuser, Chapter, Header, Featdoc \\
    External Documentation & \textbf{Explanation:} These topics focus on formal and legal documentation, such as research citations, licensing terms, error corrections, and prompt templates for fine-tuning.
    & \textbf{Explanation:} GH topics are more community-driven, focusing on structural, visual, and feature documentation, as well as tracking contributions and updates. \\
    \hline
\end{tabular}
\begin{tabular}{p{0.10\textwidth} p{0.45\textwidth} p{0.45\textwidth}}
    & \textbf{Topics:} Enable, Training   
    & \textbf{Topics:} Watchdog, Jupiter, Finetune, Lock, Bump, Setup, Lora, Proxy\\
    Training Infrastructure & \textbf{Explanation:} These topics reflect efforts to activate or configure training processes—such as initiating fine-tuning runs or modifying training-related settings—often leveraging scripts or model artifacts from GH repositories.
    & \textbf{Explanation:} GH topics deal with various tools and frameworks for setting up the basic environment—including model training utilities, version management, and task-specific components like fine-tuning—while downstream deploys these setups. \\
    \hline
\end{tabular}
\begin{tabular}{p{0.10\textwidth} p{0.45\textwidth} p{0.45\textwidth}}
    & \textbf{Topics:} Inference, Model, Evaluation 
    & \textbf{Topics:} Edit, Wandblogmodel, Baselines, Vision, Squad, Roberta, Leaderboard, Coverage, Monitorpy, Travis\\
    Validation Infrastructure & \textbf{Explanation:} These topics focus on the process of running models for inference and evaluation, providing insights into their actual performance through metrics, benchmarks, and leaderboard placements, rather than guaranteeing specific outcomes. 
    & \textbf{Explanation:} GH topics are more focused on the tools, frameworks, and pipelines used to validate models, including managing evaluation tasks, setting up benchmarks (e.g., Baselines, Squad), monitoring performance (e.g., Coverage, Monitorpy), and ensuring proper test execution (e.g., Travis). \\
    \hline
\end{tabular}
\begin{tabular}{p{0.10\textwidth} p{0.45\textwidth} p{0.45\textwidth}}
    & \textbf{Topics:}  Fast, Vocab, Tokenizationphismallpy, Language, Files, Tokenizer, Tokenizerconfigjson, Chat 
    & \textbf{Topics:} Clippy, Localdocs, Quotes, Tatoeba, Glue, alignment, Lint, Cleanup, Wiki\\
    Preprocessing & \textbf{Explanation:} These topics involve preprocessing tasks like tokenization, vocabulary setup, and preparing tokenizer files (e.g., tokenizer\_config.json) for language and chat-based models. 
    & \textbf{Explanation:} GH topics in this category emphasize tasks like removing noise (e.g., Cleanup, Quotes), ensuring syntactic correctness (Lint, Clippy), and aligning multilingual or task-specific datasets (alignment, Tatoeba, Glue, Wiki). \\
    \hline
\end{tabular}
\begin{tabular}{p{0.10\textwidth} p{0.45\textwidth} p{0.45\textwidth}}
    & \textbf{Topics:}  Main, Sync, Huggingfacehub, Safetensors, Duplicate, Model 
    & \textbf{Topics:} Develop, Xlnet, Commit, Resolve, Gitbook, Origin-big-refactor, Upstreammaster, Bigrefactor, Patch\\
    Sharing & \textbf{Explanation:} Focuses on sharing models, managing repositories, handling safe tensor formats, and preventing duplication. 
    & \textbf{Explanation:} Emphasizes repository synchronization, major refactoring, and documentation updates. \\
    \hline
\end{tabular}
\begin{tabular}{p{0.10\textwidth} p{0.45\textwidth} p{0.45\textwidth}}
    & \textbf{Topics:}  Update 
    & \textbf{Topics:} Tidy, Length, Isort, Changelog, Flake, Cleanup, Branch, Ruff, Gitignore, Woops\\
    Internal Documentation & \textbf{Explanation:} General updates related to internal documentation, possibly covering model descriptions, usage guidelines, or minor adjustments. 
    & \textbf{Explanation:} Involves documentation-related efforts for maintaining code cleanliness, formatting, changelogs, and minor refinements. \\
    \hline
\end{tabular}
\begin{tabular}{p{0.10\textwidth} p{0.45\textwidth} p{0.45\textwidth}}
    & \textbf{Topics:}  No identified topics 
    & \textbf{Topics:} Prefetch, Simplify, Feattoolkit, Path, Linted, Optimization, Proxy, Tqdm, Sort, Cuda\\
    Pipeline Performance & \textbf{Explanation:} No specific topics were found related to direct pipeline performance improvements on HF. 
    & \textbf{Explanation:}  These topics focus on optimizing training time, including prefetching data, simplifying code, improving toolkit features, optimizing CUDA operations, and enhancing overall performance. \\
    \hline
\end{tabular}
\begin{tabular}{p{0.10\textwidth} p{0.45\textwidth} p{0.45\textwidth}}
    & \textbf{Topics:}  ModelMaxLength, Config, Configjson, Configjson 
    & \textbf{Topics:} Mcli, Tweak, Global, Criteria, Timeout, UseDevOption, Interval, Anneal, Configjson, Clipping\\
    Parameter Tuning & \textbf{Explanation:} These topics focus on modifying model configuration files and setting constraints like maximum model length to optimize hyperparameter settings. 
    & \textbf{Explanation:} These topics involve adjusting model behavior through hyperparameter refinements—such as criteria, timeouts, annealing schedules, clipping thresholds, and interval settings—implemented both in code and through configuration files (e.g., configjson), particularly in downstream repositories. \\
    \hline
\end{tabular}
\begin{tabular}{p{0.10\textwidth} p{0.45\textwidth} p{0.45\textwidth}}
    & \textbf{Topics:}  Dataset 
    & \textbf{Topics:} Testdata, Samples, Customtraintxt, Tatoeba, Prompt, Arabic, Partial, Json, Template, Branch\\
    Input data & \textbf{Explanation:} Focuses on modifications to dataset handling, including loading and managing datasets for model training and inference. 
    & \textbf{Explanation:}  Covers a broader range of data-related changes, including test data updates, sample modifications, custom training data, prompt handling, language-specific adjustments (e.g., Arabic), and template structures for input data formatting. \\
    \hline
\end{tabular}
\begin{tabular}{p{0.10\textwidth} p{0.45\textwidth} p{0.45\textwidth}}
    & \textbf{Topics:}  Update 
    & \textbf{Topics:} Pandas, Legion, Sklearn, Flair, Chrome, Scann, Vulnerabilities, Tqdm, Deno, Corenlp\\
    Update dependency & \textbf{Explanation:} This topic generally refers to broad updates, which may include changes to libraries (and their versions), frameworks, or other components required for running or training the model, though the specific dependencies involved are not explicitly stated in the commit messages. 
    & \textbf{Explanation:}  Clearly focuses on updating specific dependencies and libraries, addressing security vulnerabilities, improving compatibility, and enhancing performance by upgrading tools such as Pandas, Scikit-learn (Sklearn), and CoreNLP. \\
    \hline
\end{tabular}
\begin{tabular}{p{0.10\textwidth} p{0.45\textwidth} p{0.45\textwidth}}
    & \textbf{Topics:}  No identified topics 
    & \textbf{Topics:} Submodule, Initpy, Spacy, Vllm, Pyprojecttoml, Posthog, Reqs, Cmake, Flake, Isort\\
    Add Dependency & \textbf{Explanation:}  There are no clear topics indicating the addition of new dependencies, suggesting that dependency additions may not be explicitly highlighted in the commit messages or are handled differently. 
    & \textbf{Explanation:}  These topics reflect the explicit addition of new dependencies, including machine learning libraries (Spacy, Vllm), dependency management files (Pyproject.toml, Reqs), and tools for code quality and build management (Flake, Isort, CMake). \\
    \hline
\end{tabular}
\begin{tabular}{p{0.10\textwidth} p{0.45\textwidth} p{0.45\textwidth}}
    & \textbf{Topics:}  No identified topics 
    & \textbf{Topics:} Directory, Submodule, Transformers, Scripts, Import, Dataset, File, Yaml, Merge, Deepspeed\\ 
    Remove Dependency & \textbf{Explanation:}  HF does not show specific topics related to the removal of dependencies. However, this could involve indirect changes to model configurations or related files that no longer use certain dependencies 
    & \textbf{Explanation:}  These topics indicate possible changes in removing dependencies within the GH project. Changes in Directory, Submodule, and Import suggest reorganization or removal of related code, while Scripts, File, Dataset, and Yaml might indicate the removal of code or data files linked to now-removed dependencies. Deepspeed could indicate the removal of an optimization library, and Merge suggests combining code that excludes the removed dependencies.\\
    \hline
\end{tabular}
\begin{tabular}{p{0.10\textwidth} p{0.45\textwidth} p{0.45\textwidth}}
    & \textbf{Topics:}  Release, License, Commit, Llmfoundry, Basemodel, Link 
    & \textbf{Topics:} Codeowners, Create, Unstable, Byml, Dhuang, Chore, Repo, Notes, Multilingual, Adam\\
    Project Metadata & \textbf{Explanation:}  These topics indicate changes related to project metadata, including licensing updates (License), version releases (Release), foundational model information (Basemodel, Llmfoundry), and commit tracking (Commit). Updates may also include documentation links (Link) for reference. 
    & \textbf{Explanation:}  These topics relate to project metadata updates, including ownership files (e.g., Codeowners), versioning stability (Unstable), documentation notes, contributor mentions (e.g., Dhuang, Adam), multilingual support, workflow refinements (Chore), and metadata structuring (Byml).  \\
    \hline
\end{tabular}
\begin{tabular}{p{0.10\textwidth} p{0.45\textwidth} p{0.45\textwidth}}
    & \textbf{Topic:}  Modelsafetensorsindexjson, Pytorchmodelbin, Onnx, Files, Safetensors, Huggingfacehub 
    &  \textbf{Topic:} Outputs, Piperoriginrevid, Json, Output, Merge, Meta\\
    Output Data & \textbf{Explanation:} Topics focus on how model files are saved, structured, or converted across different formats—such as PyTorch, ONNX, or Safetensors—and how these formats are managed or uploaded to repositories like the HF Hub. 
    & \textbf{Explanation:}  Topics indicate changes related to output data management, including handling JSON outputs, managing versions, and merging output files.\\
    \hline
\end{tabular}
\vspace{1em}

\begin{Summary}
{Summary}{On the downstream HF platform, updates primarily involve external documentation—such as licensing, prompt templates, and publication citations—while upstream changes on GH focus exclusively on the codebase, including large-scale refactoring, task-specific modifications, and architecture revisions, with limited overlap in the changes across the two platforms.}
\end{Summary}

\begin{figure}[t]
  \centering
  \includegraphics[width=0.8\textwidth]{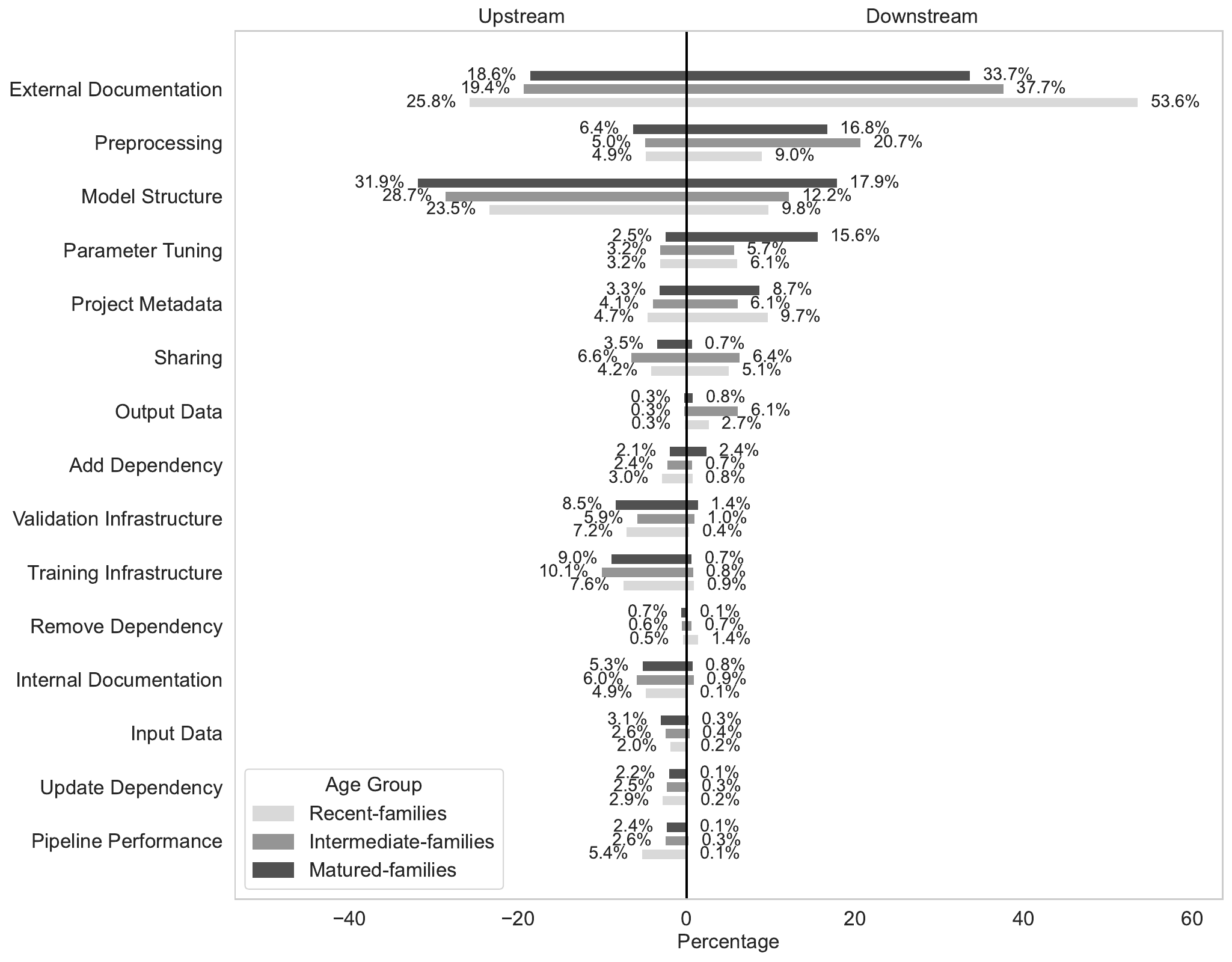} 
  \caption{Variation in the distribution of prevalent PTLM change types across model maturity stages on GH and HF, highlighting shifting emphases on external documentation, model structure, preprocessing, and training infrastructure.}
  \label{argument_1}
\end{figure}

\noindent\textbf{As models mature, the proportion of prevalent \changes evolves, with external documentation decreasing, model structure increasing, and training infrastructure and preprocessing showing differing trends across GH and HF.} As shown in Figure~\ref{argument_1}, the prevalence of external documentation, model structure, preprocessing, and training infrastructure changes over model maturity stages. Some change types show consistent trends, while others vary by commit type, with external documentation decreasing and model structure increasing as models mature.

For recent models, GH prioritizes ``external documentation" (25.8\%) and “model structure" (23.5\%), while ``training infrastructure" is lower (7.6\%). HF emphasizes ``external documentation" (53.6\%), with notable ``model structure" (9.8\%) and ``project metadata" (9.7\%) activities. The prominence of ``project metadata"—non-functional updates like versioning, licensing, or initial setup—reflects downstream efforts to improve discoverability, compliance, and reuse.

For intermediate models, GH ranks ``model structure" highest (28.7\%), followed by ``external documentation" (19.4\%) and ``training infrastructure" (10.1\%). HF maintains ``external documentation" dominance (37.7\%), with ``preprocessing" (20.7\%) and ``model structure" (12.2\%) following. These distributions indicate intermediate models are refined upstream for architecture and training setup, while downstream efforts improve accessibility via documentation and input adaptations. The recurrence of model structure” and external documentation” across both platforms shows their shared importance.

For matured models, GH sees ``model structure" (31.9\%) surpass external ``documentation" (18.6\%), while ``training infrastructure" remains at 9.0\%. HF exhibits continued decline in ``external documentation" (33.7\%), with ``model structure" (17.9\%) and ``preprocessing" (16.8\%) remaining significant. ``Preprocessing" decreases compared to intermediate models, suggesting major data-related adjustments are typically resolved earlier.

Overall, GH consistently features top activities across all age groups: ``external documentation" decreases, ``model structure" increases, and ``training infrastructure" peaks at intermediate stage. On HF, ``external documentation" declines, ``model structure" rises, ``preprocessing" is higher in intermediate models, and ``project metadata" is key in recent models. These trends highlight the complementary roles of GH and HF: GH emphasizes core engineering, HF focuses on accessibility and reuse.

Chi-square tests show statistically significant differences between GH and HF at all maturity stages. Recent models: $\chi^2 = 2675.86$ ($p < 0.0000$, df = 14), Cramér's V = 0.1961 (strong effect). Intermediate: $\chi^2 = 7391.76$ ($p < 0.0000$, df = 14), Cramér's V = 0.2698 (very strong effect). Matured: $\chi^2 = 5981.77$ ($p < 0.0000$, df = 14), Cramér's V = 0.2164 (strong effect). All results are below the Bonferroni-corrected threshold ($p < 0.0167$), confirming significant platform differences.\\

\begin{Summary}
    {Summary}{GH shows more changes to model structure, external documentation, and training infrastructure, while HF emphasizes external documentation, preprocessing, and project metadata—changes that are statistically distinct and shift with model maturity.}
\end{Summary}

\begin{figure*}[t]
\centering
\includegraphics[width=0.8\textwidth]{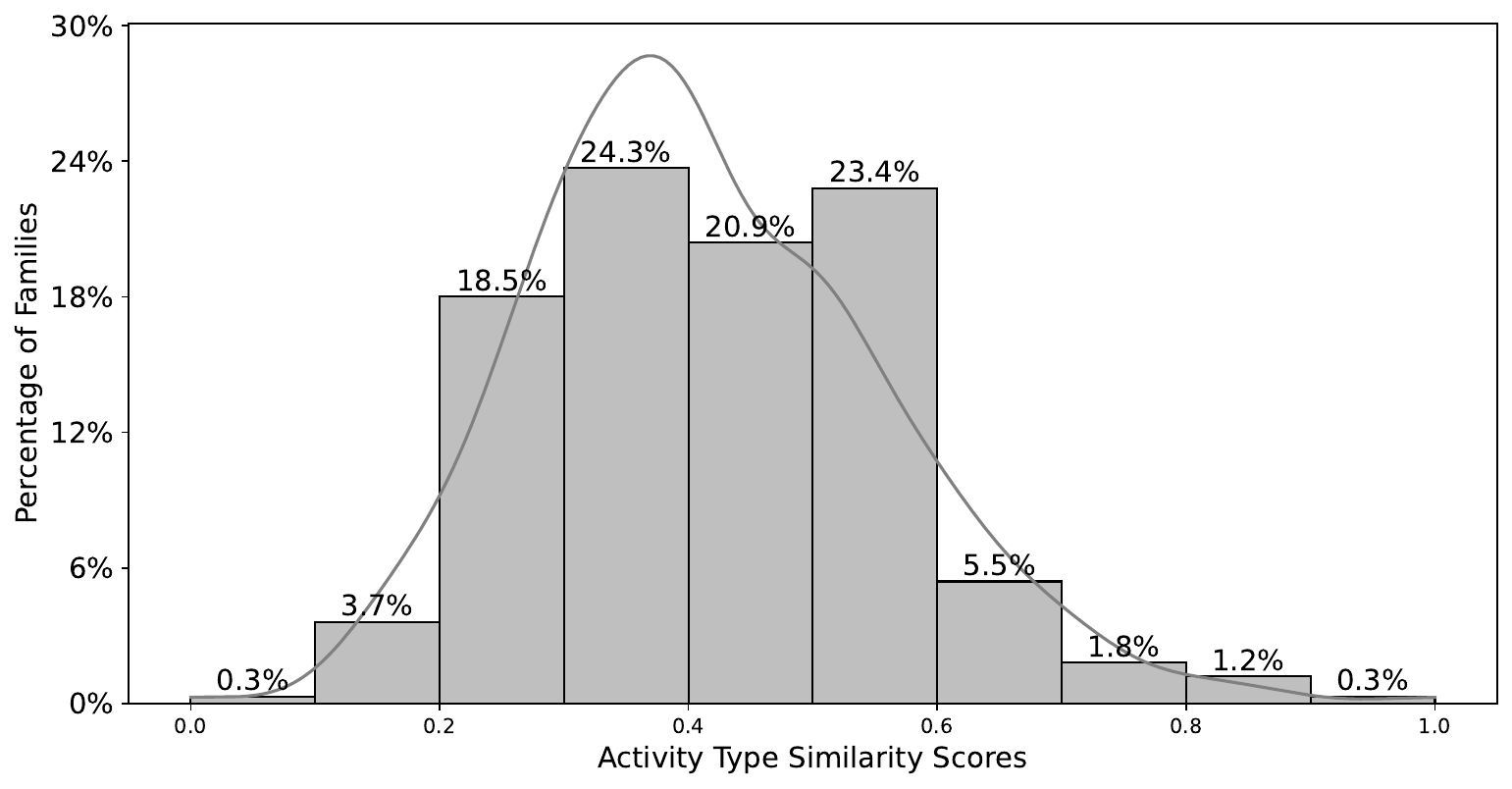}
\caption{Jaccard similarity distribution of commit change types across PTLMs, showing varying degrees of activity synchronization between GH and HF.}
\label{R02}
\end{figure*}

\noindent\textbf{PTLM families exhibit moderate to high similarity scores in their \changes across GH and HF}, with varying overlap, as shown in \Cref{R02}. The Jaccard similarity distribution indicates that 77\% of PTLM families (similarity $>$ 0.3) show similar types of changes between platforms, while a small fraction shows minimal overlap: 0.3\% (0.0–0.1) and 3.7\% (0.1–0.2). About 18.5\% of families exhibit moderate overlap (0.2–0.3), sharing many \changes but with notable differences.

The largest groups are 24.3\% (0.3–0.4) and 20.9\% (0.4–0.5), reflecting significant but not complete alignment. Over 23.4\% show high overlap (0.5–0.6), while smaller fractions show even higher similarity: 5.5\% (0.6–0.7), 1.8\% (0.7–0.8), 1.2\% (0.8–0.9), and 0.3\% (0.9–1.0), indicating nearly identical usage for a few families. Overall, GH and HF often serve complementary roles, but usage patterns vary across PTLM families.

\noindent \textbf{The similarity score between GH and HF is largely independent of the number of HF/GH commits in a family} To examine whether the observed high similarity values between GH and HF commits were influenced by the number of commits per family, we extended our analysis to include the commit volumes on both platforms. Table 1 presents the top 20 PTLM families with the highest similarity scores. These families vary substantially in commit counts—from fewer than 100 to over 16,000—yet maintain consistently high similarity values ($\geq$0.75). For instance, \textit{Family D} (16,356 commits, similarity = 0.80) and \textit{Family B} (2,942 commits, similarity = 0.87) both exhibit strong alignment between GH and HF despite their differing sizes. This suggests that the similarity measure captures genuine correspondence between upstream and downstream activities rather than artifacts of small sample sizes. 

To further confirm this observation, we performed a Mann-Whitney U test on the similarity scores of small versus large families from the whole set of families, which revealed no statistically significant difference (p=0.314). This suggests that the similarity measure captures genuine correspondence between upstream and downstream activities rather than artifacts of small sample sizes.

\begin{table}[t]
    \centering
    \caption{Top-20 families with highest similarity scores and their commit frequencies. The letters serve as placeholders for the family names.}
    \label{similarity_scores}
    \begin{tabular}{l c c c c}
        \toprule
        Family & Similarity Score & \#GH Commits & \#HF Commits & Total \#Commits \\
        \midrule
        A & 1.000000 & 2 & 6 & 8 \\
        B & 0.866667 & 2700 & 242 & 2942 \\
        C & 0.833333 & 34 & 33 & 67 \\
        D & 0.800000 & 16163 & 193 & 16356 \\
        E & 0.800000 & 128 & 151 & 279 \\
        F & 0.750000 & 14 & 6 & 20 \\
        G & 0.727273 & 104 & 36 & 140 \\
        H & 0.714286 & 19 & 31 & 50 \\
        I & 0.714286 & 38 & 18 & 56 \\
        J & 0.714286 & 757 & 561 & 1318 \\
        K & 0.714286 & 34 & 32 & 66 \\
        L & 0.700000 & 163 & 39 & 202 \\
        M & 0.700000 & 44 & 15 & 59 \\
        N & 0.666667 & 3439 & 206 & 3645 \\
        O & 0.666667 & 12 & 7 & 19 \\
        P & 0.666667 & 3 & 9 & 12 \\
        Q & 0.666667 & 1018 & 189 & 1207 \\
        R & 0.666667 & 20 & 8 & 28 \\
        S & 0.666667 & 3 & 7 & 10 \\
        T & 0.666667 & 21 & 6 & 27 \\
        \bottomrule
    \end{tabular}
\end{table}

\begin{figure}[t]
\centering
\includegraphics[width=0.8\textwidth]{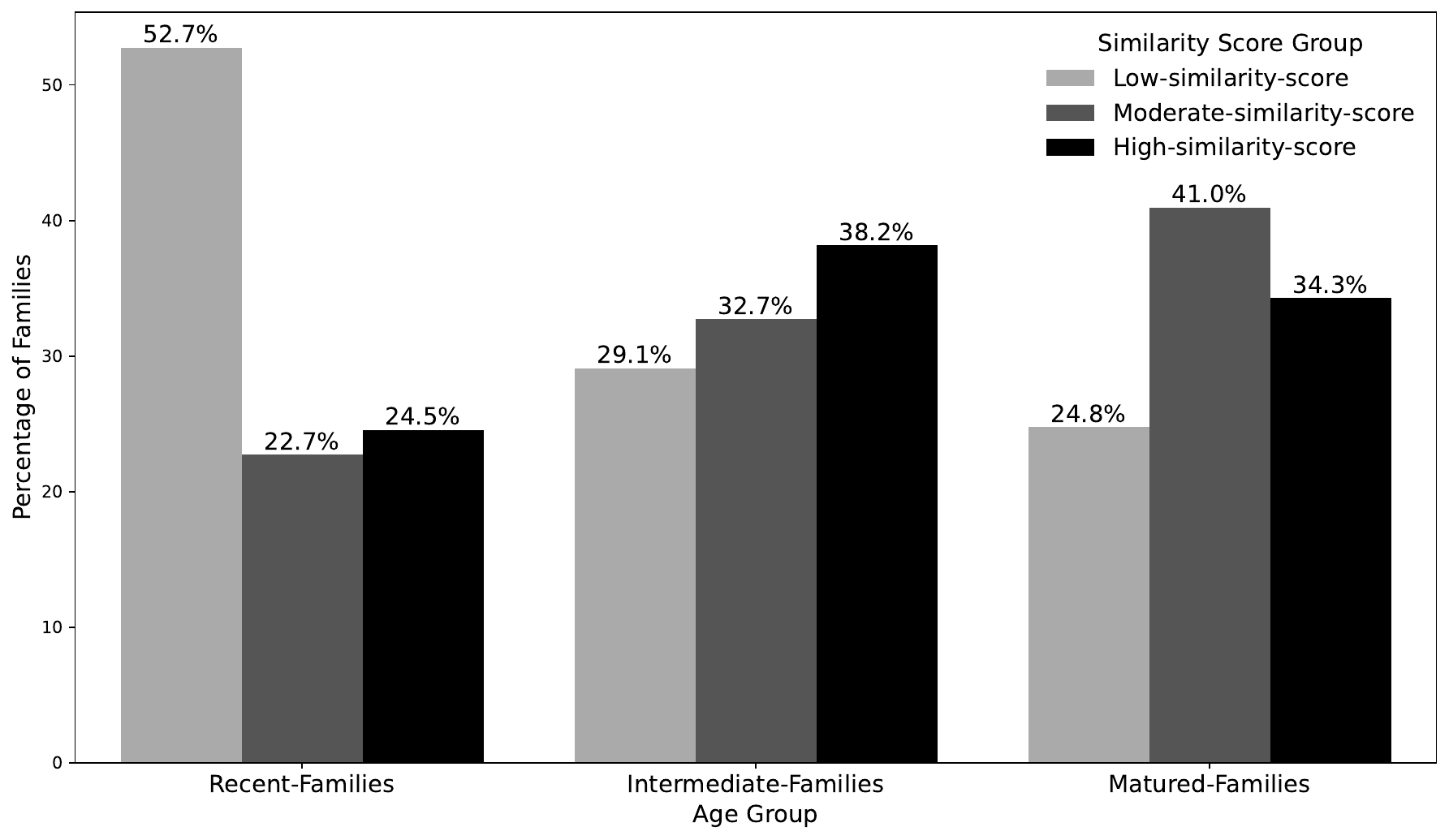}
\caption{Similarity score distribution by \project maturity stage, showing peak synchronization between GH and HF in intermediate models.}
\label{argument2_1}
\end{figure}

\noindent\textbf{Model maturity correlates with similarity scores between GH and HF, peaking in intermediate families.} To explain differences in Jaccard similarity across platforms, we analyzed PTLM families by age (Figure~\ref{argument2_1}). Recent families mostly fall in the low similarity category (52.7\%), indicating substantial divergence. In intermediate families, low similarity decreases to 29.1\%, while moderate (32.7\%) and high (38.2\%) similarity increase, reflecting greater alignment.

For matured families, the largest share (41.0\%) is moderate similarity, high similarity slightly decreases to 34.3\%, and low similarity declines to 24.8\%, indicating fewer families with strong divergence. The drop in high similarity suggests full convergence between GH and HF does not occur with maturity.

These results reveal a non-linear relationship: synchronization peaks at intermediate families, reflecting active refinement across core components, documentation, and usability. As models mature, upstream work stabilizes while downstream shifts toward maintenance and curation, increasing the proportion of moderate similarity scores.

\begin{figure}[t]
\centering
\includegraphics[width=0.8\textwidth]{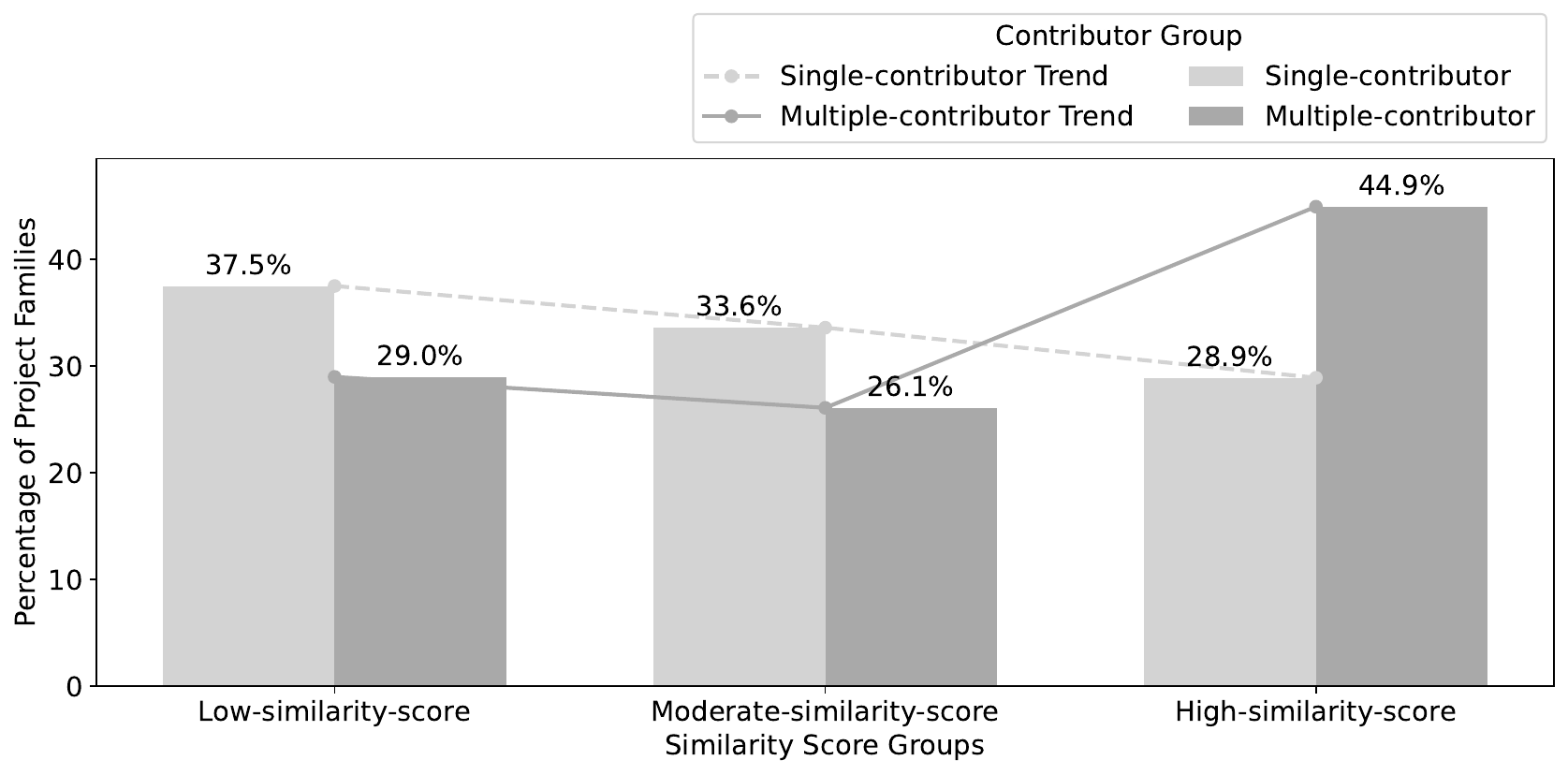}
\caption{Cross-platform contributor distribution across similarity score categories, showing that higher similarity between GH and HF activities is associated with multiple shared contributors.}
\label{argument2_2}
\end{figure}

\noindent\textbf{Similarity scores between GH and HF correlate with the number of cross-platform \contributors, with the highest multiple-author proportion (44.9\%) in PTLM families with high similarity, compared to 29.0\% in low similarity families.} The distribution of single and multiple cross-platform \contributors across similarity categories (Figure \ref{argument2_2}) shows that in low similarity families, 37.5\% have a single author and 29.0\% have multiple authors. For moderate similarity, single-author contributions decrease to 33.6\% and multiple-author contributions to 26.1\%. In high similarity families, single-author contributions drop to 28.9\%, while multiple-author contributions rise to 44.9\%, the highest observed.

A Chi-Square test confirmed a significant relationship between contributor composition and similarity score ($\chi^2 = 6.38$, p = 0.0412), with Cramér’s V = 0.140 (moderate association). While not implying causality, these results suggest that PTLM families with highly aligned change types across platforms are more often associated with multiple cross-platform contributors, reflecting coordinated development practices.

\begin{Summary}
{Summary}{Approximately 77\% of PTLM families show moderate to high similarity in commit change types across GH and HF, with the degree of similarity associated with model maturity and the presence of cross-platform contributors.}
\end{Summary}

\subsection{\textbf{RQ$_2$:} \RQb} 

\subsubsection{Motivation}\label{RQ2_motivation}
In RQ1, our analysis revealed that GH and HF show distinct commit activity types, with clear differences in focus and scope. Although some thematic overlap exists, the nature, distribution, and intensity of these \changes differ significantly between platforms. However, we still do not know how commit activities \coordinate across platforms over time. Understanding this coordination is essential for streamlining releases and optimizing GH-HF integration. We investigate key overlapping commit activities that form synchronization patterns (see \Cref{example}) and the extent of synchronization across platforms. This helps us understand how activities \coordinate across GH and HF repositories.

\subsubsection{Approach}\label{RQ1_approach}
\paragraph{Identifying and explaining factors behind synchronization patterns.} To uncover factors driving synchronization patterns across HF and GH, we analyzed combined commit histories of 325 PTLM families. We manually examined 177 families using open and closed card sorting methods~\citep{wood2008card} and automatically labeled the remaining 148 using a custom script. In the open card sorting phase, we analyzed 50 randomly sampled families to inductively create categories related to delay patterns, commit frequency, and synchronization types, based on commit timeline visualizations. We applied these categories in a closed card sorting phase to the remaining 127 manually reviewed families. We categorized the remaining 148 families automatically using the established scheme.

\noindent\textbf{Step 2.1: Selecting representative samples.} We used stratified random sampling\footnote{https://www.surveymonkey.com/mp/sample-size-calculator/}
 with a 95\% confidence level and a 5\% margin of error~\citep{singh2014sampling, cocks2013sample} to select 177 PTLM families as representative samples for manual analysis.

\noindent\textbf{Step 2.2: Visualizing the representative samples.} To examine synchronization of commit activities between GH and HF for each of the 177 families, we generated scatter plots showing commit dates on both platforms. We aggregated activities into bi-weekly intervals to assess synchronization and identify patterns. Using a two-week window balances capturing significant trends without overloading visualizations. A longer window, such as monthly, would smooth out important events, while a shorter window could clutter plots, especially for long-lived PTLM families.

\begin{figure*}[t]
\centering
\includegraphics[width=\textwidth]{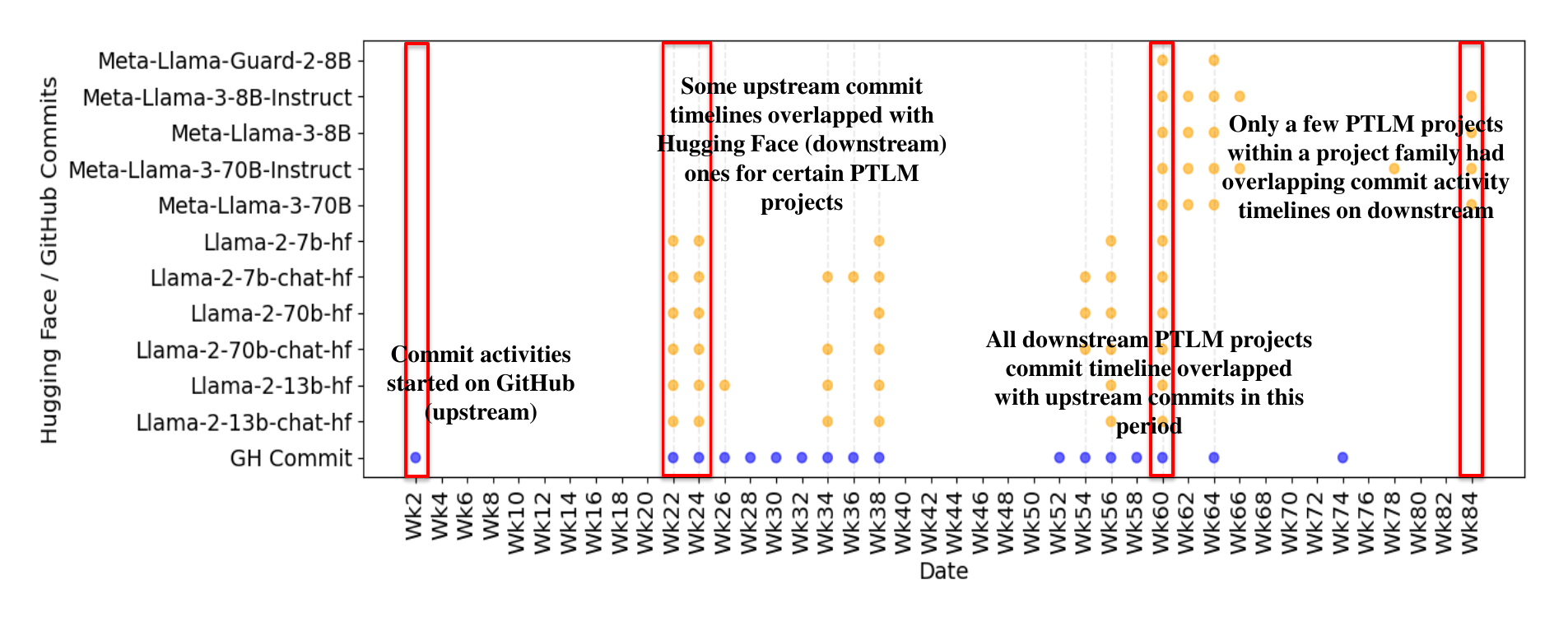}
\caption{Visualization of commit activity patterns for a PTLM family, illustrating activity overlap, synchronization types, and intensity across the upstream (GH) and downstream (HF) platforms. The blue dots represent biweekly commit activities on GitHub, while the yellow dots indicate the biweekly commit activities of different PTLM variants within the same family.}
\label{example}
\end{figure*}

We added vertical dashed lines at the start of each bi-weekly period with at least one commit on both GH and HF (see \Cref{example}). These lines help visualize co-occurrence of activities and identify synchronization or divergence in commit timing across platforms.

\noindent\textbf{Step 2.3: Selection and coding of 50 random samples.} Initially, the first and second authors independently analyzed the visualization of 50 randomly selected PTLM families from the 177 projects to identify factors contributing to the observed synchronization patterns. The process to identify these factors involved:
\begin{compactenum}
    \item Assigning descriptive phrases to observations derived from scatter plots that visualized commit activity relationships between GH and HF, helping surface underlying synchronization patterns (see \Cref{example_observation} for examples. Columns two and three show how each author described their observations.);
    
    \item Comparing, refining, and consolidating these descriptive phrases into a shared set of preliminary codes that captured synchronization characteristics (see columns four to eight in \Cref{example_observation} for examples of intensity and synchronization types, the given codes, interpretations, and final pattern names);
    
    \item Calculating inter-rater agreement between the authors for columns 4 to 7 using Cohen’s Kappa, which yielded a score of 0.73—indicating substantial agreement. This outcome provided a foundation for the closed card sorting phase to consistently categorize synchronization patterns across the remaining dataset.
\end{compactenum}

\begin{table}[t]
\centering
\begin{tabular}{|p{2cm}|p{2cm}|p{2cm}|p{0.5cm}|p{1cm}|p{0.5cm}|p{0.5cm}|p{1cm}|p{1cm}|}
\hline
\textbf{Family} & \textbf{Author 1} & \textbf{Author 2} & \textbf{Act. Lead} & \textbf{Intensity} & \textbf{Sync Types} & \textbf{Code} & \textbf{Meaning of Code} & \textbf{Pattern Name} \\
\hline
hkunlp\_instructor & HF has $\leq$3 biweekly activities but all align with GH. GH has longer activities period but sporadic in nature & Frequent sporadic GH with rare HF& None & R & CS & RCS & Rare complete synchronization & Rare synchronization \\
\hline
benjamin\_segment & HF and GH activities are irregular and partially aligned & Frequent sporadic GH with sporadic HF & HF & S & PS & SPS & Sporadic partial synchronization & Disperse synchronization \\
\hline
meta-llama\_CodeLlama & HF has $\leq$3 biweekly activities that are not aligned with GH. GH extended for a long period but with bursts of activities at times & Sporadic GH with rare HF & HF & R & AS & RAS & Rare Asynchronous & Rare Disjoint Synchronization \\
\hline
\end{tabular}
\caption{Examples of descriptive observations and consolidated codes from qualitative analysis of synchronization patterns across HF and GH. Sync: Synchronization, Act: Activity, CS: Complete Synchronization, PS: Partial Synchronization, AS: Asynchronous.}
\label{example_observation}
\end{table}

\noindent\textbf{Step 2.4: Coding of the Remaining Dataset.} After confirming inter-rater reliability in the previous step, the first author used a closed card sorting approach to code the remaining 127 sampled PTLM families based on the established categories. The substantial agreement observed in the initial phase, reflected by a Cohen’s Kappa score of 0.73, justified this method. Studies with Kappa scores 0.61 and above commonly permit single-rater coding in similar contexts~\citep{el1998benchmarking}. Furthermore, no new codes emerged during this phase, demonstrating consistency with the predefined framework and ensuring a uniform identification of synchronization patterns across the dataset.

\noindent\textbf{Step 2.5: Naming identified patterns.} After coding the dataset, we assigned names to the codes based on three aspects: (1) \textit{activity lead}, indicating which platform initiated activity within a bi-weekly period (e.g., \Cref{example} shows GH starting in week 2, while HF begins in week 22); (2) \textit{synchronization type}, reflecting overlap of activity periods between platforms (e.g., \Cref{example} shows weeks 22–24 where $>$50\% of PTLMs overlap with GH, followed by an 8-week HF burst, with some HF commits outside GH timelines, indicating partial synchronization); (3) \textit{intensity}, capturing concentration and propagation of updates across PTLMs (e.g., \Cref{example} shows a four-week overlap, an 8-week HF burst with $>$90\% PTLMs, sporadic activity, then a 12-week inactivity period, indicating sporadic intensity).

We combined \textit{synchronization type} and \textit{intensity} to define a specific pattern for each family (\Cref{example_observation}). We used \textit{activity lead} as an independent indicator to show which platform usually starts activity.

\noindent\textbf{Step 2.6: Automatic coding and naming of remaining samples.}
To extend analysis to the remaining 148 families, we developed a script~\citep{replication} that processes commit histories, merges datasets, excludes manually coded families, standardizes timestamps, and organizes commit activities into bi-weekly periods.

\begin{itemize}
    \item \textbf{Identifying activity lead:} The script determines which platform starts commits in the initial bi-weekly period (e.g., Intermittent figure on GH\footnote{https://github.com/SAILResearch/replication-25-synchronization-Patterns/tree/main/examples\%20of\%20each\%20synchronization\%20pattern} shows GH leading for 52 weeks before HF activity). We segment commit histories and analyze sequences to detect GH-first, HF-first, or simultaneous starts (\Cref{activity-lead}).
    \item \textbf{Identifying overlapping timelines:} The script categorizes commit trends based on temporal overlap: (1) consistent commits on both platforms, (2) partial alignment, (3) concentrated on one platform, or (4) no clear relationship (example in Frequent figure on GH\footnote{https://github.com/SAILResearch/replication-25-synchronization-Patterns/tree/main/examples\%20of\%20each\%20synchronization\%20pattern}). We detail the algorithm in \Cref{commit_overlap}.
    \item \textbf{Identifying frequency of commit activity:} The script evaluates overlap frequency: (1) few scattered HF commits, (2) consistent long-term alignment, or (3) irregular activity. Frequent figure on GH\footnote{https://github.com/SAILResearch/replication-25-synchronization-Patterns/tree/main/examples\%20of\%20each\%20synchronization\%20pattern} illustrates consistent overlap, and we detail the algorithm in \Cref{commit_intensity}.
\end{itemize}

\noindent\textbf{Step 2.7: Validating and applying the script.}
We validated the script by re-coding the 177 manually annotated families (\Cref{example_observation}). It achieved Cohen’s Kappa scores of 1.00 for lag (column 4) and 0.96 for pattern codes (column 7). The script misclassified three families (labeling “\emph{sporadic intensity} (S)” as “\emph{rare intensity} (R)”) due to commit fluctuations near intensity boundaries. We corrected these manually, then applied the script to the remaining 148 families.

After coding, we assigned meaningful names to each project (\Cref{example_observation}, column 9), corresponding to distinct commit trends and showing how updates unfold across platforms.

\paragraph{Explaining identified synchronization patterns.}
We treated synchronization type and intensity as dimensions defining GH-HF synchronization patterns. These dimensions capture temporal dynamics and distributional characteristics: where activities begin, temporal overlap, and frequency of switches. They represent the core synchronization behaviors of upstream and downstream platforms. We illustrate resulting patterns using PTLM family examples (\Cref{RQ2_results}) to classify patterns and understand release workflows.

\subsubsection{Results.}\label{RQ2_results}
\noindent\textbf{There are three codes for intensity, four codes for synchronization types, and three codes for lag (delay).} Since intensity and synchronization types were combined to form synchronization patterns, some combinations are theoretically implausible or absent in practice. For example, the asynchronous (AS) code lacks overlapping activity, making combinations such as (F, AS, CS) unlikely, while the cross-variant synchronization (VS) typically appears as sporadic and simultaneous. We report only patterns that are empirically grounded in observed commit synchronization. \Cref{factors} summarizes these characteristics.

\renewcommand{\thetable}{\arabic{table}}
\begin{table}[t]
\centering
\begin{tabular}{|>{\raggedright}m{4cm}|m{5cm}|m{4cm}|}
\hline
\textbf{Intensity} & \textbf{Synchronization type} & \textbf{Lag (delay)} \\
\hline
Complete Synchronization (CS) & Rare (R) & GH First \\
\hline
Partial Synchronization (PS) & Frequent (F) & HF First \\
\hline
Cross-variant Synchronization (VS) & Sporadic (S) & Simultaneous Activities \\
\hline
 & No Synchronization (AS) & \\
\hline
\end{tabular}
\caption{synchronization pattern characteristics in PTLMs: Intensity (frequency of activity), synchronization type (overlapping of activities), and Lag (delay) (order of updates)} 
\label{factors}
\end{table}

\noindent\textbf{\intensity:} This characteristic captures how often commit activities occur across GH and HF. We identified three levels—\textit{rare}, \textit{sporadic}, and \textit{frequent}—which are illustrated below with examples.

\begin{itemize}
    \item \textbf{Rare \Intensity:}\\
    This occurs when commit activities are infrequent on one or both platforms, often due to model stability, limited updates, or lack of maintenance interest. Examples: MBZUAI/LaMini-Flan-T5-783M (HF\footnote{https://huggingface.co/MBZUAI/LaMini-Flan-T5-783M} vs GH\footnote{https://github.com/mbzuai-nlp/lamini-lm}) and cais/HarmBench-Llama-2-13b-cls (HF\footnote{https://huggingface.co/cais/HarmBench-Llama-2-13b-cls} vs GH\footnote{https://github.com/centerforaisafety/HarmBench}).\\
    
    \item \textbf{Sporadic \Intensity:} \\
    Characterized by irregular bursts of updates on one platform while the other remains stable. This reflects sprint-based development and platform-specific priorities. Examples: alisawuffles/roberta-large-wanli (HF\footnote{https://huggingface.co/alisawuffles/roberta-large-wanli} vs GH\footnote{https://github.com/alisawuffles/wanli}), Babelscape/wikineural-multilingual-ner (HF\footnote{https://huggingface.co/Babelscape/wikineural-multilingual-ner} vs GH\footnote{https://github.com/Babelscape/wikineural}), aubmindlab/bert-base-arabertv02 (HF\footnote{https://huggingface.co/aubmindlab/bert-base-arabertv02} vs GH\footnote{https://github.com/aub-mind/arabert}). \\

    \item \textbf{Frequent \Intensity:} \\
    Involves consistent updates across both platforms, often sustained over multiple periods. This indicates tight synchronization, shared contributors, and mechanisms such as synchronized deployment cycles, automated update pipelines, or coordinated development workflows. Examples: 01-ai/Yi-1.5-34B-Chat (HF\footnote{https://huggingface.co/01-ai/Yi-1.5-34B-Chat} vs GH\footnote{https://github.com/01-ai/Yi}), BAAI/llm-embedder (HF\footnote{https://huggingface.co/BAAI/llm-embedder} vs GH\footnote{https://github.com/FlagOpen/FlagEmbedding}).
\end{itemize}

\noindent\textbf{Synchronization Type:} This describes how commit activities on GH and HF align over time, reflecting the synchronization of activities across platforms. Inspired by agile practices like dependency management and team synchronization (e.g., daily meetings, sprint planning) \citep{strode2012coordination}, we identified four synchronization types: complete, partial, cross-variant, and none.

\begin{itemize}
    \item \textbf{Complete synchronization (CS):} \\
    Commit activities are consistently time-aligned across GH and HF, showing deliberate synchronization—e.g., code updates on GH and documentation on HF within the same period. Examples: MBZUAI/LaMini-Flan-T5-783M (HF\footnote{https://huggingface.co/MBZUAI/LaMini-Flan-T5-783M}, GH\footnote{https://github.com/mbzuai-nlp/lamini-lm}), 01-ai/Yi-1.5-34B-Chat-16K (HF\footnote{https://huggingface.co/01-ai/Yi-1.5-34B-Chat-16K}, GH\footnote{https://github.com/01-ai/yi}), and cais/HarmBench-Llama-2-13b-cls (HF\footnote{https://huggingface.co/cais/HarmBench-Llama-2-13b-cls}, GH\footnote{https://github.com/centerforaisafety/HarmBench}).

    \item \textbf{Partial synchronization (PS):} \\
    Commit activities align at some points but diverge at others, reflecting intermittent synchronization due to shifting priorities or platform focus. Examples: bigcode/starcoder2-15b (HF\footnote{https://huggingface.co/bigcode/starcoder2-15b}, GH\footnote{https://github.com/bigcode-project/starcoder2}), google-bert/bert-large-uncased-whole-word-masking (HF\footnote{https://huggingface.co/google-bert/bert-large-uncased-whole-word-masking}, GH\footnote{https://github.com/google-research/bert}), and dbmdz/bert-base-german-cased (HF\footnote{https://huggingface.co/dbmdz/bert-base-german-cased}, GH\footnote{https://github.com/dbmdz/berts})).

    \item \textbf{Cross-variant synchronization (VS):} \\
    Synchronization occurs across multiple HF model variants but are not temporally aligned with the corresponding GH repository, reflecting internal HF synchronization independent of GH updates. Examples: google/switch-base-128 (HF\footnote{https://huggingface.co/google/switch-base-128}, GH\footnote{https://github.com/google-research/t5x}), intfloat/multilingual-e5-large (HF\footnote{https://huggingface.co/intfloat/multilingual-e5-large}, GH\footnote{https://github.com/intfloat/SimKGC}), and jinaai/jina-embeddings-v2-base-en (HF\footnote{https://huggingface.co/jinaai/jina-embeddings-v2-base-en}, GH\footnote{https://github.com/jina-ai/finetuner}).

    \item \textbf{No synchronization (AS):} \\
    Activities on GH and HF, or across HF variants, operate independently without alignment, possibly due to different maintainers or missing linked repositories. Examples: CAMeL-Lab/bert-base-arabic-camelbert-mix-ner (HF\footnote{https://huggingface.co/CAMeL-Lab/bert-base-arabic-camelbert-mix-ner}, GH\footnote{https://github.com/CAMeL-Lab/CAMeLBERT}), csebuetnlp/banglat5\_nmt\_bn\_en (HF\footnote{https://huggingface.co/csebuetnlp/banglat5\_nmt\_bn\_en}, GH\footnote{https://github.com/csebuetnlp/banglanmt}), and dandelin/vilt-b32-mlm (HF\footnote{https://huggingface.co/dandelin/vilt-b32-mlm}, GH\footnote{https://github.com/dandelin/ViLT}).
\end{itemize}

\noindent\textbf{Lag (delay):} This refers to the order in which PTLM commit activities first appear on GH and HF. A lag occurs when updates appear on one platform significantly earlier than the other, creating a temporal gap. We identified three types of workflow lag:

\begin{itemize}
    \item \textbf{GH First:} Here, model-related commits begin on GH before being uploaded to HF. This mirrors software workflows where code is first developed upstream before distribution. In PTLM development, GitHub typically hosts training routines or bug fixes, while HF is used later for broader accessibility. For instance, google-bert/bert-large-uncased-whole-word-masking was released on GH\footnote{https://github.com/google-research/bert} in 2019 and appeared on HF\footnote{https://huggingface.co/google-bert/bert-large-uncased-whole-word-masking} only in 2022. Similar patterns were found for dbmdz/bert-base-german-cased (HF\footnote{https://huggingface.co/dbmdz/bert-base-german-cased} vs GH\footnote{https://github.com/dbmdz/berts}) and tner/roberta-large-ontonotes5 (HF\footnote{https://huggingface.co/tner/roberta-large-ontonotes5} vs GH\footnote{https://github.com/asahi417/tner}). 

    \item \textbf{HF First:} In this pattern, models are published on HF before any related GitHub activity. While the initial development may occur privately or internally, HF provides early public access. Code appears later, often due to internal release policies. For example, Efficient-Large-Model/VILA1.5-13b appeared first on HF\footnote{https://huggingface.co/Efficient-Large-Model/VILA1.5-13b}, with a bulk GitHub commit added later\footnote{https://github.com/NVLabs/VILA}. Similarly, LumiOpen/Poro-34B was on HF\footnote{https://huggingface.co/LumiOpen/Poro-34B} a full year before appearing on GH\footnote{https://github.com/LumiOpen/evaluation}. In contrast, MBZUAI/LaMini-Flan-T5-783M had a short delay between the platforms (HF\footnote{https://huggingface.co/MBZUAI/LaMini-Flan-T5-783M} vs GH\footnote{https://github.com/mbzuai-nlp/lamini-lm}). 

    \item \textbf{\Simultaneous Activity:} Commits occur on both platforms within about two weeks, showing no strong order. Updates may differ but reflect synchronized activity. This pattern may result from agile workflows or automation pipelines that coordinate releases across platforms. Examples include 01-ai/Yi-1.5-34B-Chat-16K (HF\footnote{https://huggingface.co/01-ai/Yi-1.5-34B-Chat-16K} vs GH\footnote{https://github.com/01-ai/yi}), bigcode/starcoder2-15b (HF\footnote{https://huggingface.co/bigcode/starcoder2-15b} vs GH\footnote{https://github.com/bigcode-project/starcoder2}), and cais/HarmBench-Llama-2-13b-cls (HF\footnote{https://huggingface.co/cais/HarmBench-Llama-2-13b-cls} vs GH\footnote{https://github.com/centerforaisafety/HarmBench}). 
\end{itemize}

\begin{Summary}{Summary}{We identified three aspects of cross-platform commit behavior—lag, synchronization type, and intensity. Lag measures the time delay between corresponding commits, synchronization type evaluates how activities on GH and HF align over time, and intensity assesses the frequency and consistency of commits on both platforms. Together, these aspects capture key elements of workflow synchronization, reflecting practices that contribute to efficient PTLM release strategies}
\end{Summary}

\noindent\textbf{We identified eight distinct synchronization patterns in commit activities during the release process of PTLMs, based on the two characteristics defined earlier.} These patterns reflect varying levels of synchronization between activities across both ecosystems. Each pattern is illustrated with examples from development activities and is presented in order—beginning with complete synchronization, followed by partial synchronization, and concluding with those that show no synchronization at all.

\noindent\rule{\linewidth}{0.4pt}

\noindent\textbf{\RAP (i.e., Rare Complete Synchronization (RCS))}

\noindent\rule{\linewidth}{0.4pt}

\noindent\textbf{Definition:} This pattern represents \emph{rare but complete synchronization}, where one platform exhibits infrequent commit activity—typically across only two to three periods—while the other maintains a more continuous or extended activity timeline. Despite the disparity in the frequency of commits, each commit on the less active platform is completely overlapped periodically with commits on the more active one.

\noindent\textbf{Explanation:} This pattern may result from several factors beyond deliberate synchronization. Developers might prioritize one platform—often GH for core development—while treating HF as a secondary publishing endpoint, pushing only essential updates. Alternatively, the pattern could stem from limited contributor availability or a lack of ownership across platforms, possibly reflecting a passive or indifferent maintenance approach. While synchronization is technically complete when it happens, its rarity raises significant release engineering concerns: users may unknowingly rely on outdated or incomplete models, reducing reproducibility and potentially undermining trust in the model’s reliability and support.

\noindent\textbf{Notable Instances:} Examples of this pattern can be found in these \project in our analysis, Tner Roberta, Lnyuan Distilbert, and Medicalai ClinicalBert \projects, with an example in the Rare Synchronization figure on GH\footnote{https://github.com/eyinlojuoluwa/Synchronization-Patterns/tree/main/Synchronization\_Examples}. These PTLM families demonstrate that, despite the rarity of synchronization, each commit activity in one platform is fully aligned with activities in the other, reflecting coordinated but \emph{sparse synchronization}.


\noindent\rule{\linewidth}{0.4pt}

\noindent\textbf{\IAP (i.e., Sporadic Complete Synchronization (SCC))}

\noindent\rule{\linewidth}{0.4pt}

\noindent\textbf{Definition:} The \IAP is characterized by commit activities occurring at irregular intervals across five or more periods on either HF or GH, where each activity is \emph{completely coordinated} with corresponding commit activities on the other platform. This pattern differs from the \emph{\RAP} due to a longer duration of activity and greater intensity of synchronization across platforms.

\noindent\textbf{Explanation:} This pattern reflects irregular but recurring synchronization. Commit activities on HF span five or more periods but are spaced unevenly, often aligning fully with GH updates when they occur. This may suggest that contributors update HF in bursts—perhaps following internal milestones or after bundling multiple changes. Alternatively, it could reflect loosely structured workflows, where contributors act independently with minimal synchronization planning. Though synchronization exists, the irregular timing may still delay important updates, leading to possible version drift. For release engineering, this poses moderate risk—users may not receive updates promptly, impacting consistency and deployment accuracy.

\noindent\textbf{Notable Instances:} Examples of \project exhibiting the \IAP in our study include Facebook ESM2, EleutherAI GPT, and Unsloth Models. A typical example of this pattern can be found in the Intermittent Synchronization figure on GH\footnote{https://github.com/SAILResearch/replication-25-synchronization-Patterns/tree/main/examples\%20of\%20each\%20synchronization\%20pattern}, showing how synchronization is maintained, though irregular, between the commit activities on HF and GH.


\noindent\rule{\linewidth}{0.4pt}

\noindent\textbf{\FAP (i.e., Frequent Complete Synchronization (FCC))}

\noindent\rule{\linewidth}{0.4pt}

\noindent\textbf{Definition:} This pattern is characterized by regularly occurring commit activities on both HF and GH for at least three PTLMs within a PTLM family, extending over five or more periods, with complete synchronization maintained throughout.

\noindent\textbf{Explanation:} This pattern reflects a highly disciplined and well-structured release practice. Commit activities on HF occur consistently across five or more consecutive periods, maintaining full synchronization with those on GH. Such regular synchronization could indicate a mature project workflow—possibly driven by strong collaboration among contributors, automated CI/CD pipelines, or well-defined release schedules. In projects with multiple PTLMs, the alignment across all PTLMs in the same period further emphasizes deliberate, synchronized effort. For release engineering, this pattern is ideal: it ensures users can reliably access up-to-date versions across both platforms, reducing confusion, version drift, and maintenance ambiguity. It also boosts confidence in the model’s reliability and long-term support.

\noindent\textbf{Notable Instances:} Examples of \project exhibiting the \FAP in our study include BAAI\_BGE\_models and Maidalun1020 Models. A typical example is available in the Frequent Synchronization figure on GH\footnote{https://github.com/SAILResearch/replication-25-synchronization-Patterns/tree/main/examples\%20of\%20each\%20synchronization\%20pattern}, where synchronization between the \project commit activities on HF and GH is maintained regularly over multiple periods.


\noindent\rule{\linewidth}{0.4pt}

\noindent\textbf{\DAP (i.e., Sporadic Partial Synchronization (SPC))}

\noindent\rule{\linewidth}{0.4pt}

\noindent\textbf{Definition:} The \DAP is characterized by the partial synchronization of commit activities between GH and HF, where some commit activities on one platform occur simultaneously with those on the other, while other commit activities take place independently. This synchronization pattern often involves commit activities on HF extending beyond the active period of commit activities on GH.

\noindent\textbf{Explanation:} This synchronization pattern reflects a transitional or handoff-like behavior between platforms. In many cases, one platform—typically GH—initiates commit activities and maintains them for several periods. As activity on GH declines or ceases, commit activities begin or intensify on HF. The overlap between the two platforms is often brief and limited to only a few PTLMs, with the majority of HF activity occurring independently after GH activity has ended. This raises practical concerns: if changes made on GH during its active period are not fully replicated on HF, users accessing the models later may receive incomplete or outdated versions. Additionally, the source and nature of the updates on HF—whether performed locally, manually, or from external automation—remain unclear, especially when there’s no ongoing synchronization with GH. This partial and time-shifted synchronization may stem from strategic migration, contributor preference, or toolchain differences, but it risks fragmenting the model’s update history and undermining reproducibility.

\noindent\textbf{Notable Instances:} Examples of families exhibiting the \DAP in our study include Cardiffnlp Ttweeteval, Facebook Galactica, and Meta Llama. A typical example is given available in the Dispersed Synchronization figure on GH\footnote{https://github.com/SAILResearch/replication-25-synchronization-Patterns/tree/main/examples\%20of\%20each\%20synchronization\%20pattern}, where the synchronization and independence of commit activities across platforms are evident.



\noindent\rule{\linewidth}{0.4pt}

\noindent\textbf{\SAP (i.e., Rare Partial Synchronization (RPC))} 

\noindent\rule{\linewidth}{0.4pt}

\noindent\textbf{Definition:} This pattern is characterized by brief periods of commit activities on either platform, typically lasting no more than three periods per PTLM. These commit activity periods are spread over a long duration, with extended gaps between them.

\noindent\textbf{Explanation:} This pattern reflects rare and short-lived commit activities across platforms, typically lasting no more than three periods per PTLM, and often spread out with long gaps in between. Commit activities on GH may occur early and infrequently, while HF may exhibit a few rare updates, sometimes appearing only in the final periods. This temporal separation could suggest that the model development was done locally, with contributors pushing finalized code to GH and later uploading the model to HF. In such cases, updates on HF may involve post-release tasks such as documentation or minor adjustments that do not require corresponding changes on GH. While this pattern may seem suboptimal from a release engineering standpoint—due to the limited synchronization and visibility—it may not necessarily indicate poor practices if the development process is well-managed behind the scenes. However, the lack of consistent synchronization and the brevity of activity periods could hinder reproducibility and traceability for end users.

\noindent\textbf{Notable Instances:} Examples of families exhibiting this pattern in our study include Prithivida Parrot, EmergentMethods Gliner, and Wukevin TCR-BERT. A typical example is available in the Sparse Synchronization figure on GH\footnote{https://github.com/SAILResearch/replication-25-synchronization-Patterns/tree/main/examples\%20of\%20each\%20synchronization\%20pattern}.


\noindent\rule{\linewidth}{0.4pt}

\noindent\textbf{\DPAP (i.e., Frequent Partial Synchronization (FPC))}

\noindent\rule{\linewidth}{0.4pt}

\noindent\textbf{Definition:} This pattern is characterized by two distinct phases of commit activities: a phase of scattered and sporadic commit activities with limited synchronization with commit activities on the other platform, followed by a phase of at least five consecutive periods of frequent and coordinated commit activities. These phases can also occur in reverse order.

\noindent\textbf{Explanation:} This pattern reflects a dynamic shift in the synchronization of commit activities between platforms. In the early phase, commit activities are scattered and sporadic, with limited or no alignment across platforms. This may reflect a lack of centralized planning, too many loosely involved contributors, or experimentation without a structured release process. Later, the project enters a more stable phase with frequent and well-coordinated commits on both platforms, sustained over at least five periods. This shift may result from a reduced contributor base, clearer team roles, or the adoption of a more structured release strategy. In some cases, the order is reversed—starting with strong synchronization and later becoming less structured, possibly due to reduced interest or changing priorities. Unlike the \SAP, which shows brief and isolated activity, or the \DAP, where one platform continues after the other, this pattern is marked by a clear transition between sporadic and coordinated phases. It highlights how synchronization practices can evolve as projects mature or as team dynamics shift.

\noindent\textbf{Notable Instances:} Examples of families following this pattern in our study include BigScience Workshop, Mistral-Inference, and JackFram Llama. A typical example of this pattern is available in the Dense Partial Synchronization figure on GH\footnote{https://github.com/SAILResearch/replication-25-synchronization-Patterns/tree/main/examples\%20of\%20each\%20synchronization\%20pattern}.

\noindent\rule{\linewidth}{0.4pt}

\noindent\textbf{\SDP (i.e., Sporadic Asynchronous (SAS))}

\noindent\rule{\linewidth}{0.4pt}

\noindent\textbf{Definition:} The \SDP is characterized by commit activities occurring at different periods on each platform, with no synchronization between commit activities on GH and commit activities on HF.

\noindent\textbf{Explanation:} This pattern reflects a lack of synchronization in commit activities between GH and HF. Commits appear first on one platform and later on the other, often in alternating phases without overlapping periods. This may suggest a sequential workflow where development or updates are first completed and stabilized on one platform—often HF—and only later pushed to GH, possibly after internal testing or review. Such behavior could stem from workflow preferences, platform-specific roles, or a manual syncing process. While this approach allows for flexibility and separation of concerns, it carries risks such as version drift or delayed updates across platforms. From a release engineering perspective, it isn’t inherently flawed, but it may indicate inefficiencies or a lack of automation in maintaining consistency. The absence of overlap may also point to limited resources, fragmented teams, or unintegrated tools. Unlike other patterns, this one shows strictly alternating and non-overlapping activity phases, suggesting a reactive or staged update strategy rather than continuous integration across platforms.

\noindent\textbf{Notable Instances:} Examples of PTLM families exhibiting this pattern include Microsoft BiomedNLP, Alisawuffles Roberta, and Facebook Hubert. A typical example of this pattern is available in the Sporadic Disjoint Synchronization figure on \footnote{https://github.com/SAILResearch/replication-25-synchronization-Patterns/tree/main/examples\%20of\%20each\%20synchronization\%20pattern}.


\noindent\rule{\linewidth}{0.4pt}

\noindent\textbf{\RDP (i.e., Rare Asynchronous (RAS))}

\noindent\rule{\linewidth}{0.4pt}

\noindent\textbf{Definition:} This pattern is characterized by commit activities on one platform being very rare, occurring in no more than three periods per PTLM, and showing no synchronization with commit activities on the other platform.

\noindent\textbf{Explanation:} This pattern shows that commit activities on one platform are very limited and completely uncoordinated with those on the other. In some cases, one platform finishes all its activity before the other even begins. If commit activity appears first on HF, it may suggest that models were trained and uploaded before the training code or supporting resources were made available on GH. Conversely, if GH activity comes first, it could mean the training code was shared and stabilized before any model was uploaded to HF. This sequential and uncoordinated process could be risky from a release engineering standpoint. Users might access the model before understanding how it was built, or access the training code without access to the corresponding model, leading to inconsistencies, confusion, or trust issues. While not necessarily wrong, this approach lacks transparency and may reflect poor release planning or a fragmented development process. In contrast to patterns with even minimal synchronization, this one highlights a complete disconnect between platforms, suggesting missed opportunities for integrated, reliable, and user-aligned releases.

\noindent\textbf{Notable Instances:} Examples of PTLM families exhibiting this pattern in our study include Facebook Roberta, Google Muril, and Meta-llama CodeLlama. A typical example of this pattern is available in the Rare Disjoint Synchronization figure on GH\footnote{https://github.com/SAILResearch/replication-25-synchronization-Patterns/tree/main/examples\%20of\%20each\%20synchronization\%20pattern}.\\


\begin{Summary}
    {Summary}{We identified eight distinct synchronization patterns in commit activities of PTLMs on HF and GH: \emph{Dense partial synchronization}, \emph{Disperse synchronization}, \emph{Frequent synchronization}, \emph{Intermittent synchronization}, \emph{Rare synchronization}, Rare disjoint, \emph{Sparse synchronization}, and \emph{Sporadic disjoint}, which help practitioners understand how commit activities align or diverge across platforms, enabling them to identify synchronization challenges and optimize workflows for smoother PTLM development across HF and GH.}
\end{Summary}

\subsection{\textbf{RQ$_3$:} \RQc} 

\subsubsection{Motivation}\label{RQ3_motivation}
In RQ2, we identified distinct synchronization patterns showing how PTLM families manage commits across GH and HF. To examine how these patterns distribute and evolve across families of different ages, we analyze their prevalence within PTLMs at varying maturity stages. This analysis helps practitioners assess whether their coordination aligns with common patterns and adjust practices to improve cross-platform consistency, responsiveness, or contributor alignment. We also investigate how patterns shift in early versus later stages, particularly how younger and older PTLM families transition between synchronization types, and whether contributor numbers and commit activity associate with these transitions. Understanding these dynamics helps model owners anticipate trends, identify weaknesses in current practices, and supports platform maintainers and researchers in improving synchronization and documentation strategies.

\subsubsection{Approach}\label{RQ2_approach}
\paragraph{Exploring the relationship between change types and synchronization patterns on GH and HF.} We analyzed the relationship between synchronization patterns and change types separately for GitHub and Hugging Face. For GitHub, we merged labeled commit data with synchronization patterns by PTLM family. Since each family has a single synchronization pattern, we grouped data by pattern and change type, counted unique commit messages, and generated a matrix showing the proportion of each change type per pattern. We visualized this matrix as a heatmap.

For Hugging Face, we used a model-level approach because each PTLM family may contain multiple models. We normalized change type counts per model, merged with synchronization patterns, and aggregated by computing the median proportion of each change type across models in a pattern. Median aggregation mitigates the influence of outliers. We visualized the resulting matrix using a log-scaled heatmap with original values as annotations.

To assess whether change type distributions differ between GH and HF for each pattern, we grouped the 15 fine-grained change types into four broader categories following \citet{bhatia2023towards}: maintenance (preprocessing, parameter tuning, model structure), meta program (documentation, sharing, validation infrastructure), data (input/output, project metadata), and dependency management (adding, removing, updating dependencies). This aggregation reduces dimensionality, avoids sparse contingency tables, and ensures reliable chi-square tests. For each pattern, we created a 2×4 table of category-level counts and applied a chi-square test of independence to check for significant differences between platforms.

\noindent\paragraph{Exploring lags (delays) between platforms.} We calculated the proportion of PTLM families in each lag category, converted to percentages, and visualized the results with bar plots. We then examined lag prevalence across project age groups using the same approach. These findings highlight how promptly updates occur across platforms and indicate which platform typically initiates activity, providing a basis for improving release timing.

\noindent\paragraph{Exploring synchronization patterns by PTLM family maturity.} We applied the same calculation and visualization approach to analyze the eight synchronization patterns. We examined the proportion of families exhibiting each pattern across age groups, helping developers and maintainers reflect on how synchronization evolves and adjust practices to ensure consistent, timely updates between GH and HF as PTLMs mature.

\noindent\paragraph{Exploring the correlation between PTLM counts and \contributor counts on synchronization patterns.} To assess whether scale affects coordination, we:
\begin{enumerate}
    \item Categorized each family as “Single-release” (one PTLM) or “Multiple-release” (more than one PTLM) and visualized distributions using a symmetric bidirectional bar chart.
    \item Counted unique contributors per family and platform and used a boxplot to show platform-specific synchronization trends.
    \item Applied the Spearman rank correlation coefficient~\citep{zar1972significance} to evaluate relationships between contributor counts and temporal synchronization.
\end{enumerate}
To enable rank-based analysis, we quantified synchronization by computing, for each family, the average number of biweekly intervals in which GH and HF activity overlapped. This overlap measure converted the nominal synchronization patterns into an ordinal scale, where higher values correspond to stronger coordination. Using this ranked variable, we applied Spearman correlation to assess whether larger teams exhibit tighter synchronization, informing best practices for scaling multi-platform release workflows.

\noindent\paragraph{Exploring the time taken for families across different lags (delays), synchronization patterns, and different maturity levels to communicate changes between GH and HF.} To achieve this, we: 
\begin{enumerate}
    \item Measured the time it takes for a \project to perform corresponding activities on the second platform after previously completing them on the first. This helps us understand the frequency and promptness of updates across platforms.
    \item Examined whether these synchronization times significantly differ across various lags (delays) using the Kruskal-Wallis test—a nonparametric statistical test suitable for comparing three or more independent groups (GH-First, HF-First, Simultaneous in our case) when the data is not normally distributed~\citep{mckight2010kruskal}. The Shapiro-Wilk test~\citep{yazici2007comparison} confirmed non-normality (p = 0.00), justifying the use of the Kruskal-Wallis test. When significant differences were observed (p $<$ 0.05), we applied a post-hoc Dunn’s test with Bonferroni correction\footnote{https://help.easymedstat.com/support/solutions/articles/77000536997-dunn-bonferroni-post-hoc-test} to identify pairwise group differences.
    \item Calculated the average time differences (in days) between initial commit activities on one platform and the appearance of corresponding commits on the other within each synchronization pattern. This reflects the temporal synchronization behavior between platforms.
    \item Tested whether these time intervals vary significantly across synchronization patterns using the same Kruskal–Wallis test approach, followed by Dunn-Bonferroni post-hoc tests where appropriate.
    \item Assessed the average time taken by different maturity groups (based on PTLM family age) to coordinate changes across platforms to explore the effects of maturity on synchronization efficiency.
    \item Used the Kruskal-Wallis test to evaluate differences in these synchronization times across maturity levels, followed by Dunn’s test with Bonferroni correction when significant differences were found.
\end{enumerate} 
This multidimensional analysis offers insight into how synchronization timing varies by lags (delays), structural synchronization types, and project maturity, informing strategies for improving cross-platform synchronization in release engineering.

\subsubsection{Results.}\label{RQ3_results}
\begin{figure*}[t]
\centering
\includegraphics[width=0.8\textwidth]{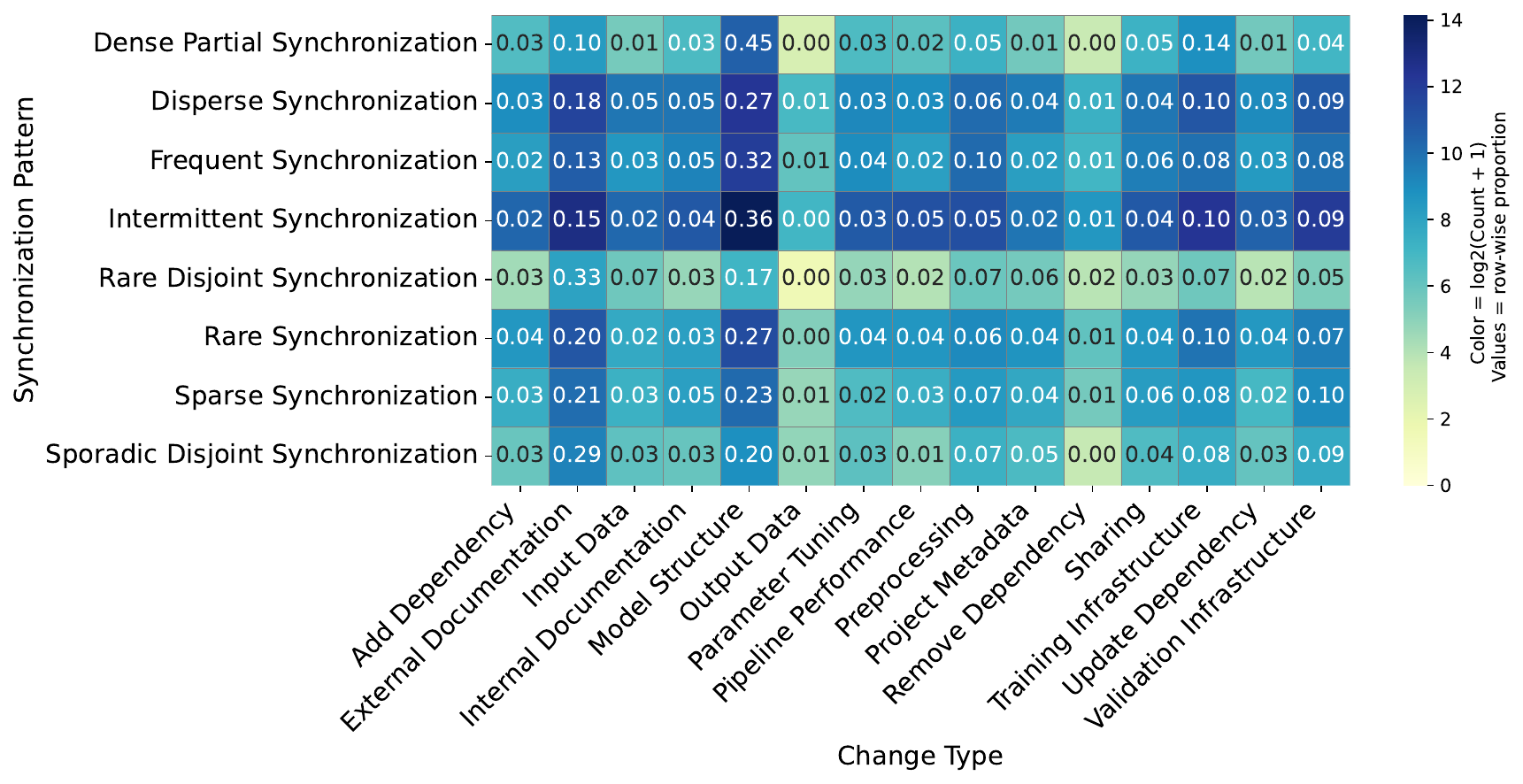}
\caption{Distribution of change types across synchronization patterns on GH. Cell color shows $\log_2$(count + 1) of commits per (pattern, label) pair; text shows row-wise proportion of each change type within a pattern.}
\label{sync_vs_change_GH}
\end{figure*}

\begin{figure*}[t]
\centering
\includegraphics[width=0.8\textwidth]{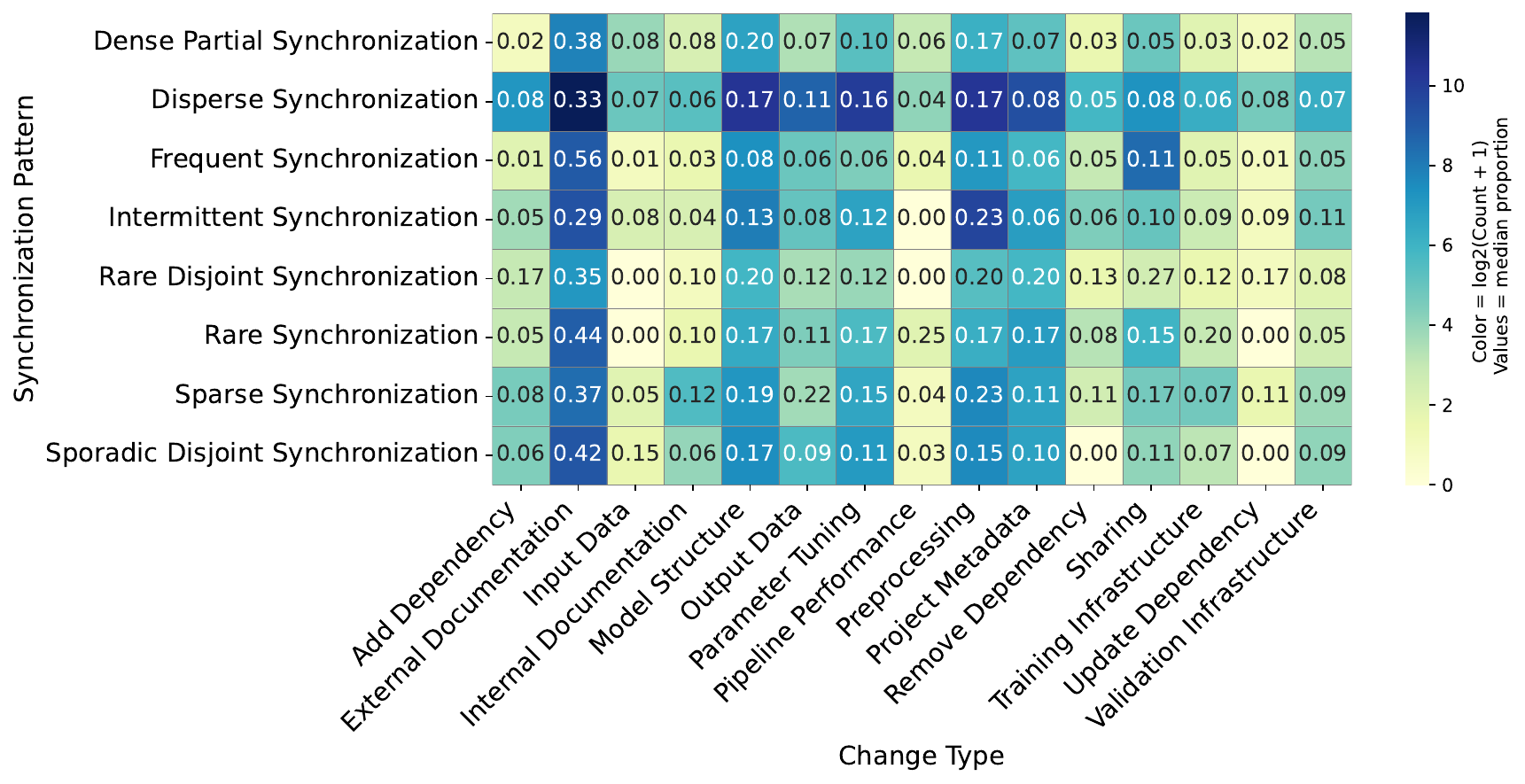}
\caption{Median proportion of change types across synchronization patterns on HF. Cell color shows $\log_{2}(\text{count}+1)$ of raw occurrences; text reports the median proportion of each change type within the pattern.}
\label{sync_vs_change_HF}
\end{figure*}

\noindent\textbf{Most synchronization patterns (six out of eight) focus on model structure on GH, while HF emphasizes external documentation, preprocessing, and model structure.} The visualizations in \Cref{sync_vs_change_GH} and~\Cref{sync_vs_change_HF} confirm RQ1’s findings: GitHub maintains a stable distribution of change types across patterns, whereas Hugging Face shows more variation, with at least four patterns diverging from the platform-wide trend—reflecting a pattern-sensitive approach to downstream updates.

As shown in \Cref{sync_vs_change_GH}, model structure dominates six synchronization patterns on GitHub—including Dense Partial (0.45), Intermittent (0.36), and Frequent (0.32) Synchronization—indicating a strong focus on defining and refining architecture. External documentation and training infrastructure frequently appear among the top three, with documentation leading in Disperse and Sparse, and training infrastructure consistently ranking third elsewhere. Rare Disjoint Synchronization is the only pattern where external documentation (0.33) surpasses model structure, suggesting a shift toward clarity in less coordinated releases.

On HF, external documentation leads all synchronization patterns, from 0.29 in Intermittent to 0.56 in Frequent Synchronization, reflecting Hugging Face’s downstream role. Secondary change types vary: preprocessing appears in Intermittent, Sparse, Frequent, and Rare Disjoint; output data and sharing dominate Disperse, Sparse, and Rare Disjoint. Rare Synchronization prioritizes pipeline performance (0.25) and training infrastructure (0.20), while Sporadic Disjoint emphasizes input data—highlighting the data-centric focus of downstream refinement (\Cref{sync_vs_change_HF}).

We constructed a 2 (platforms: GitHub, HF) × 4 (change categories) contingency table to test whether platform differences are significant within each pattern. All p-values were $<$ 0.05, showing that change type distributions differ meaningfully between platforms. Frequent Synchronization showed the strongest difference ($\chi^2$ = 698.33, p $<$ 0.001), followed by Disperse (411.25) and Rare (390.11). Even smaller chi-square values for Sporadic Disjoint (44.26) and Rare Disjoint (13.27) remained significant, indicating meaningful platform-specific behaviors across all coordination levels.

Overall, GitHub and HF maintain distinct roles across synchronization patterns. Regardless of coordination tightness, practitioners consistently use each platform differently, confirming a stable separation of development and sharing responsibilities.\\

\begin{Summary}
    {Summary}{The upstream–downstream divide is reflected in the dominance of structural changes on GH and documentation-focused changes on HF, with statistically significant differences in change type distributions across all synchronization patterns.}
\end{Summary}

\begin{figure*}[t]
\centering
\includegraphics[width=0.8\textwidth]{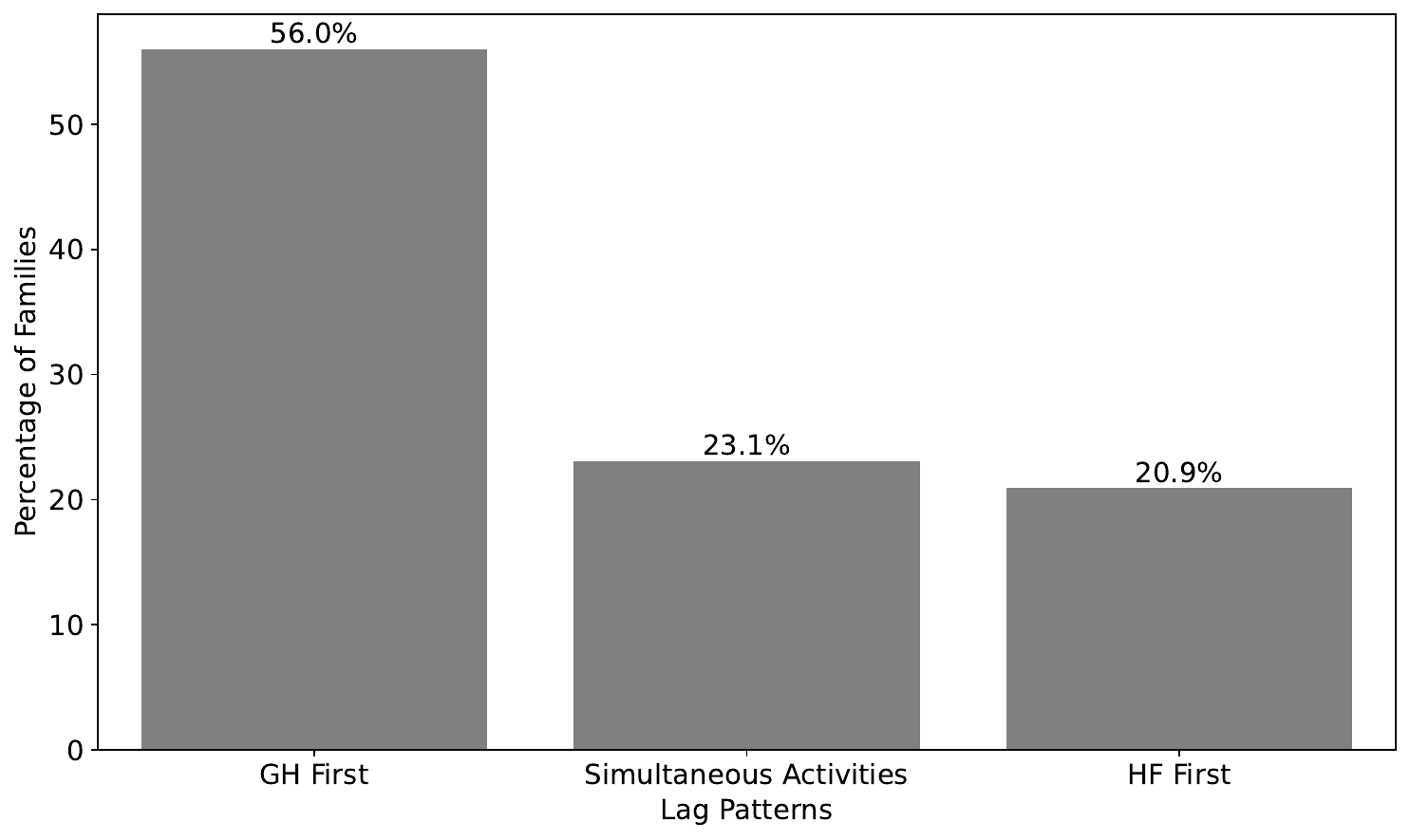}
\caption{Distribution of lags (delays) across PTLM families on HF.}
\label{lagged}
\end{figure*}

\noindent\textbf{Most projects (56\%) start commits on GH, compared to 20.9\% on HF and 23.1\% on both platforms simultaneously.} \Cref{lagged} shows the distribution of lags across PTLM families, indicating GH remains the primary platform for initiating development, especially for older projects predating Hugging Face. For example, stanfordnlp/stanza\footnote{https://huggingface.co/facebook/fasttext-language-identification} and facebook/fasttext-language-identification\footnote{https://huggingface.co/facebook/fasttext-language-identification} started on GH years before corresponding HF commits. This pattern aligns with upstream–downstream workflows in software engineering, where core development occurs upstream (GH) before flowing downstream to distribution platforms like HF \citep{lin2022upstream}.

Analysis of 39 Microsoft-hosted PTLMs shows over 75\% started development on GitHub. During an informal discussion at ICSE 2025, the first author learned that Microsoft conducts multiple internal review cycles and cross-functional meetings before publishing to HF, explaining why commits often appear first on GH, especially for large organizations.

The 23.1\% of families committing simultaneously on both platforms indicate a shift toward more synchronized workflows. Discussions on HuggingFace forums\footnote{\url{https://discuss.huggingface.co/t/github-repo-and-hugging-face-repo-sync/114697}} reflect growing interest in coordinating GH and HF repositories. The 20.9\% starting on HF first may prioritize early model availability without immediately open-sourcing training code. Overall, while GH-first remains dominant—especially for older projects—recent practices embrace tighter GH–HF integration or HF-first releases, reflecting increased awareness of synchronization and distribution needs.

\begin{figure*}[t]
\centering
\includegraphics[width=0.8\textwidth]{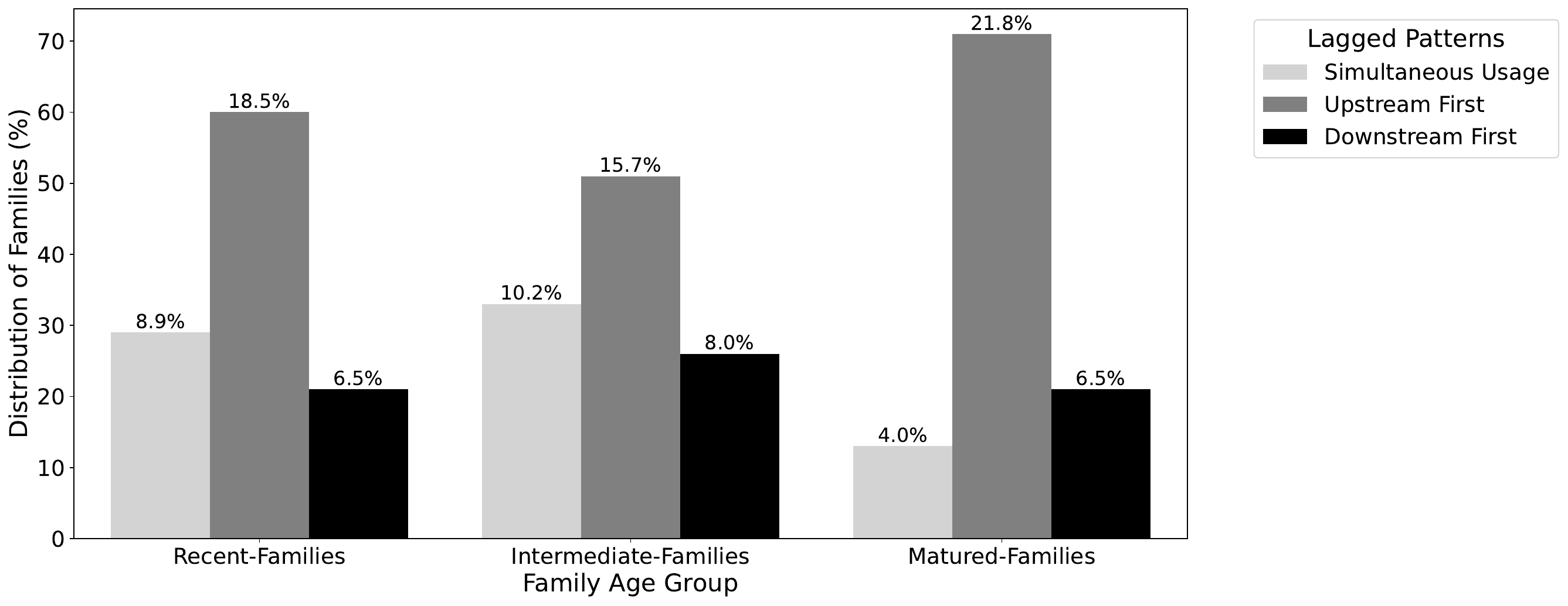}
\caption{The distribution of lagged patterns among HF PTLM families of different age groups, normalized by the number of families (\%) in each group}
\label{activity_lag}
\end{figure*}

\noindent\textbf{Matured PTLM families contribute the most to the GH-first pattern, with 21.8\% starting commits on GH.} As shown in \Cref{activity_lag}, recent (18.5\%) and intermediate (15.7\%) families also follow this pattern, confirming GH as the most common starting point across all maturity stages.

The intermediate group shows the highest rate of simultaneous commits (10.2\%), declining to 4.0\% in the matured group, indicating cross-platform synchronization peaks in mid-stage development. Recent projects do not lead in simultaneous activity; many developers may train models locally in private GH repositories and only publish to HF once stable. GH-first timestamps reflect creation dates rather than public release dates.

These results suggest synchronization practices evolve with project age but not linearly, emphasizing the need for workflows accommodating phased releases and varied public visibility strategies.\\

\begin{Summary}
    {Summary}{Most projects (56\%) begin their commit activities on GH—compared to 20.9\% on HF and 23.1\% initiating on both platforms within the same biweekly period—with matured PTLM families contributing most to the GH-first, accounting for 21.8\% of GH-first.}
\end{Summary}

\begin{figure*}[t]
\centering
\includegraphics[width=0.8\textwidth]{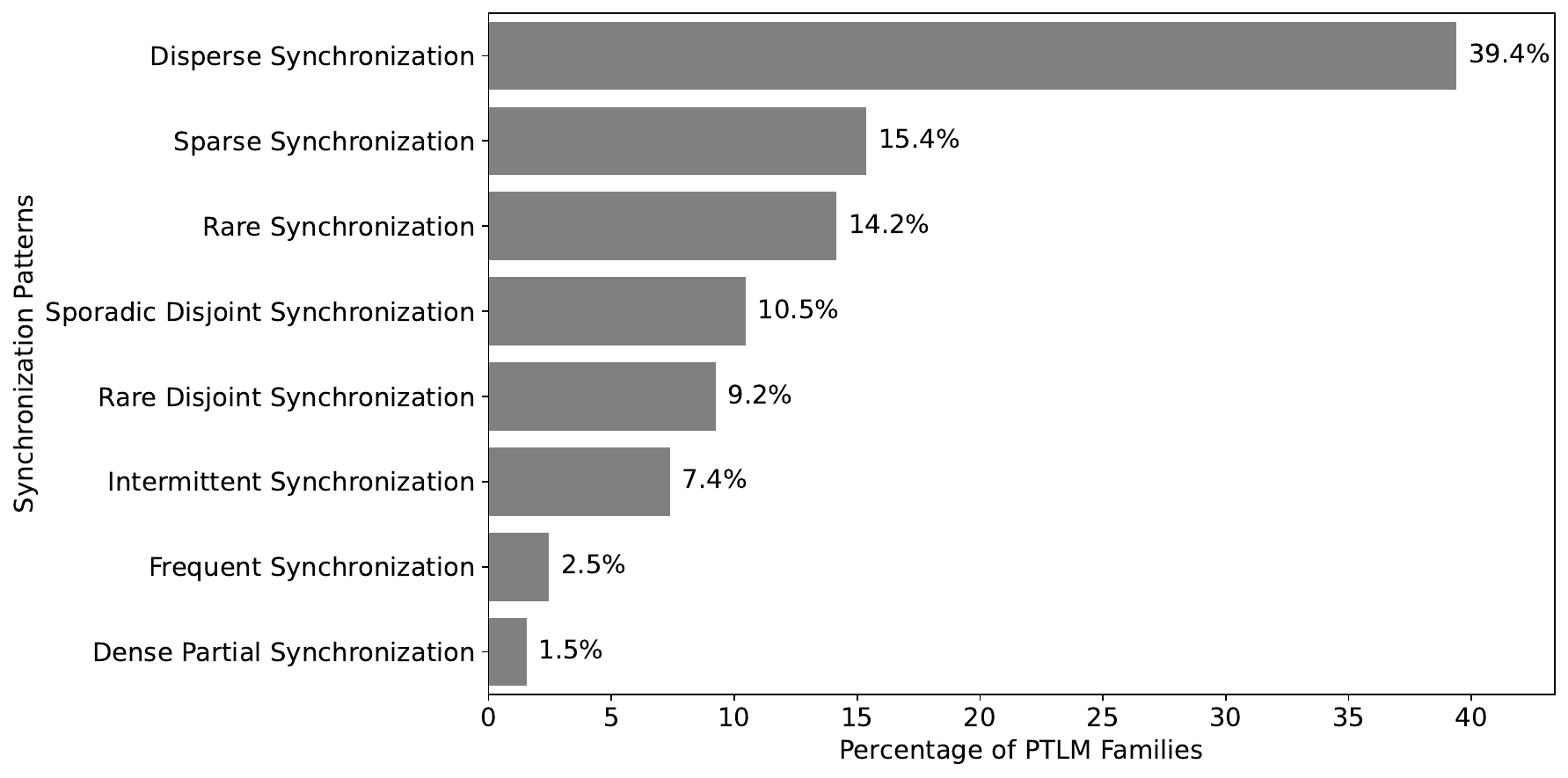}
\caption{The distribution of synchronization patterns among the HF PTLM families}
\label{sync_pattern}
\end{figure*}

\noindent\textbf{\emph{Disperse synchronization} (39.4\%), \emph{Sparse synchronization} (15.4\%), and \emph{Rare synchronization} (14.2\%) dominate PTLM commit activities between GH and HF.} \Cref{sync_pattern} shows the distribution of patterns. \emph{Disperse synchronization} indicates partial alignment, where some HF commits match GH while others occur independently. \emph{Sparse synchronization} reflects infrequent, loosely overlapping commits, and \emph{Rare synchronization} shows limited but fully aligned updates lacking regularity.

\emph{Dense partial synchronization} (1.5\%) and \emph{Frequent synchronization} (2.5\%) are the least common. The former represents transitions between irregular and sustained synchronization, and the latter reflects consistent, coordinated commits over time. The rarity of these patterns highlights challenges in release engineering and limited adoption of cross-platform synchronization, despite potential benefits for downstream accessibility and reliability of PTLMs.

\begin{figure*}[t]
\centering
\includegraphics[width=0.8\textwidth]{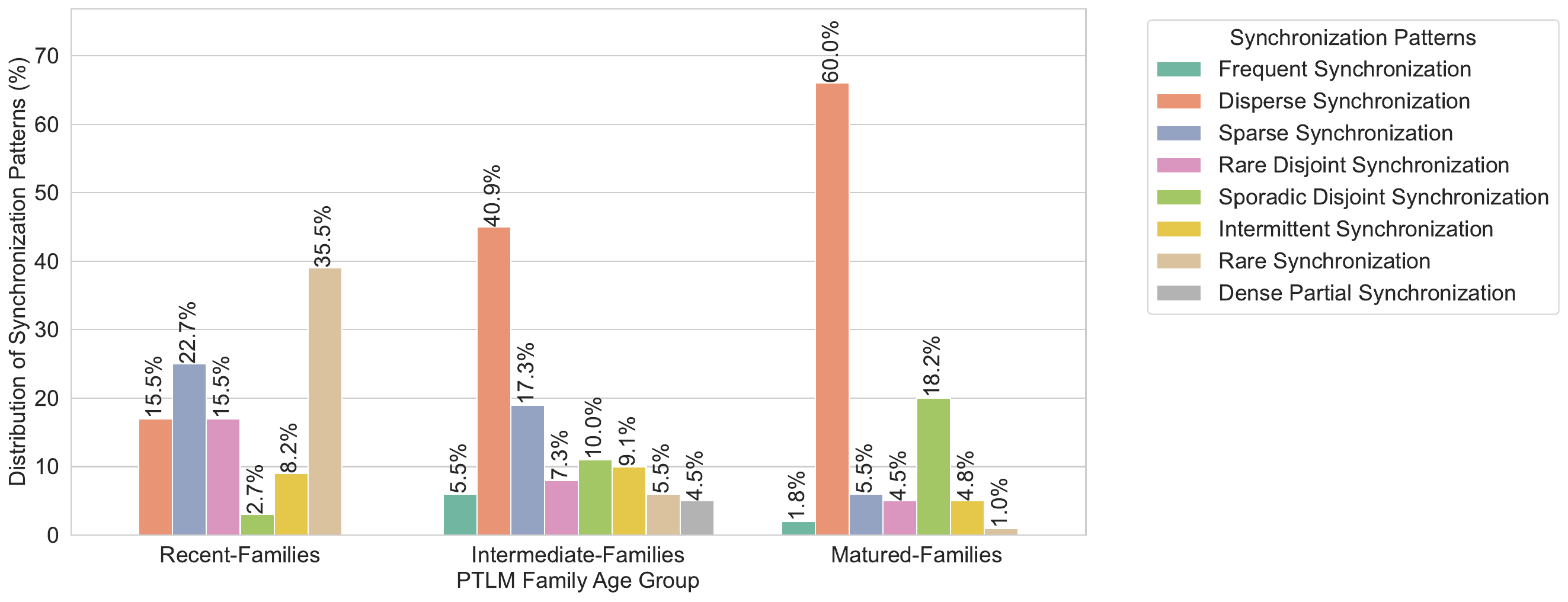}
\caption{Distribution of prevalent synchronization patterns across maturity levels, normalized by number of \project (\%) in each age group.}
\label{activity_sync_ages}
\end{figure*}

\noindent\textbf{\emph{Disperse synchronization} occurred most frequently in intermediate and matured family age groups, while \emph{Rare synchronization} dominated the recent family age group.} \Cref{activity_sync_ages} shows shifts in synchronization as PTLM families mature. \emph{Rare synchronization} drops from 35.5\% in recent families to 1.0\% in matured ones, and \emph{Sparse synchronization} decreases from 22.7\% to 5.5\%, suggesting early alignment efforts fade over time. Meanwhile, less structured patterns rise—\emph{Disperse synchronization} from 15.5\% to 60\%, and \emph{Sporadic Disjoint} from 2.7\% to 18.2\%—indicating more fragmented commit activities. Since HF serves as the main user access point, these patterns raise concerns about the timeliness and completeness of updates.

Well-coordinated patterns remain rare. \emph{Frequent synchronization} occurs in 5.5\% of intermediate families and drops to 1.8\% in matured ones; \emph{Dense partial synchronization} appears in 4.5\% of intermediate families and vanishes in matured families. Achieving and sustaining synchronization becomes harder as complexity grows, reflecting shifting priorities or maintenance fatigue. These findings highlight the need to reinforce synchronization best practices early and integrate lightweight habits into standard workflows, especially for long-term maintenance.\\

\begin{Summary}
    {Summary}{\emph{Disperse synchronization} (39.4\%), \emph{Sparse synchronization} (15.4\%), and \emph{Rare synchronization} (14.2\%) are the most prevalent synchronization patterns observed in PTLM commit activities across GH and HF, with \emph{Disperse synchronization} occurring most frequently in the intermediate and matured family groups, and \emph{Rare synchronization} dominating in the recent family age group.}
\end{Summary}

\begin{figure}[t]
\centering
\includegraphics[width=0.8\textwidth]{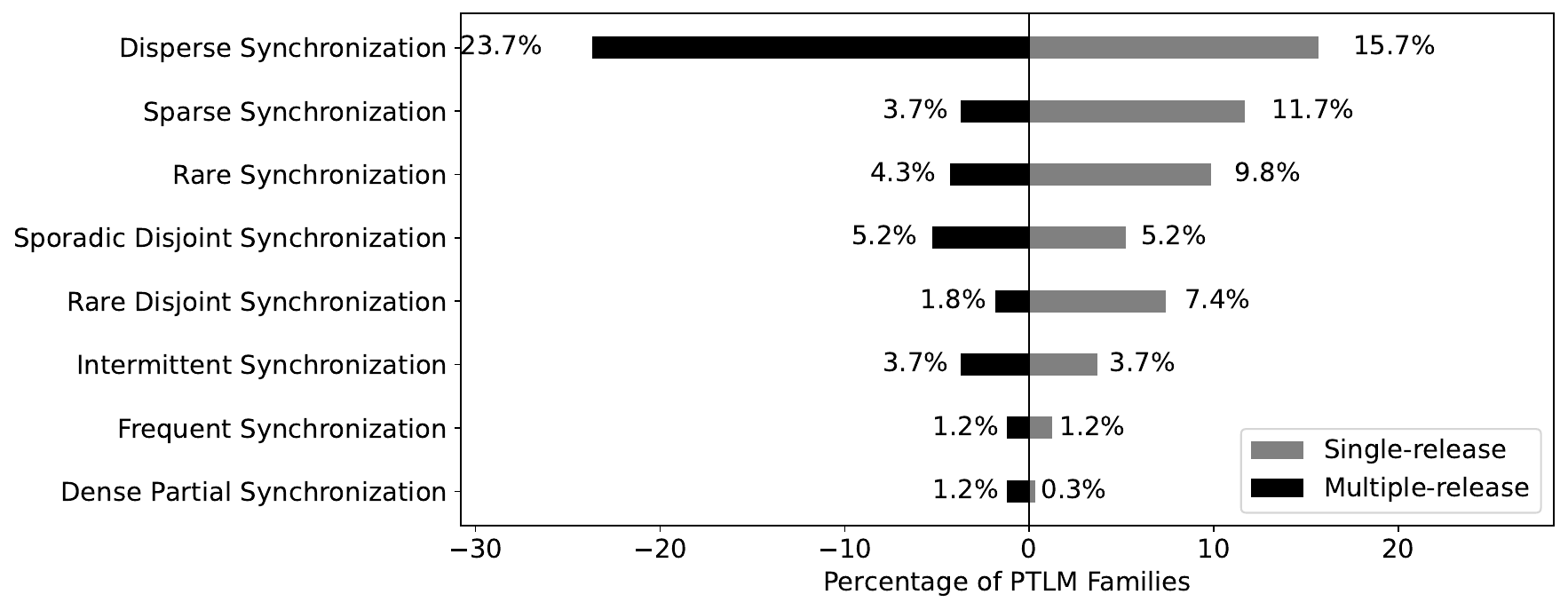}
\caption{Proportion of synchronization patterns in single-PTLM and multiple-PTLM families, with percentages adding up to 100\% across both categories.}
\label{argument3_1}
\end{figure}

\noindent\textbf{Synchronization patterns correlate with the number of PTLMs within a family.} \Cref{argument3_1} shows synchronization patterns across single-PTLM and multiple-PTLM families. \emph{Disperse synchronization} dominates, especially in multiple-PTLM families (23.7\%) versus single-PTLM families (15.7\%), indicating that partial synchronization—where some GH and HF activities overlap while others remain independent—is more common with multiple models. This reflects the challenges of coordinating multiple variants without implying inefficiency, suggesting flexible release strategies for each model's lifecycle.

\emph{Sparse synchronization} occurs more in single-PTLM families (11.7\%) than in multiple-PTLM families (3.7\%), showing that single-model projects experience isolated bursts with long inactivity. Similarly, \emph{Rare Disjoint} and \emph{Rare synchronization} are more frequent in single-PTLM families (7.4\% and 9.8\%) than multiple-PTLM families (1.8\% and 4.3\%), reflecting low-frequency or infrequent full synchronization.

Patterns such as \emph{Sporadic Disjoint}, \emph{Intermittent synchronization}, and \emph{Frequent synchronization} appear at low and balanced frequencies (1.2–5.2\%) across both family types. \emph{Dense partial synchronization}—a transition from sporadic to consistent synchronization—occurs slightly more in multiple-PTLM families (1.2\%) than in single-PTLM families (0.3\%).

These findings show that PTLM family complexity shapes synchronization behavior: multiple-variant families exhibit higher synchronization diversity and intensity, while single-variant families align with intermittent or minimal synchronization.

\begin{figure}[t]
\centering
\includegraphics[width=0.8\textwidth]{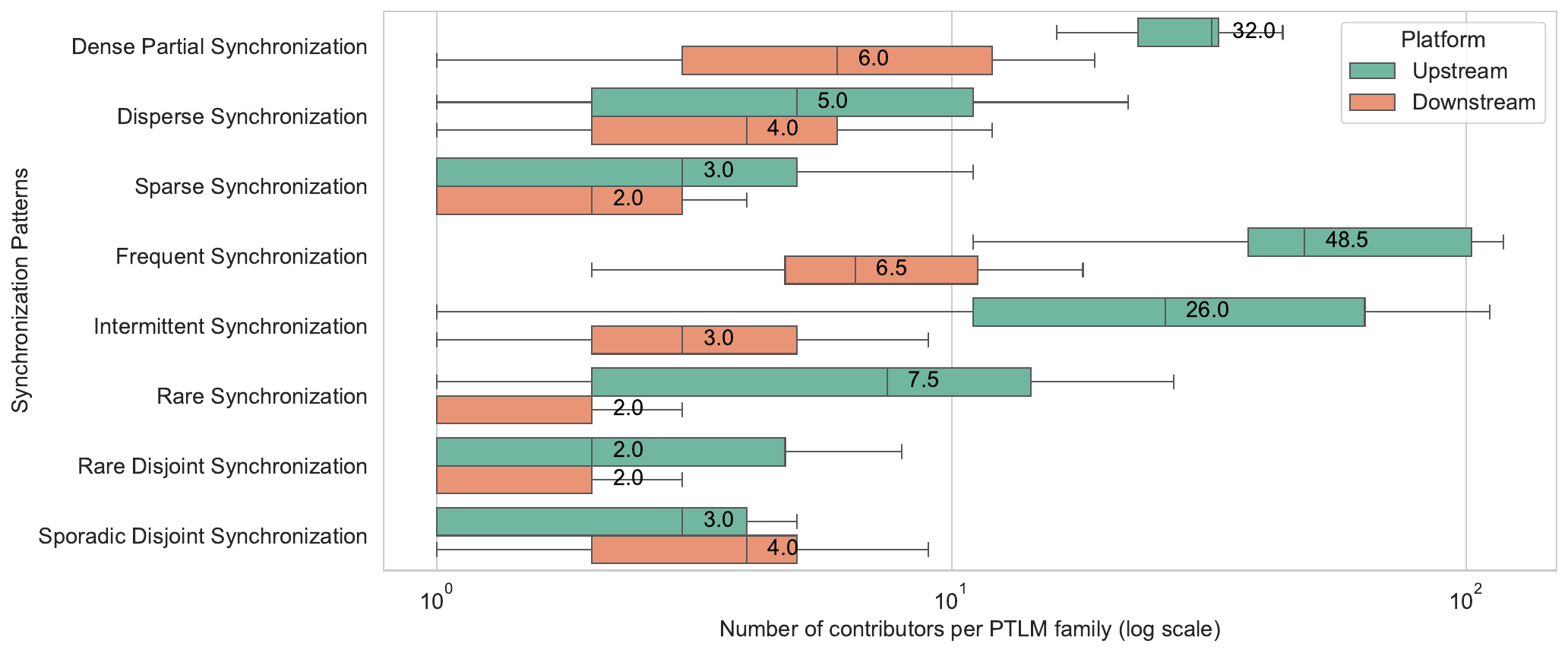}
\caption{Boxplot showing the distribution of the average (median) number of contributors per PTLM family for each synchronization pattern. The median value is annotated within each box for comparison.}
\label{argument3_2}
\end{figure}

\noindent\textbf{Synchronization patterns correlate with contributor counts, but more contributors do not always correlate to structured synchronization.} \Cref{argument3_2} shows contributor distributions across patterns. In Rare, Intermittent, and \emph{Frequent synchronization}, GH consistently has higher counts ($\approx$7.5, 26.0, 48.5) than HF ($\approx$2.0, 3.0, 6.5), indicating that larger GH teams drive updates, with smaller HF teams mirroring these efforts.

In \emph{Disperse synchronization}, counts are balanced (5.0 GH vs. 4.0 HF), reflecting partial coordination. \emph{Sparse synchronization} shows low counts on both platforms (3.0 and 2.0), consistent with minimal collaboration. \emph{Dense partial synchronization} shows GH leading early (32.0) with HF contributing later (6.0). \emph{Sporadic Disjoint} and \emph{Rare Disjoint} exhibit low counts on both platforms ($\approx$3–4 and 2.0), reflecting disjointed efforts.

Contributor distributions highlight that GH often leads development with larger teams, while HF supports or mirrors activity depending on the synchronization pattern.

Spearman correlations reveal strong, significant negative associations (HF: -0.8743, $p < 0.001$; GH: -0.9701, $p < 0.001$), indicating that larger teams tend to produce dispersed or loosely coordinated commits. This suggests that more contributors may correlate with fragmented synchronization due to parallel work streams or decentralized release responsibilities, challenging the assumption that more contributors ensure tightly coordinated releases.\\

\begin{Summary}
    {Summary}{synchronization patterns in PTLM families are shaped by both the number of model variants (PTLMs) and contributors, with multiple-PTLM families exhibiting more diverse and intense synchronization behaviors, while an increase in contributors—especially downstream—correlates with more fragmented and less structured synchronization.}
\end{Summary}

\noindent\textbf{Families that start on HF take longer to perform activities on GH, while simultaneous users switch faster.} Although some PTLM families use both platforms simultaneously, we found no evidence of continuous integration automatically syncing commits between GH and HF. The average delay of 9.32 days suggests synchronization is managed manually or through non-automated workflows.

Families starting on HF take an average of 22.35 days to engage on GH, while GH-first families take 15.82 days to begin on HF. A Kruskal-Wallis test shows significant differences among the three groups—HF First, GH First, and Simultaneous—with all pairwise comparisons significant, indicating that platform engagement patterns affect synchronization speed.

\noindent\textbf{PTLM families with \emph{Frequent} and \emph{Intermittent synchronization} engage the other platform faster, while \emph{Rare Disjoint} and \emph{Sporadic Disjoint} show longer delays.} Engagement times vary across synchronization patterns. \emph{Frequent synchronization} is the fastest (0.87 days), followed by \emph{Intermittent} (1.17 days) and \emph{Dense partial} (3.68 days), likely due to automated tools or dedicated cross-platform teams. In contrast, \emph{Sparse} (26.98 days), \emph{Rare Disjoint} (109.39 days), and \emph{Sporadic Disjoint} (127.39 days) experience longer delays from unstructured synchronization, slowing updates and engagement on the second platform.

\setcounter{table}{2}
\begin{table}[t]
\centering
\begin{tabular}{|c|c|c|c|c|c|c|c|c|}
\hline
\textbf{} & \textbf{DPS} & \textbf{DS} & \textbf{FS} & \textbf{IS} & \textbf{RD} & \textbf{RS} & \textbf{SS} & \textbf{SD} \\
\hline
\textbf{DPS} & -- &  &  &  &  &  &  &  \\
\hline
\textbf{DS} & $<$ & -- &  &  &  &  &  &  \\
\hline
\textbf{FS} & $<$ & $<$ & -- &  &  &  &  &  \\
\hline
\textbf{IS} & $<$ & $<$ & x & -- &  &  &  &  \\
\hline
\textbf{RD} & x & $<$ & $<$ & x & -- &  &  &  \\
\hline
\textbf{RS} & $<$ & $<$ & $<$ & $<$ & $<$ & -- &  &  \\
\hline
\textbf{SS} & $<$ & x & $<$ & $<$ & $<$ & $<$ & -- &  \\
\hline
\textbf{SD} & $<$ & $<$ & $<$ & $<$ & $<$ & x & $<$ & -- \\
\hline
\end{tabular}
\caption{Statistical test of average time taken by each \model family to reflect changes between the two platforms in each synchronization pattern. DPS: Dense partial synchronization, DS: Dispersed synchronization, FS: Frequent synchronization, IS: Intermittent synchronization, RS: Rare synchronization, RD: Rare Disjoint, SS: Sparse synchronization, SD: Sporadic Disjoint. 
The symbol ``--” means self-comparison; ``$<$” denotes $p<0.05$; ``x” indicates no significant difference.}
\label{alignment_comparison}
\end{table}

\Cref{alignment_comparison} shows the statistical significance of these differences using Dunn’s post-hoc test with Bonferroni correction. Most patterns show significant differences in time to act on the other platform, though some do not: \emph{Frequent vs. Intermittent}, \emph{Frequent vs. Rare}, \emph{Intermittent vs. Rare}, \emph{Dense partial vs. Disperse}, \emph{Dense partial vs. Rare}, and \emph{Rare Disjoint vs. Sporadic Disjoint}.\\

\noindent\textbf{Time to communicate changes between GH and HF increases with project maturity, with matured families experiencing the longest delays (24 days).} Recent and intermediate PTLM families take, on average, 9 days to synchronize changes, while matured families average 24 days. A Kruskal-Wallis test confirms significant differences across age groups: Intermediate vs. Matured (p $<$ 0.05), Intermediate vs. Recent (p $<$ 0.05), and Matured vs. Recent (p $<$ 0.05). These results indicate that recent families synchronize \changes faster than intermediate and matured families.\\

\begin{Summary}
    {Summary}{PTLMs using both platforms experience an average synchronization delay of over 9 days, with \emph{Frequent synchronization} families updating the fastest (0.87 days) and \emph{Sparse synchronization}, \emph{Rare Disjoint}, and \emph{Sporadic Disjoint} families showing increasingly longer delays (26.98, 109.39, and 127.39 days, respectively), indicating significant variation in update times and revealing that mature projects tend to have more stable synchronization and slower cross-platform communication.}
\end{Summary}

\section{Discussion and Implication} \label{discussion}
\subsection{Discussion}
The increasing availability of PTLMs has sparked growing interest in model repositories and release engineering practices \citep{min2023recent}. HF, as one of the largest hosting platforms for these models \citep{jiang2023empirical}, plays a central role in shaping their release processes. We argue that these repositories represent only one link in the broader supply chain of AI model development. Our study focuses on the synchronization between model repositories on HF and their upstream code repositories on GH. By examining the types of changes in the commit activities, how these commit are synchronized across these platforms, and the dynamics of these synchronization patterns, we offer a comprehensive view of cross-repository PTLM development. This analysis is grounded in 904 PTLMs linked to 325 GH repositories, encompassing 140,000 GH commits and 17,000 HF commits.

To contextualize our analysis, we begin by distinguishing between synchronization and coordination to clarify cross-platform activity between GH and HF. Synchronization refers to the timing of commits across platforms, while coordination also involves managing interdependencies, aligning goals, and maintaining shared understanding among contributors. For PTLMs, coordination is difficult to assess directly due to the complexity of their supply chain, which spans datasets, model weights, configurations, licenses, and distributed contributors. These layers introduce coordination challenges that are not easily visible through code alone. As a result, we focus on synchronization as a practical first step. 

Building on this framing, our findings confirm a clear division of labor between GH and HF, consistent with their roles in the PTLM release workflow. PTLM families tend to exhibit moderate-to-high similarity in change types across platforms, suggesting some level of synchronization. At the high level of change taxonomy, the most prevalent types on GH are model structure, external documentation, and training infrastructure, while HF prioritizes external documentation, model structure, and preprocessing. Despite this apparent overlap—particularly in model structure and documentation—our topic-level analysis reveals that each platform emphasizes different facets of these shared categories. On GH, model structure changes are primarily code-driven, involving large-scale refactoring (BigRefactor), task-specific updates (MultipleChoiceTask), and restructuring of training scripts and infrastructure (e.g., Makefile, Directory). In contrast, HF’s model structure topics focus on uploading model weights, specifying architecture components, and integrating models into the HF ecosystem. A similar divergence appears in external documentation: GH emphasizes community-facing documentation such as structural layouts and feature explanations, while HF focuses on formal artifacts like licensing terms, research citations, and fine-tuning prompt templates. These differences indicate that even when PTLM updates fall under the same broad categories, the platforms serve distinct functions within the release pipeline. However, synchronization across these roles remains partial and temporally irregular, as indicated by the prevalence of the \emph{Disperse synchronization pattern} found in RQ3.

These platform distinctions are further influenced by the stage of PTLM family maturity. In the early stages of PTLM families, commit activities are often concentrated on a single platform—typically GitHub—while activity on Hugging Face is minimal. When updates do occur on both platforms at this stage, they tend to be tightly aligned in time, resulting in \emph{Rare synchronization patterns} that are minimal but fully synchronized. As projects mature, the volume of activity across both platforms increases, but these updates become more fragmented, with only a subset of changes occurring in close temporal proximity. This leads to the prevalence of \emph{Disperse synchronization} in intermediate and mature PTLM families, where synchronization exists but is partial and inconsistently timed. This trend may reflect challenges similar to those observed in large-scale software systems, where misaligned planning practices—such as fragmented specification, prioritization, estimation, and allocation—hinder inter-team coordination and lead to delays or redundant work \citep{bick2017coordination}. Accordingly, the average time lag between cross-platform updates increases with maturity, reaching 109 days in a case like \emph{Rare Disjoint Synchronization}. Because our dataset focuses on models with over 10,000 downloads, it likely captures PTLMs that are already in their intermediate or mature stages, as more than 50\% of the PTLMs in our dataset are over two years old. This may bias our observations toward Disperse synchronization patterns—where activities often begin on one platform, particularly GH, before gradually extending to HF. This trend may reflect that many mature models were originally maintained on GH before later adopting HF for broader distribution.

In addition to platform roles and maturity, contributor behavior adds another layer of complexity to synchronization dynamics across platforms. While it might be intuitive to assume that larger contributor bases facilitate smoother synchronization, our findings indicate the opposite: PTLM families with more contributors—particularly on HF—tend to exhibit more fragmented synchronization patterns. Although PTLM families with high change similarity scores also show a greater proportion of multi-author contributions across GH and HF, this does not consistently translate into synchronized activity. This apparent contradiction reflects known challenges in large-scale distributed collaboration. Prior research has shown that, despite the benefits of developer coordination—such as improved software quality—teams composed of diverse and widely distributed contributors often face obstacles such as time zone differences, strategic misalignment, limited knowledge sharing, geographical distance, awareness gaps, and organizational boundaries \citep{suali2017developers}. These factors may interrupt the continuity of updates across platforms, leading to disjointed contributions even when collaborative intent exists. In the context of PTLMs—where contributors are drawn from research institutions, industry, and open-source communities—these barriers likely contribute to the prevalent \emph{Disperse synchronization pattern} observed in our study.

These patterns of fragmentation culminate in significant delays. We observed that lags averaging 15.8 days for GH-first updates and exceeding 100 days in disjoint not merely statistical artifacts; they carry tangible implications for the PTLM development process. Such delays may lead to version drift between the upstream code and the downstream model, introducing several critical challenges. First, this fragmentation threatens reproducibility, as users cannot replicate model results if the training code has diverged from the released weights. Second, they enable silent performance regressions, where bug fixes or optimizations on GitHub fail to reach the model on Hugging Face, leaving users with outdated or suboptimal artifacts. Finally, they contribute to trust erosion, as discrepancies between platforms obscure the authoritative source of truth. Addressing these issues requires future work on diagnostic tools to detect harmful drift, synchronization policies defining when and how updates should propagate, and provenance mechanisms to explicitly link each model artifact to its originating commit.

While our study documents the current state of synchronization, the prevalence of patterns like \emph{Disperse} and \emph{Rare} Synchronization reveals a significant gap in the tooling and methodologies available for cross-platform model management. This opens several avenues for future research aimed at moving from descriptive analysis to prescriptive techniques that can actively support synchronization and coordination across repositories.

\subsection{The nature of synchronization: intentional divergence vs. neglect}
While our study focuses on the temporal alignment of commits across GitHub (upstream) and Hugging Face (downstream), timestamps alone cannot explain why certain changes are synchronized or omitted. To understand the nature of these gaps, we conducted a targeted qualitative examination of instances where upstream development continued despite a cessation of downstream activity for at least three months. By manually reviewing 80 commits filtered for core artifact keywords (e.g., tokenizer, weights, config) within these divergence windows, we identified two distinct patterns of behavior.

\textbf{Intentional divergence} occurs when platforms serve different roles, making synchronization unnecessary. Our analysis revealed that the majority of divergent commits fall into this category. For instance, in the DeepPavlov\_rubert family, we observed a GitHub commit titled ``fact retrieval refactoring (\#30)"—which involved extensive code restructuring, cleaning paths, and updating tests—occurring alongside a Hugging Face commit that merely modified tokenizer\_config.json. These updates are independent: the upstream refactoring did not alter the exported model artifacts, and the downstream change did not impact the training codebase. Similarly, documentation edits or CI/CD configuration changes on GitHub are appropriate on one platform but unnecessary on the other.

\textbf{Neglectful drift}, however, presents a more concerning pattern. In our sample, we identified multiple cases where critical GitHub commits affecting core model functionality were never propagated to Hugging Face. For example, in the lmsys\_fastchat family, commits implementing fundamental fixes—such as ``fix tokenizer of chatglm2 (\#2711)" and ``fix: inconsistent tokenization by llama tokenizer (\#3006)"—were never synchronized downstream. Similarly, the EleutherAI\_neox family saw architecture-level modifications like ``add intermediate\_size to gpt-neox models (\#1212)" that were not reflected in the downstream configurations. We also observed cases in the Open-Orca\_OpenOrca family where commits explicitly requiring new weights—such as ``fix untrained tokens (\#1771)" and ``llama3 dpo (\#1610)"—were not followed by a model re-upload. These are not minor updates but core corrections to tokenization, configuration integrity, and model weights that directly impact how a model processes input and generates output.

These two patterns show that high synchronization frequency does not necessarily imply meaningful synchronization. Some repositories exhibited frequent cross-platform updates dominated by low-impact documentation changes, while others with lower synchronization frequency contained critical unsynchronized modifications. This highlights the need for a semantic interpretation of commits:
\begin{itemize}
    \item \textbf{Independent updates} are platform-specific changes that do not require propagation (e.g., documentation edits, CI configuration, refactoring).
    \item \textbf{Interdependent updates} are changes that logically require synchronization to maintain consistency (e.g., model architecture, tokenizer definitions, training code).
\end{itemize}
Distinguishing these categories at a large scale is the logical next step building on our synchronization study, this paper provides the essential foundation for that future work by quantifying the frequency and duration of synchronization gaps. While our current qualitative exploration gives some indications that neglectful drift of interdependent updates is a real problem—potentially leaving downstream users with outdated models—more research is needed to fully quantify the impact of this phenomenon.

\subsection{How the observed synchronization patterns relate to versioning practices}
Although our study does not directly analyze model versioning, the observed synchronization patterns between GH and HF—particularly \textit{Disperse}, \textit{Sparse}, and \textit{Rare} Synchronization—raise practical concerns about the consistency and traceability of PTLM versions across platforms. For instance, in the case of the aubmindlab/bert-base-arabert\footnote{https://huggingface.co/aubmindlab/bert-base-arabert} model, the initial GitHub repository \footnote{https://github.com/aub-mind/arabert} was created on February 21, 2020, while the corresponding Hugging Face model page was established shortly afterward, on February 27, 2020. Notably, on March 12, 2020, a critical update was made on GitHub that corrected the model training procedure by changing the training step parameter from a hardcoded \textit{TRAIN\_STEPS} to the variable \textit{num\_train\_steps}. This adjustment, while syntactically minor, could significantly alter the resulting model’s behavior: if the number of training steps changed, the model might become either undertrained or overfitted, leading to degraded or altered performance.

However, Hugging Face did not reflect any corresponding update in the model weights or documentation until much later—specifically, on July 7, 2020, with an update to the \textit{tf\_model.h5} file. Even then, there is no clear indication that this new file incorporates the March 12 fix, nor is there any metadata explicitly connecting that model weight to the earlier upstream change. This disconnect exemplifies the risks posed by loosely coordinated synchronization: downstream users accessing the Hugging Face model in the interim would have relied on potentially outdated or suboptimal weights, without any visibility into important upstream corrections.

Our findings, coupled with this example, emphasize a broader challenge: when synchronization between upstream and downstream platforms is irregular or delayed, it becomes exceedingly difficult to track what constitutes a new ``version” of a model. Without a semantic versioning scheme or provenance metadata—such as change logs, compatibility information, or training parameters—users cannot reliably determine whether they are using a model version that reflects critical updates. As argued in our prior work \citep{ajibode2025towards}, the multidimensional nature of PTLMs (e.g., changes to architecture, data, or training procedure) makes one-dimensional version numbers inadequate. Thus, better synchronization practices must go hand in hand with versioning standards to ensure reproducibility, user trust, and correct downstream adoption.

\subsection{Implications}
Our findings offer practical implications for different stakeholders involved in the development and maintenance of PTLMs across platforms. In particular, they provide actionable directions for producers, consumers, and researchers aiming to improve synchronization, transparency, and reliability in multi-platform model development.

\noindent \textbf{Producers (Upstream pre-trained model developers and contributors):}
\begin{itemize}[label = $\bullet$, labelsep=1em]
    \item Synchronization between GitHub and HF is often delayed—especially in mature projects—suggesting that contributors should proactively structure release responsibilities and timelines across platforms.
    \item Our findings show that large contributor teams do not necessarily ensure synchronization, highlighting the need for clearer role assignments and well-defined update ownership to avoid fragmented releases.
    \item Synchronizing key updates—such as documentation, checkpoints, and variant releases—across both platforms can help reduce drift and improve the consistency of PTLM families.
\end{itemize}

\noindent \textbf{Consumers (Downstream users of pre-trained models):}
\begin{itemize}[label = $\bullet$, labelsep=1em]
    \item Users should be cautious of assuming that both platforms reflect the latest model state. Delays and partial updates are common, which may result in outdated or inconsistent model versions.
    \item Since GitHub emphasizes technical implementation and HF focuses on user-facing artifacts, users should consult both sources to fully understand a model’s operational state, version lineage, and intended use.
\end{itemize}

\noindent \textbf{Researchers:}
\begin{itemize}[label = $\bullet$, labelsep=1em]
    \item \textbf{Developing Synchronization-Assisting Tools:} The prevalence of Disperse and Sparse Synchronization patterns, characterized by long delays (e.g., 19+ days), indicates a critical need for automation. Future work should focus on designing and evaluating tools that can: (i) automatically detect drift between a GH repository and its corresponding HF models; (ii) generate actionable recommendations that guide developers on when and how to propagate changes from GH to HF or vice versa, thereby resolving detected drift; and (iii) implement CI/CD pipelines for model releases that treat model weights, code, and documentation as interconnected artifacts requiring coordinated deployment.
    
    \item \textbf{Model Versioning and Provenance:} The disconnect between platforms, as illustrated in our versioning case study, highlights the inadequacy of current versioning schemes, thereby confirming the critical need for a Semantic Versioning (SemVer) framework to impose a clear contractual structure (MAJOR.MINOR.PATCH) that communicates the impact and compatibility of PTLM changes to downstream users. Researchers should investigate (i) semantic versioning schemes for PTLMs that encode changes in training data, model weights, architecture, and performance—reflecting the multi-dimensional changes observed in commit activity—and (ii) provenance tracking frameworks that explicitly link a model version on HF to the specific commit(s) on GH that produced it, addressing the traceability issues inherent in Rare Disjoint patterns.
    
    \item \textbf{Coordination Mechanisms for Distributed Teams:} Our finding that more contributors are associated with less synchronization contradicts simple intuition and warrants deeper study. Future research could (i) conduct ethnographic studies or surveys to understand the social and technical barriers that prevent cross-platform coordination; (ii) identify successful coordination strategies used by projects exhibiting Frequent Synchronization patterns and distill them into best practices; and (iii) explore the role of ``platform ambassadors"—contributors active on both GH and HF—and how they facilitate synchronization
    
    \item \textbf{Impact of Synchronization on Model Quality:} The core, unverified assumption is that better synchronization leads to better models. Researchers should empirically investigate whether tighter synchronization correlates with improved model performance, fewer bugs, or higher user trust, and how different synchronization patterns affect the usability, reproducibility, and maintainability of PTLM families over time.
    
    \item \textbf{Generalization to Other Domains:} While this study focused on NLP models, a direct extension is to replicate the analysis in other ML domains (e.g., Computer Vision, Reinforcement Learning) to determine whether similar change types and synchronization patterns hold, or whether domain-specific practices lead to distinct synchronization dynamics.
\end{itemize}

\section{Threats to Validity}\label{ttv}
\subsection{Internal Validity}
A potential threat to internal validity stems from our decision to apply a 10,000-download threshold when selecting models. While this filter helped us focus on high-quality, widely adopted PTLMs, it may have resulted in the underrepresentation of less popular or emerging models—especially those that share a GitHub repository with the included models but fall below the threshold. This incompleteness in capturing full PTLM families could obscure differences in release engineering behaviors across a wider spectrum of model popularity and maintenance practices. Nonetheless, we chose this threshold to improve the reliability of our analysis by ensuring adequate development activity and documentation. More importantly, this decision was grounded in the need for thorough manual analysis across all research questions. Specifically, we manually verified and curated GitHub links for 971 repositories, labeled 1,600 commits for RQ1, and reviewed 177 synchronization patterns for RQ2. Future studies may revisit this trade-off to capture broader PTLM families, including those with lower visibility or niche use cases.

Another threat comes from the automated filtering of GH links. By retaining only links containing either the owner’s name or model name, we aimed to reduce noise and ensure the dataset's relevance. However, this approach may have inadvertently excluded models with unconventional repository naming structures, potentially limiting the completeness of the dataset.

Additionally, the selection of the top 10 most significant change topics for each change type posed a potential threat. In cases where fewer than 10 topics were available—which occurred only on HF—we selected all the available topics. When no topic was available, also only on HF, we used ``No identified instances” as a placeholder and proceeded with the analysis using only GH data. While this ensured consistency in the number of topics considered across platforms, it may have limited the representation of specific types of model maintenance or release activities on HF, such as bug fixes or weight uploads. However, this limitation did not affect the core results of our comparison, as our approach maintained alignment in topic coverage between the two platforms.

Finally, the choice of a bi-weekly interval for aggregating commit activities may have affected the granularity of the analysis. While this interval balanced capturing trends and avoiding clutter, it could misclassify commit patterns in projects with more frequent or slower updates. For instance, a project with concentrated commits in a short time span might be misclassified as ``rare" or ``infrequent." Nevertheless, the bi-weekly window was necessary to capture meaningful synchronization events without overwhelming the visualization, especially for long-lived projects. Despite its limitations, this approach was essential for clear and interpretable results.

While the use of average model age to estimate a PTLM family’s maturity is justified by our finding that 70.9\% of families release all models within the same week, 75\% within four weeks and 86\% within 26 weeks, we acknowledge that this measure may not fully capture maturity for a small subset of families that release multiple models at once but then remain inactive for almost a year or more before releasing additional models. In such cases, the average age could under-represent families that experience long gaps between release cycles. Alternative metrics, such as combining update frequency with average age, may better capture maturity in those few instances. Nevertheless, given the predominance of short release cycles in our dataset, we believe that the average age remains an appropriate and balanced measure for the purposes of this study.

Furthermore, the semantic interpretation of the extracted topics presents a potential threat to validity. Since our topic modeling approach (BERTopic) relies on statistical co-occurrence patterns within technical commit messages, the resulting keywords reflect the raw, granular vocabulary of practitioners rather than a manually curated taxonomy. Consequently, certain keywords may appear contextually ambiguous when viewed in isolation. For instance, terms usually associated with training activities (e.g., `Finetune') may appear in structural categories due to their use in file naming (e.g., finetune.sh), and specific entity names (e.g., `AlexNet', `Chrome') often act as proxies for broader categories. Additionally, we observed instances of generic or noisy keywords, such as 'Transformers' in documentation or `Proxy' in infrastructure, which offer limited explanatory value. We deliberately chose to retain these terms to preserve the authentic, unfiltered vocabulary of MLOps practitioners, rather than manually sanitizing the data, though we acknowledge that this approach requires interpreting keywords within their specific operational contexts.

\subsection{External Validity} A key threat to external validity is the focus on NLP models, which may limit the generalizability of our findings to other domains such as Computer Vision or Reinforcement Learning. Different development practices, repository structures, and documentation standards across these domains could lead to distinct synchronization patterns. While the three synchronization characteristics we examined—lag (delay), synchronization types, and intensity—are likely to be present in other domains, different combinations or interactions of these characteristics may arise. For example, a domain might exhibit high-intensity synchronization with minimal lag but still follow a distinct synchronization type not observed in NLP. Since our methodology is tailored to NLP models, its direct application to other domains may require adjustments to account for domain-specific practices. Nevertheless, our approach offers a structured framework that can be adapted for future studies on model release and synchronization in other fields.

\subsection{Construct Validity} A potential limitation arises from the use of a large language model (LLM) for commit labeling, as the model may introduce inherent biases in its outputs. While we mitigated random variations by repeating the labeling process 10 times per platform and averaging the results, this approach does not eliminate potential systematic biases inherent in the LLM's pretraining data. To improve the reliability of the analysis, a substantial portion of the labeled data was manually reviewed and validated by the first and second authors. Consequently, some classifications may still reflect the model's biases rather than the true nature of the commits. Future work could address this limitation more systematically by incorporating human-in-the-loop validation for bias detection and correction.

Another limitation is our use of the taxonomy from \citet{bhatia2023towards}, originally designed for general ML applications, which may not fully capture PTLM-specific change types—such as tokenizer updates, adapter integration, or fine-tuning data format changes. While this framework served as a useful starting point and has been applied in similar contexts (e.g., \citep{castano2024machine}), we acknowledge that some relevant changes unique to LLM development may be misclassified or overlooked. Future work could refine this taxonomy to better reflect PTLM-specific development activities.

\section{Conclusion}\label{conclusion}
This study investigated the synchronization of commit activities between GH and HF platforms for PTLMs, using a mixed-methods approach. We addressed three main research questions: the types of commit activity changes across platforms, synchronization patterns, and the dynamics of these synchronization patterns across PTLM families.

Our findings show that while GH commits tend to emphasize training infrastructure, model structure, and external documentation, HF activities are more focused on preprocessing steps, external documentation, and model structure. While 77\% of projects show moderate to high similarity in commit activity across platforms, this distribution of changes aligns with the roles these platforms play—GH often driving upstream development and infrastructure setup, while HF supports downstream adaptation and usage refinement.

We identified eight synchronization patterns, with the most common being \emph{Dispersed synchronization} (39.4\%), where changes take an average of 19 days to propagate between platforms. The synchronization pattern evolves as projects mature, with newer projects exhibiting \emph{Rare synchronization} and more mature ones shifting toward Dispersed synchronization. Our analysis indicates that while more contributors can increase activity, it does not guarantee better synchronization, especially as PTLM families mature and contributor engagement declines.

These insights highlight the need for better synchronization and approaches that support synchronization, particularly for PTLM families at different maturity levels.

\section*{Data Availability}
\label{sec:availability}
The datasets generated and analyzed during this study are available in the replication package~\citep{replication}.
\section*{Funding} 
This research was supported by the NSERC Discovery Grant RGPIN-2025-04654.
\section*{Ethical Approval} This study does not involve human participants or animals.
\section*{Informed Consent} No human subjects were involved in this study.
\section*{Conflicts of Interests/Competing Interests}
The authors declare that they have no known competing interests or personal relationships that could have (appeared to) influenced the work reported in this article.
\section*{Author Contributions}
\begin{itemize}
    \item Adekunle Ajibode: Conceptualization, Data Collection, Methodology, Data Analysis, Writing – Original Draft.
    \item Abdul Ali Bangash: Methodology, Data Validation, Writing – Review \& Editing.
    \item Oussama Ben Sghaier: Data Validation, Writing – Review \& Editing.
    \item Bram Adams: Supervision, Writing – Review \& Editing, Conceptual Guidance, Research Direction.
    \item Ahmed E. Hassan: Supervision, Research Direction.
\end{itemize}


\begin{thebibliography}{72}
\providecommand{\natexlab}[1]{#1}
\providecommand{\url}[1]{\texttt{#1}}
\expandafter\ifx\csname urlstyle\endcsname\relax
  \providecommand{\doi}[1]{doi: #1}\else
  \providecommand{\doi}{doi: \begingroup \urlstyle{rm}\Url}\fi

\bibitem[Adrian(2016)]{adrian}
Bridgwater Adrian.
\newblock What is upstream/downstream software?
\newblock \url{https://www.computerweekly.com/blog/Open-Source-Insider/What-is-upstream-downstream-software}, 2016.
\newblock Accessed: 2025-04-03.

\bibitem[Ait et~al.(2025)Ait, C{\'a}novas~Izquierdo, and Cabot]{ait2025suitability}
Adem Ait, Javier~Luis C{\'a}novas~Izquierdo, and Jordi Cabot.
\newblock On the suitability of hugging face hub for empirical studies.
\newblock \emph{Empirical Software Engineering}, 30\penalty0 (2):\penalty0 1--48, 2025.

\bibitem[Ajibode(2025)]{replication}
Adekunle Ajibode.
\newblock Synchronization patterns between github and hugging face, 2025.
\newblock URL \url{https://github.com/SAILResearch/replication-25-synchronization-Patterns}.
\newblock GitHub repository.

\bibitem[Ajibode et~al.(2025)Ajibode, Bangash, Cogo, Adams, and Hassan]{ajibode2025towards}
Adekunle Ajibode, Abdul~Ali Bangash, Filipe~R Cogo, Bram Adams, and Ahmed~E Hassan.
\newblock Towards semantic versioning of open pre-trained language model releases on hugging face.
\newblock \emph{Empirical Software Engineering}, 30\penalty0 (3):\penalty0 1--63, 2025.

\bibitem[Akoglu(2018)]{akoglu2018user}
Haldun Akoglu.
\newblock User's guide to correlation coefficients.
\newblock \emph{Turkish journal of emergency medicine}, 18\penalty0 (3):\penalty0 91--93, 2018.

\bibitem[Aubry et~al.(2023)Aubry, Quaintenne, Dupuy, Francesiaz, Guillemain, and Caizergues]{aubry2023using}
Philippe Aubry, Gwena{\"e}l Quaintenne, Jeremy Dupuy, Charlotte Francesiaz, Matthieu Guillemain, and Alain Caizergues.
\newblock On using stratified two-stage sampling for large-scale multispecies surveys.
\newblock \emph{Ecological Informatics}, 77:\penalty0 102229, 2023.

\bibitem[Berntzen et~al.(2021)Berntzen, Stray, and Moe]{berntzen2021coordination}
Marthe Berntzen, Viktoria Stray, and Nils~Brede Moe.
\newblock Coordination strategies: managing inter-team coordination challenges in large-scale agile.
\newblock In \emph{International Conference on Agile Software Development}, pages 140--156. Springer, 2021.

\bibitem[Bhatia et~al.(2023)Bhatia, Eghan, Grichi, Cavanagh, Jiang, and Adams]{bhatia2023towards}
Aaditya Bhatia, Ellis~E Eghan, Manel Grichi, William~G Cavanagh, Zhen~Ming Jiang, and Bram Adams.
\newblock Towards a change taxonomy for machine learning pipelines: Empirical study of ml pipelines and forks related to academic publications.
\newblock \emph{Empirical Software Engineering}, 28\penalty0 (3):\penalty0 60, 2023.

\bibitem[Bick et~al.(2017)Bick, Spohrer, Hoda, Scheerer, and Heinzl]{bick2017coordination}
Saskia Bick, Kai Spohrer, Rashina Hoda, Alexander Scheerer, and Armin Heinzl.
\newblock Coordination challenges in large-scale software development: a case study of planning misalignment in hybrid settings.
\newblock \emph{IEEE Transactions on Software Engineering}, 44\penalty0 (10):\penalty0 932--950, 2017.

\bibitem[Bock et~al.(2022)Bock, Hunsen, Joblin, and Apel]{bock2022synchronous}
Thomas Bock, Claus Hunsen, Mitchell Joblin, and Sven Apel.
\newblock Synchronous development in open-source projects: A higher-level perspective.
\newblock \emph{Automated Software Engineering}, 29\penalty0 (1):\penalty0 3, 2022.

\bibitem[Casta{\~n}o et~al.(2024{\natexlab{a}})Casta{\~n}o, Caba{\~n}as, Salmer{\'o}n, Lo, and Mart{\'\i}nez-Fern{\'a}ndez]{castano2024machine}
Joel Casta{\~n}o, Rafael Caba{\~n}as, Antonio Salmer{\'o}n, David Lo, and Silverio Mart{\'\i}nez-Fern{\'a}ndez.
\newblock How do machine learning models change?
\newblock \emph{arXiv preprint arXiv:2411.09645}, 2024{\natexlab{a}}.

\bibitem[Casta{\~n}o et~al.(2024{\natexlab{b}})Casta{\~n}o, Mart{\'\i}nez-Fern{\'a}ndez, Franch, and Bogner]{castano2024analyzing}
Joel Casta{\~n}o, Silverio Mart{\'\i}nez-Fern{\'a}ndez, Xavier Franch, and Justus Bogner.
\newblock Analyzing the evolution and maintenance of ml models on hugging face.
\newblock In \emph{2024 IEEE/ACM 21st International Conference on Mining Software Repositories (MSR)}, pages 607--618. IEEE, 2024{\natexlab{b}}.

\bibitem[Chagnon et~al.(2024)Chagnon, Pandolfi, Donatelli, and Ushizima]{chagnon2024benchmarking}
Eric Chagnon, Ronald Pandolfi, Jeffrey Donatelli, and Daniela Ushizima.
\newblock Benchmarking topic models on scientific articles using berteley.
\newblock \emph{Natural Language Processing Journal}, 6:\penalty0 100044, 2024.

\bibitem[Ciraci et~al.(2012)Ciraci, S{\"o}zer, and Tekinerdogan]{ciraci2012approach}
Selim Ciraci, Hasan S{\"o}zer, and Bedir Tekinerdogan.
\newblock An approach for detecting inconsistencies between behavioral models of the software architecture and the code.
\newblock In \emph{2012 IEEE 36th Annual Computer Software and Applications Conference}, pages 257--266. IEEE, 2012.

\bibitem[Cocks and Torgerson(2013)]{cocks2013sample}
Kim Cocks and David~J Torgerson.
\newblock Sample size calculations for pilot randomized trials: a confidence interval approach.
\newblock \emph{Journal of clinical epidemiology}, 66\penalty0 (2):\penalty0 197--201, 2013.

\bibitem[Dabbish et~al.(2012)Dabbish, Stuart, Tsay, and Herbsleb]{dabbish2012social}
Laura Dabbish, Colleen Stuart, Jason Tsay, and Jim Herbsleb.
\newblock Social coding in github: transparency and collaboration in an open software repository.
\newblock In \emph{Proceedings of the ACM 2012 conference on computer supported cooperative work}, pages 1277--1286, 2012.

\bibitem[Devlin et~al.(2018)Devlin, Chang, Lee, and Toutanova]{devlin2018bert}
Jacob Devlin, Ming-Wei Chang, Kenton Lee, and Kristina Toutanova.
\newblock Bert: Pre-training of deep bidirectional transformers for language understanding.
\newblock \emph{arXiv preprint arXiv:1810.04805}, 2018.

\bibitem[Diamantopoulos et~al.(2023)Diamantopoulos, Nastos, and Symeonidis]{diamantopoulos2023semantically}
Themistoklis Diamantopoulos, Dimitrios-Nikitas Nastos, and Andreas Symeonidis.
\newblock Semantically-enriched jira issue tracking data.
\newblock In \emph{2023 IEEE/ACM 20th International Conference on Mining Software Repositories (MSR)}, pages 218--222. IEEE, 2023.

\bibitem[Donald et~al.(2025)Donald, Galanopoulos, Curry, Mu{\~n}oz, Ullah, Waskow, Kalra, Saxena, and Iqbal]{donald2025semantic}
Andy Donald, Apostolos Galanopoulos, Edward Curry, Emir Mu{\~n}oz, Ihsan Ullah, MA~Waskow, Manan Kalra, Sagar Saxena, and Talha Iqbal.
\newblock A semantic approach for linked model, data, and dataspace cards.
\newblock \emph{IEEE Access}, 2025.

\bibitem[El~Emam(1998)]{el1998benchmarking}
Khaled El~Emam.
\newblock \emph{Benchmarking kappa for software process assessment reliability studies}.
\newblock Fraunhofer-IESE, 1998.

\bibitem[Foundjem and Adams(2021)]{foundjem2021release}
Armstrong Foundjem and Bram Adams.
\newblock Release synchronization in software ecosystems: Empirical study on openstack.
\newblock \emph{Empirical Software Engineering}, 26:\penalty0 1--50, 2021.

\bibitem[Giuffrida and Dittrich(2015)]{giuffrida2015conceptual}
Rosalba Giuffrida and Yvonne Dittrich.
\newblock A conceptual framework to study the role of communication through social software for coordination in globally-distributed software teams.
\newblock \emph{Information and Software Technology}, 63:\penalty0 11--30, 2015.

\bibitem[Google(2024)]{google2024gemini}
Google.
\newblock Gemini api pricing, 2024.
\newblock URL \url{https://ai.google.dev/pricing}.
\newblock Accessed on December 9, 2024.

\bibitem[Grootendorst(2022)]{grootendorst2022bertopic}
Maarten Grootendorst.
\newblock Bertopic: Neural topic modeling with a class-based tf-idf procedure.
\newblock \emph{arXiv preprint arXiv:2203.05794}, 2022.

\bibitem[Gu et~al.(2023)Gu, He, and Zhou]{gu2023self}
Haiqiao Gu, Hao He, and Minghui Zhou.
\newblock Self-admitted library migrations in java, javascript, and python packaging ecosystems: A comparative study.
\newblock In \emph{2023 IEEE International Conference on Software Analysis, Evolution and Reengineering (SANER)}, pages 627--638. IEEE, 2023.

\bibitem[Herbsleb et~al.(2001)Herbsleb, Mockus, Finholt, and Grinter]{herbsleb2001empirical}
James~D Herbsleb, Audris Mockus, Thomas~A Finholt, and Rebecca~E Grinter.
\newblock An empirical study of global software development: distance and speed.
\newblock In \emph{Proceedings of the 23rd International Conference on Software Engineering. ICSE 2001}, pages 81--90. IEEE, 2001.

\bibitem[Heri{\v{c}}ko and {\v{S}}umak(2023)]{herivcko2023commit}
Tja{\v{s}}a Heri{\v{c}}ko and Bo{\v{s}}tjan {\v{S}}umak.
\newblock Commit classification into software maintenance activities: A systematic literature review.
\newblock In \emph{2023 IEEE 47th Annual Computers, Software, and Applications Conference (COMPSAC)}, pages 1646--1651. IEEE, 2023.

\bibitem[Hindle et~al.(2008)Hindle, German, and Holt]{hindle2008large}
Abram Hindle, Daniel~M German, and Ric Holt.
\newblock What do large commits tell us? a taxonomical study of large commits.
\newblock In \emph{Proceedings of the 2008 international working conference on Mining software repositories}, pages 99--108, 2008.

\bibitem[Hou et~al.(2024)Hou, Zhao, Liu, Yang, Wang, Li, Luo, Lo, Grundy, and Wang]{hou2024large}
Xinyi Hou, Yanjie Zhao, Yue Liu, Zhou Yang, Kailong Wang, Li~Li, Xiapu Luo, David Lo, John Grundy, and Haoyu Wang.
\newblock Large language models for software engineering: A systematic literature review.
\newblock \emph{ACM Transactions on Software Engineering and Methodology}, 33\penalty0 (8):\penalty0 1--79, 2024.

\bibitem[Ivchenko and Honov(1998)]{ivchenko1998jaccard}
GI~Ivchenko and SA~Honov.
\newblock On the jaccard similarity test.
\newblock \emph{Journal of Mathematical Sciences}, 88:\penalty0 789--794, 1998.

\bibitem[Janke and M{\"a}der(2024)]{janke20247}
Mario Janke and Patrick M{\"a}der.
\newblock 7 dimensions of software change patterns.
\newblock \emph{Scientific Reports}, 14\penalty0 (1):\penalty0 6141, 2024.

\bibitem[Jiang et~al.(2023{\natexlab{a}})Jiang, Jones, Yasmin, Synovic, Sashti, Chen, Thiruvathukal, Tian, and Davis]{jiang2023peatmoss}
Wenxin Jiang, Jason Jones, Jerin Yasmin, Nicholas Synovic, Rajeev Sashti, Sophie Chen, George~K Thiruvathukal, Yuan Tian, and James~C Davis.
\newblock Peatmoss: Mining pre-trained models in open-source software.
\newblock \emph{arXiv preprint arXiv:2310.03620}, 2023{\natexlab{a}}.

\bibitem[Jiang et~al.(2023{\natexlab{b}})Jiang, Synovic, Hyatt, Schorlemmer, Sethi, Lu, Thiruvathukal, and Davis]{jiang2023empirical}
Wenxin Jiang, Nicholas Synovic, Matt Hyatt, Taylor~R Schorlemmer, Rohan Sethi, Yung-Hsiang Lu, George~K Thiruvathukal, and James~C Davis.
\newblock An empirical study of pre-trained model reuse in the hugging face deep learning model registry.
\newblock In \emph{2023 IEEE/ACM 45th International Conference on Software Engineering (ICSE)}, pages 2463--2475. IEEE, 2023{\natexlab{b}}.

\bibitem[Kanaparan and Strode(2025)]{kanaparan2025investigating}
Geetha Kanaparan and Diane~E Strode.
\newblock Investigating the relationship between coordination strategy and coordination effectiveness in agile software development projects.
\newblock \emph{Information and Software Technology}, 182:\penalty0 107708, 2025.

\bibitem[Krause-Glau et~al.(2022)Krause-Glau, Bader, and Hasselbring]{krause2022collaborative}
Alexander Krause-Glau, Marcel Bader, and Wilhelm Hasselbring.
\newblock Collaborative software visualization for program comprehension.
\newblock In \emph{2022 Working Conference on Software Visualization (VISSOFT)}, pages 75--86. IEEE, 2022.

\bibitem[Kraut and Streeter(1995)]{kraut1995coordination}
Robert~E Kraut and Lynn~A Streeter.
\newblock Coordination in software development.
\newblock \emph{Communications of the ACM}, 38\penalty0 (3):\penalty0 69--81, 1995.

\bibitem[Li and Maedche(2012)]{li2012formulating}
Ye~Li and Alexander Maedche.
\newblock Formulating effective coordination strategies in agile global software development teams.
\newblock 2012.

\bibitem[Lin et~al.(2022)Lin, Zhang, Adams, and Hassan]{lin2022upstream}
Jiahuei Lin, Haoxiang Zhang, Bram Adams, and Ahmed~E Hassan.
\newblock Upstream bug management in linux distributions: An empirical study of debian and fedora practices.
\newblock \emph{Empirical Software Engineering}, 27\penalty0 (6):\penalty0 134, 2022.

\bibitem[Lin et~al.(2013)Lin, Ma, and Chen]{lin2013empirical}
Sihai Lin, Yutao Ma, and Jianxun Chen.
\newblock Empirical evidence on developer's commit activity for open-source software projects.
\newblock In \emph{Seke}, volume~13, pages 455--460, 2013.

\bibitem[Liu et~al.(2019)Liu, Ott, Goyal, Du, Joshi, Chen, Levy, Lewis, Zettlemoyer, and Stoyanov]{liu2019roberta}
Yinhan Liu, Myle Ott, Naman Goyal, Jingfei Du, Mandar Joshi, Danqi Chen, Omer Levy, Mike Lewis, Luke Zettlemoyer, and Veselin Stoyanov.
\newblock Roberta: A robustly optimized bert pretraining approach.
\newblock \emph{arXiv preprint arXiv:1907.11692}, 2019.

\bibitem[Loeliger and McCullough(2012)]{loeliger2012version}
Jon Loeliger and Matthew McCullough.
\newblock \emph{Version Control with Git: Powerful tools and techniques for collaborative software development}.
\newblock " O'Reilly Media, Inc.", 2012.

\bibitem[Magelinski et~al.(2022)Magelinski, Ng, and Carley]{magelinski2022synchronized}
Thomas Magelinski, Lynnette Ng, and Kathleen Carley.
\newblock A synchronized action framework for detection of coordination on social media.
\newblock \emph{Journal of Online Trust and Safety}, 1\penalty0 (2), 2022.

\bibitem[Malone and Crowston(1994)]{malone1994interdisciplinary}
Thomas~W Malone and Kevin Crowston.
\newblock The interdisciplinary study of coordination.
\newblock \emph{ACM Computing Surveys (CSUR)}, 26\penalty0 (1):\penalty0 87--119, 1994.

\bibitem[Manakul et~al.(2023)Manakul, Liusie, and Gales]{manakul2023selfcheckgpt}
Potsawee Manakul, Adian Liusie, and Mark~JF Gales.
\newblock Selfcheckgpt: Zero-resource black-box hallucination detection for generative large language models.
\newblock \emph{arXiv preprint arXiv:2303.08896}, 2023.

\bibitem[Mao(2020)]{mao2020survey}
Huanru~Henry Mao.
\newblock A survey on self-supervised pre-training for sequential transfer learning in neural networks.
\newblock \emph{arXiv preprint arXiv:2007.00800}, 2020.

\bibitem[McHugh(2013)]{mchugh2013chi}
Mary~L McHugh.
\newblock The chi-square test of independence.
\newblock \emph{Biochemia medica}, 23\penalty0 (2):\penalty0 143--149, 2013.

\bibitem[McKight and Najab(2010)]{mckight2010kruskal}
Patrick~E McKight and Julius Najab.
\newblock Kruskal-wallis test.
\newblock \emph{The corsini encyclopedia of psychology}, pages 1--1, 2010.

\bibitem[Min et~al.(2023)Min, Ross, Sulem, Veyseh, Nguyen, Sainz, Agirre, Heintz, and Roth]{min2023recent}
Bonan Min, Hayley Ross, Elior Sulem, Amir Pouran~Ben Veyseh, Thien~Huu Nguyen, Oscar Sainz, Eneko Agirre, Ilana Heintz, and Dan Roth.
\newblock Recent advances in natural language processing via large pre-trained language models: A survey.
\newblock \emph{ACM Computing Surveys}, 56\penalty0 (2):\penalty0 1--40, 2023.

\bibitem[Mockus and Votta(2000)]{mockus2000identifying}
Mockus and Votta.
\newblock Identifying reasons for software changes using historic databases.
\newblock In \emph{Proceedings 2000 international conference on software maintenance}, pages 120--130. IEEE, 2000.

\bibitem[Mockus et~al.(2000)Mockus, Fielding, and Herbsleb]{mockus2000case}
Audris Mockus, Roy~T Fielding, and James Herbsleb.
\newblock A case study of open source software development: the apache server.
\newblock In \emph{Proceedings of the 22nd international conference on Software engineering}, pages 263--272, 2000.

\bibitem[OpenAI(2023)]{openai2023gpt}
R~OpenAI.
\newblock Gpt-4 technical report. arxiv 2303.08774.
\newblock \emph{View in Article}, 2:\penalty0 13, 2023.

\bibitem[Oreamuno et~al.(2024)Oreamuno, Khan, Bangash, Stinson, and Adams]{oreamuno2024state}
Ernesto~Lang Oreamuno, Rohan~Faiyaz Khan, Abdul~Ali Bangash, Catherine Stinson, and Bram Adams.
\newblock The state of documentation practices of third-party machine learning models and datasets.
\newblock \emph{IEEE Software}, 41\penalty0 (5):\penalty0 52--59, 2024.

\bibitem[P{\'e}rez et~al.(2020)P{\'e}rez, D{\'\i}az, Garcia-Martin, and Tabuenca]{perez2020systematic}
Jorge P{\'e}rez, Jessica D{\'\i}az, Javier Garcia-Martin, and Bernardo Tabuenca.
\newblock Systematic literature reviews in software engineering—enhancement of the study selection process using cohen’s kappa statistic.
\newblock \emph{Journal of Systems and Software}, 168:\penalty0 110657, 2020.

\bibitem[Ray et~al.(2013)Ray, Kim, Person, and Rungta]{ray2013detecting}
Baishakhi Ray, Miryung Kim, Suzette Person, and Neha Rungta.
\newblock Detecting and characterizing semantic inconsistencies in ported code.
\newblock In \emph{2013 28th IEEE/ACM International Conference on Automated Software Engineering (ASE)}, pages 367--377. IEEE, 2013.

\bibitem[Singh and Masuku(2014)]{singh2014sampling}
Ajay~S Singh and Micah~B Masuku.
\newblock Sampling techniques \& determination of sample size in applied statistics research: An overview.
\newblock \emph{International Journal of economics, commerce and management}, 2\penalty0 (11):\penalty0 1--22, 2014.

\bibitem[Stalnaker et~al.(2025)Stalnaker, Wintersgill, Chaparro, Heymann, Di~Penta, German, and Poshyvanyk]{stalnaker2025ml}
Trevor Stalnaker, Nathan Wintersgill, Oscar Chaparro, Laura~A Heymann, Massimiliano Di~Penta, Daniel~M German, and Denys Poshyvanyk.
\newblock The ml supply chain in the era of software 2.0: Lessons learned from hugging face.
\newblock \emph{arXiv preprint arXiv:2502.04484}, 2025.

\bibitem[Stray and Moe(2020)]{stray2020understanding}
Viktoria Stray and Nils~Brede Moe.
\newblock Understanding coordination in global software engineering: A mixed-methods study on the use of meetings and slack.
\newblock \emph{Journal of Systems and Software}, 170:\penalty0 110717, 2020.

\bibitem[Strode and Huff(2015)]{strode2015coordination}
Diane~E Strode and Sid~L Huff.
\newblock A coordination perspective on agile software development.
\newblock In \emph{Modern Techniques for Successful IT Project Management}, pages 64--96. IGI Global Scientific Publishing, 2015.

\bibitem[Strode et~al.(2012)Strode, Huff, Hope, and Link]{strode2012coordination}
Diane~E Strode, Sid~L Huff, Beverley Hope, and Sebastian Link.
\newblock Coordination in co-located agile software development projects.
\newblock \emph{Journal of Systems and Software}, 85\penalty0 (6):\penalty0 1222--1238, 2012.

\bibitem[Stulova et~al.(2020)Stulova, Blasi, Gorla, and Nierstrasz]{stulova2020towards}
Nataliia Stulova, Arianna Blasi, Alessandra Gorla, and Oscar Nierstrasz.
\newblock Towards detecting inconsistent comments in java source code automatically.
\newblock In \emph{2020 IEEE 20th international working conference on source code analysis and manipulation (SCAM)}, pages 65--69. IEEE, 2020.

\bibitem[Suali et~al.(2017)Suali, Fauzi, Sobri, and Nasir]{suali2017developers}
AJ~Suali, SSM Fauzi, WAWM Sobri, and MHNM Nasir.
\newblock Developers' coordination issues and its impact on software quality: A systematic review.
\newblock In \emph{2017 3rd International Conference on Science in Information Technology (ICSITech)}, pages 659--663. IEEE, 2017.

\bibitem[Talukder et~al.(2017)Talukder, Senapathi, and Buchan]{talukder2017coordination}
ABM Talukder, Mali Senapathi, and Jim Buchan.
\newblock Coordination in distributed agile software development: a systematic review.
\newblock 2017.

\bibitem[Tian et~al.(2022)Tian, Zhang, Stol, Jiang, and Liu]{tian2022makes}
Yingchen Tian, Yuxia Zhang, Klaas-Jan Stol, Lin Jiang, and Hui Liu.
\newblock What makes a good commit message?
\newblock In \emph{Proceedings of the 44th International Conference on Software Engineering}, pages 2389--2401, 2022.

\bibitem[Vieira et~al.(2010)Vieira, Kaymak, and Sousa]{vieira2010cohen}
Susana~M Vieira, Uzay Kaymak, and Jo{\~a}o~MC Sousa.
\newblock Cohen's kappa coefficient as a performance measure for feature selection.
\newblock In \emph{International conference on fuzzy systems}, pages 1--8. IEEE, 2010.

\bibitem[Wang et~al.(2022)Wang, Li, Wu, Hovy, and Sun]{wang2022pre}
Haifeng Wang, Jiwei Li, Hua Wu, Eduard Hovy, and Yu~Sun.
\newblock Pre-trained language models and their applications.
\newblock \emph{Engineering}, 2022.

\bibitem[Wang et~al.(2025)Wang, Zhao, Hou, and Wang]{wang2025large}
Shenao Wang, Yanjie Zhao, Xinyi Hou, and Haoyu Wang.
\newblock Large language model supply chain: A research agenda.
\newblock \emph{ACM Transactions on Software Engineering and Methodology}, 34\penalty0 (5):\penalty0 1--46, 2025.

\bibitem[Wood and Wood(2008)]{wood2008card}
Jed~R Wood and Larry~E Wood.
\newblock Card sorting: current practices and beyond.
\newblock \emph{Journal of Usability Studies}, 4\penalty0 (1):\penalty0 1--6, 2008.

\bibitem[Yan et~al.(2016)Yan, Fu, Zhang, Yang, Xu, and Kymer]{yan2016automatically}
Meng Yan, Ying Fu, Xiaohong Zhang, Dan Yang, Ling Xu, and Jeffrey~D Kymer.
\newblock Automatically classifying software changes via discriminative topic model: Supporting multi-category and cross-project.
\newblock \emph{Journal of Systems and Software}, 113:\penalty0 296--308, 2016.

\bibitem[Yazici and Yolacan(2007)]{yazici2007comparison}
Berna Yazici and Senay Yolacan.
\newblock A comparison of various tests of normality.
\newblock \emph{Journal of statistical computation and simulation}, 77\penalty0 (2):\penalty0 175--183, 2007.

\bibitem[Zar(1972)]{zar1972significance}
Jerrold~H Zar.
\newblock Significance testing of the spearman rank correlation coefficient.
\newblock \emph{Journal of the American Statistical Association}, 67\penalty0 (339):\penalty0 578--580, 1972.

\bibitem[Zhao et~al.(2023)Zhao, Zhou, Li, Tang, Wang, Hou, Min, Zhang, Zhang, Dong, et~al.]{zhao2023survey}
Wayne~Xin Zhao, Kun Zhou, Junyi Li, Tianyi Tang, Xiaolei Wang, Yupeng Hou, Yingqian Min, Beichen Zhang, Junjie Zhang, Zican Dong, et~al.
\newblock A survey of large language models.
\newblock \emph{arXiv preprint arXiv:2303.18223}, 2023.

\bibitem[Zhao et~al.(2024)Zhao, Chen, Bangash, Adams, and Hassan]{zhao2024empirical}
Zhimin Zhao, Yihao Chen, Abdul~Ali Bangash, Bram Adams, and Ahmed~E Hassan.
\newblock An empirical study of challenges in machine learning asset management.
\newblock \emph{Empirical Software Engineering}, 29\penalty0 (4):\penalty0 98, 2024.

\end{thebibliography}

\newpage
\appendix
\renewcommand{\thesection}{Appendix~\Alph{section}} 

\section{Prompt structure used for the commit activities classification of PTLM families.}
\begin{figure}[ht]
\centering
\begin{minipage}{0.95\textwidth}
\begin{lstlisting}
You are a helpful assistant tasked with classifying commit messages. 
Classify each commit message with one of the below classifications and return the
results as a JSON array of objects, where each object has two keys: "commit" (the
commit message) and "label" (the classification label).
Note: Don't skip any commit without labeling, even if there are multiple 
occurrences.

Categories:
- External Documentation: Changes to end-user documentation (e.g., first draft of model card)
- Internal Documentation: Changes explaining code workings internally (e.g., update merges.txt)
- Model Structure: Structural changes to the model's code (e.g., add new sentencetransformer model)
- Training Infrastructure: Changes affecting the model training logic (e.g., iter_0600000)
- Preprocessing: Changes related to data manipulation before it reaches model training (e.g., add preprocessing file)
- Parameter Tuning: Adjustments to hardcoded hyperparameters within the ML pipeline (e.g., update config for hf transformers)
- Pipeline Performance: Modifications enhancing ML run-time pipeline efficiency (e.g., fix rows mixing)
- Validation Infrastructure: Modifications to components evaluating model performance (e.g., update benchmark comparisons to add openchat and jackalope)
- Input Data: Changes to logic for loading or ingesting external data (e.g., correct pre-training data)
- Output Data: Modifications to how output data is stored
- Sharing: Changes that enable better collaboration (e.g., from git://github.com/01-ai/yi.git/commit/043a50fe5a24ce5fbd31d171ae9aeb4d2600accf)
- Project Metadata: Changes to metadata about the data used by the ML pipeline (e.g., initial commit)
- Add Dependency: Introduction of a new dependency (e.g., allow flax)
- Remove Dependency: Removal of an existing dependency (e.g., delete flax)
- Update Dependency: Updates to the metadata of an existing dependency (e.g., update checkpoint for transformers>=4.29 (#4))

Please classify the following commit messages:
\end{lstlisting}
\label{prompt}
\end{minipage}
\end{figure}

\newpage
\section{Heuristics for synchronization pattern Labeling} \label{algorithms_appendix}
\subsection{Algorithm for detecting activity lead between GH and HF}\label{activity-lead}
\begin{algorithm}[ht]
\caption{Detecting Activity Lead Across Upstream (GH) and Downstream (HF)}
\begin{algorithmic}[1]
\State \textbf{Input:} Commit data with timestamps and source labels
\State \textbf{Output:} Activity lead label: \texttt{Upstream First}, \texttt{Downstream First}, \texttt{Simultaneous Activities}, or \texttt{No Data}

\State Remove rows with missing commit dates
\If{no valid commit dates remain}
    \State \Return \texttt{No Data}
\EndIf

\State Let \texttt{min\_date} be the earliest commit date
\State Assign each commit to a biweekly period:
\[
\texttt{biweek} = \left\lfloor \frac{(\texttt{date} - \texttt{min\_date}).\texttt{days}}{14} \right\rfloor + 1
\]

\State Construct a table that maps commit presence to each platform for every biweekly period

\For{each biweekly period in ascending order}
    \State Let \texttt{gh\_activity} = True if \texttt{GH Commit} has activity
    \State Let \texttt{PTLM\_project\_activity} = True if any PTLM\_project has activity

    \If{\texttt{gh\_activity} and \texttt{PTLM\_project\_activity}}
        \State \Return \texttt{Simultaneous Activity}
    \ElsIf{\texttt{gh\_activity} and not \texttt{PTLM\_project\_activity}}
        \State \Return \texttt{GH First}
    \ElsIf{not \texttt{gh\_activity} and \texttt{PTLM\_project\_activity}}
        \State \Return \texttt{HF First}
    \EndIf
\EndFor

\State \Return \texttt{No Data}
\end{algorithmic}
\end{algorithm}

\newpage
\subsection{Algorithm for detecting overlapping of commit activity between GH and HF}\label{commit_overlap}
\begin{algorithm}[ht]
\caption{Categorizing commit overlap between GH and HF PTLMs}
\begin{algorithmic}[1]
\State \textbf{Input:} Commit data with timestamps and model names
\State \textbf{Output:} Overlap type: \texttt{CC}, \texttt{PC}, \texttt{VC}, \texttt{AS}, or \texttt{No Data}

\State Remove rows with missing commit dates
\If{no valid commit dates remain}
    \State \Return \texttt{No Data}
\EndIf

\State Let \texttt{min\_date} be the earliest commit date
\State Assign each commit to a biweekly period:
\[
\texttt{biweek} = \left\lfloor \frac{(\texttt{date} - \texttt{min\_date}).\texttt{days}}{14} \right\rfloor + 1
\]

\State Identify periods with GH activity: \texttt{github\_periods}
\State Identify all unique PTLM names

\State Set \texttt{all\_PTLM\_project\_overlap\_with\_GH} to true
\For{each PTLM}
    \State Get \texttt{timeline} for that PTLM
    \If{\texttt{timeline} not fully within \texttt{github\_periods}}
        \State Set \texttt{all\_PTLM\_project\_overlap\_with\_GH} to false
        \State \textbf{break}
    \EndIf
\EndFor
\If{\texttt{all\_PTLM\_project\_overlap\_with\_GH} is true}
    \State \Return \texttt{CC}
\EndIf

\State Identify all PTLM periods
\State Compute \texttt{overlap\_periods} between PTLM and GH periods
\If{\texttt{overlap\_periods} is not empty}
    \State \Return \texttt{PC}
\EndIf

\State For each biweekly period, count unique PTLMs active
\State Identify periods with at least 3 unique PTLMs
\If{there are 3 or more such periods and none of them overlap with GH periods}
    \State \Return \texttt{VC}
\EndIf

\State \Return \texttt{AS}
\end{algorithmic}
\end{algorithm}

\newpage
\subsection{Algorithm for detecting the intensity of commit activity between GH and HF}\label{commit_intensity}
\begin{algorithm}[ht]
\caption{Categorizing intensity of coordination between GH and HF PTLM\_project}
\begin{algorithmic}[1]
\State \textbf{Input:} Commit data with timestamps and PTLM\_project names
\State \textbf{Output:} Intensity label: \texttt{R}, \texttt{F}, \texttt{S}, or \texttt{No Data}

\State Remove rows with missing commit dates
\If{no valid commit dates remain}
    \State \Return \texttt{No Data}
\EndIf

\State Let \texttt{min\_date} be the earliest commit date
\State Assign each commit to a biweekly period:
\[
\texttt{biweek} = \left\lfloor \frac{(\texttt{date} - \texttt{min\_date}).\texttt{days}}{14} \right\rfloor+1
\]

\State Identify periods with GH activity: \texttt{github\_periods}
\State Identify all unique PTLM\_project names: \texttt{PTLM\_project}

\For{each \texttt{PTLM\_project} in \texttt{PTLM\_projects}}
    \State Get \texttt{timeline\_periods} for that PTLM\_project
    \If{the number of \texttt{timeline\_periods} is greater than 3}
        \State \textbf{break}
    \EndIf
\EndFor

\If{all PTLM\_project have activity within 3 or fewer biweekly periods}
    \State \Return \texttt{R}
\EndIf

\For{each \texttt{PTLM\_project} in \texttt{PTLM\_projects}}
    \State Get \texttt{timeline\_periods} for that PTLM\_project
    \State Compute \texttt{overlap\_periods} = intersection of \texttt{PTLM\_project\_periods} and \texttt{github\_periods}
    \If{\texttt{overlap\_periods} is empty}
        \State \textbf{continue}
    \EndIf

    \State Initialize \texttt{consecutive\_count} = 1
    \State Set \texttt{last\_overlap} = first element in \texttt{overlap\_periods}

    \For{each \texttt{current\_overlap} in \texttt{overlap\_periods} starting from second}
        \If{\texttt{current\_overlap} == \texttt{last\_overlap} + 1}
            \State Increment \texttt{consecutive\_count}
        \Else
            \State Reset \texttt{consecutive\_count} to 1
        \EndIf
        \If{\texttt{consecutive\_count} $\geq$ 5}
            \State \Return \texttt{F}
        \EndIf
        \State Update \texttt{last\_overlap} = \texttt{current\_overlap}
    \EndFor
\EndFor

\State \Return \texttt{S}
\end{algorithmic}
\end{algorithm}

\end{document}